\documentclass[10pt]{article} 
\RequirePackage{lmodern}
\RequirePackage{anyfontsize}

\usepackage{amsmath,amsthm,amssymb}
\usepackage{xcolor}
\usepackage{algpseudocode, algorithm}
\usepackage[top=2cm, bottom=2cm, left=2.5cm, right=2.5cm]{geometry}
\usepackage{url}
\usepackage{booktabs}
\usepackage[colorlinks=true,bookmarks=false,citecolor=blue,urlcolor=blue]{hyperref} 
\usepackage{comment}
\usepackage{caption}
\usepackage{float}
\usepackage{subcaption}
\usepackage{graphicx}
\usepackage{enumitem}
\usepackage[backend=biber, maxbibnames=99]{biblatex}
\usepackage{mathtools}
\usepackage{authblk}
\usepackage{bbm}
\usepackage{etoolbox}
\usepackage{multirow}
\usepackage{fancyhdr}
\usepackage[normalem]{ulem} 
\usepackage{multicol}

\newtheorem{theo}{Theorem}[section]
\newtheorem{prop}{Proposition}[section]
\newtheorem{corro}{Corollary}[section]
\newtheorem{rque}{Remark}[section]

\newtheorem{lemma}{Lemma}[section]
\newtheorem{assump}{Assumptions}[section]
\usepackage[most]{tcolorbox}

\newtcolorbox{gs-remark}{
  colback=red!5,      
  colframe=red!75!black, 
  boxrule=0.8pt,      
  arc=4pt,            
  left=6pt,
  right=6pt,
  top=6pt,
  bottom=6pt
}

\newcommand{\ud}{\mathrm d}

\newcounter{numberassumption}
\renewcommand\thenumberassumption{(A-\arabic{numberassumption})}

\def \N{\mathbb{N}}
\def \R{\mathbb{R}}

\def \E{\mathbb{E}}

\def \Fc{{\cal F}}

\def \N{\mathbb{N}}
\def \R{\mathbb{R}}

\def \E{\mathbb{E}}

\def \Fc{{\cal F}}

\newcommand{\PP}{\mathbb{P}}

\newcommand{\ZZ}{\mathbb{Z}}

\ifdefined\revclean
\DeclareRobustCommand{\red}[1]{\ifhmode\unskip\fi}

\else
\DeclareRobustCommand{\red}[1]{{\color{red}#1}}

\fi

\author[$\sharp$$\flat$]{\normalsize Yadh Hafsi}
\author[$\ddagger$]{\normalsize Fabrizio Lillo} 
\author[$\star$]{\normalsize Grégoire Szymanski}
\affil[$\sharp$]{CMAP, École Polytechnique, Paris, France}
\affil[$\flat$]{Deepfin Research, London, UK}
\affil[$\ddagger$]{Scuola Normale Superiore, Pisa, Italy}
\affil[$\star$]{DMATH, Université du Luxembourg, Belval Esch-sur-Alzette, Luxembourg}

\def\1{\mbox{1\hspace{-0.25em}l}}

\def\beqs{\begin{eqnarray*}}
\def\enqs{\end{eqnarray*}}
\def\beq{\begin{eqnarray}}
\def\enq{\end{eqnarray}}

\def\1{\mbox{1\hspace{-0.25em}l}}

\def\beqs{\begin{eqnarray*}}
\def\enqs{\end{eqnarray*}}
\def\beq{\begin{eqnarray}}
\def\enq{\end{eqnarray}}

\title{Calibration of Passive Market Impact with a Queue-Sensitive Propagator Model \vspace{0.31cm}}

\begin{document}

\maketitle

\begin{abstract}
We develop a framework for calibrating passive market impact, defined as the price response to liquidity-providing limit orders, in contrast to the impact of liquidity-consuming market orders.
We build on the model of \textcite{ouazzaniChahdiRosenbaumSzymanskiPassive}, which combines Hawkes-driven market-order flow with queue-reactive limit-order and cancellation dynamics. Within this model, we show that the price decomposes into a universal propagator term driven by signed order-flow imbalance and a state-dependent correction that captures the dependence of price sensitivity on volume at the best quotes. This decomposition links limit-order-book microstructure to standard propagator models and yields a regression on two observable signals: the propagator signal and the queue-weighted order-flow imbalance. We calibrate the model on high-frequency order-book data and find that the passive-impact term explains substantial variation in prices beyond the classical propagator, improving both model fit and the accuracy of predicted impact curves.
  
\end{abstract}\hspace{10pt}

\section{Introduction}
\label{sec:introduction}
Market impact is the effect of trading activity on prices. It is both a key component of transaction costs and a central mechanism of price formation, describing how order flow and liquidity are incorporated into prices at high frequency. Most of the literature studies the impact of aggressive orders, namely market orders that consume liquidity at the best quotes. A standard representation is the propagator framework, in which the mid-price is written as a linear response to past signed order flow \cite{bouchaud2003fluctuations, jaisson2015market}. In its simplest form,
$$
   P_t = P_0 + \kappa \int_0^t \xi(t-s)\,\mathrm{d}D_s + M_t,
$$
for all $t \geq 0$,
where $N^a$ and $N^b$ count buy and sell market orders, $D := N^a - N^b$ is the cumulative buy-sell imbalance, $\xi$ is a decay kernel, $\kappa$ measures impact per unit trade, and the noise term $M$ is a martingale with $M_0=0$ whose increments are uncorrelated with the order flow. In this representation, each trade has an immediate effect on the price, while the kernel $\xi$ describes how this effect evolves over time. This framework has been widely used both in empirical work and in models of optimal execution. More elaborate impact models, including extensions that account for persistent order flow and liquidity dynamics, are reviewed in \textcite{bouchaud2009markets}. \\

A natural microstructural foundation for propagator models emerges when market orders are modeled by Hawkes processes \cite{hawkes1971spectra, bacry2015hawkes, bacry2015market}. In that case, the propagator kernel can be explicitly linked to the excitation kernel of order flow \cite{jaisson2015market}. This explains the emergence of several well-known empirical features, including persistence in signed order flow and concave impact curves \cite{jaisson2015limit}. As a result, the Hawkes-propagator framework has become a standard benchmark for aggressive market impact models. In this setting, \cite{jaisson2015market} shows that the price can be seen as the expected long-run signed order-flow imbalance given the information available at time $t$,
\begin{equation}
    \label{prop_expt}
P_t=P_0+\kappa\lim_{T\to\infty}\mathbb E[D_T\mid\mathcal F_t],
\end{equation}
and Section~\ref{subsec:classical-propagator} recalls how this definition yields the propagator representation.\\

\begin{figure}[htb!]

  \centering
  \includegraphics[width=0.7\textwidth]{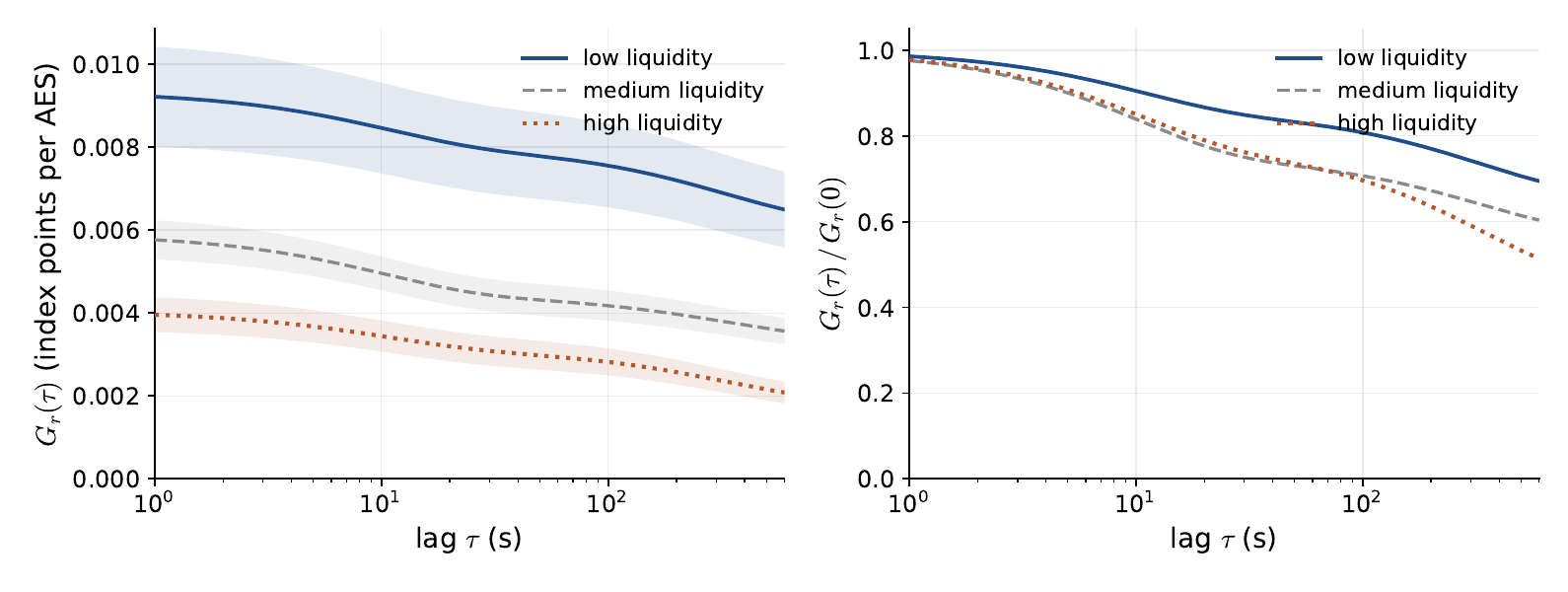}
  \caption{Left: response $G_r(\tau) = \kappa_0^r+\kappa_1^r\bar q_r
+\sum_{j=1}^{2}w_j^r e^{-\beta_j\tau}$ of the weighted mid to a unit aggressive trade in each liquidity regime (indicative bands), calibrated using the methodology of Section~\ref{sec:calibration}. Right: the same curves normalized by their initial value.}
  \label{fig:three-propagators}
\end{figure}

In this benchmark, every aggressive order has the same price sensitivity $\kappa$, independently of the state of the limit order book. Figure~\ref{fig:three-propagators} illustrates why this may be restrictive. We group the trading sessions of our E-mini S\&P 500 futures sample into three liquidity regimes according to the average volume available at the best quotes and plot, for each regime, the estimated price response to a unit aggressive trade (the construction is given in Section~\ref{sec:passive-impact-curves}). We observe that the response is larger when the book is thinner, which motivates quantifying how the price response to aggressive orders depends on available liquidity, and developing a price-formation model that reproduces this dependence.

A large share of trading activity is, however, passive rather than aggressive. In many execution settings, orders are submitted as limit orders and remain in the book pending execution. Limit orders do not consume liquidity directly. Instead, they modify the volume available at the best quotes and thereby affect the price response to subsequent market orders \cite{cetin2004liquidity,huang2015simulating,cont2014price,lehalle2017limit}. This liquidity-price-response link is also central to classical microstructure models \cite{kyle1985continuous}. This raises a natural question: can passive impact be studied within the same general framework as aggressive impact, or does it require a distinct model? The model of Ouazzani Chahdi, Rosenbaum, and Szymanski \cite{ouazzaniChahdiRosenbaumSzymanskiPassive} provides a tractable way to answer this question. Their framework combines Hawkes-driven market-order arrivals with queue-reactive dynamics for limit orders and cancellations. Its key feature is that instantaneous price sensitivity depends on the book state rather than being constant, through a mapping $\kappa:\mathcal Z\to\mathbb R_+$, where $\mathcal Z$ is the state space of the order-book variables determining price sensitivity. Thicker queues imply lower sensitivity. Within this model, the price is the infinite-horizon conditional expectation of future signed order flow, weighted by the pre-trade book state. A natural representation of that state uses the queue sizes at the best quotes, which are most relevant to the immediate price response to market orders. The corresponding state vectors are
$$Z_s^a
=
(q_{1,s}^a,\ldots,q_{\ell,s}^a,
 q_{1,s}^b,\ldots,q_{\ell,s}^b),
\qquad
Z_s^b
=
(q_{1,s}^b,\ldots,q_{\ell,s}^b,
 q_{1,s}^a,\ldots,q_{\ell,s}^a).$$
Here, $\ell \geq 1$ is the number of retained levels, and $q_{i,s}^a$ and $q_{i,s}^b$ denote the queue sizes at the $i$-th best ask and bid levels, respectively. In each vector, the queues on the side consumed by the incoming market order appear first. The same long-horizon construction as in the constant-sensitivity benchmark then yields 
\begin{equation}\label{eq:price_intro}
   P_t
   = P_0 + \lim_{T\to\infty}
     \mathbb{E}\left[
        \int_0^T \kappa(Z_{s-}^a)\,\mathrm{d}N_s^a
        -
        \int_0^T \kappa(Z_{s-}^b)\,\mathrm{d}N_s^b
        \;\Big|\;\mathcal{F}_t
     \right].
\end{equation}
This expression is derived precisely in \cite[Section~2]{ouazzaniChahdiRosenbaumSzymanskiPassive} as a generalization of \eqref{prop_expt}. The limit $T\to\infty$ incorporates the model-implied future order-flow response conditional on $\mathcal F_t$ and is the state-dependent analogue of the constant-sensitivity Hawkes limit above. The theory of Section~\ref{sec:model-passive-liquidity} uses only the best queue on the consumed side, and Section~\ref{sec:empirical-specification} considers richer book states. \\

This formulation captures a simple mechanism: incoming market orders have a smaller price effect when liquidity is abundant and a larger effect when liquidity is scarce. Passive orders matter because they alter the queue sizes governing this response. The main theoretical question then is how this queue-dependent model relates to the classical propagator representation. We show that, under mild conditions on the queue dynamics, the price admits a decomposition of the form
\begin{equation}\label{eq:decomp_intro}
   P_t = P_0
   + \bar\kappa \int_0^t \xi(t-s)\,\mathrm{d}D_s
   + \;\mathcal{R}_t,
\quad \forall t \geq 0,
\end{equation}
where $\bar\kappa$ is the stationary mean of the price-sensitivity function, $\xi$ is the same Hawkes-based propagator kernel as in the aggressive model, and $\mathcal{R}_t$ is a correction term that captures the interaction between queue fluctuations and order flow at time $t$. Passive impact is thus not a separate signed-flow term. The propagator term represents the average response to signed market-order imbalance under typical liquidity conditions, whereas $\mathcal R_t$ captures deviations from that benchmark generated by queue-state fluctuations. Theorem~\ref{thm:general_propagator} formalizes this interpretation. \\

Direct calibration is difficult because the price depends jointly on queue states, future Hawkes flow, and anticipated queue-weighted trades. The decomposition above separates the model into objects that can be estimated or constructed from high-frequency event data. In the one-level case $\ell=1$, using only the best ask and bid queues and no cross terms, an affine specification $\kappa(q)=\kappa_0+\kappa_1 q$ gives $P_t=P_0+\kappa_0\int_0^t\xi(t-s)\,\mathrm dD_s+\kappa_1(H_t+J_t)$ (Corollary~\ref{cor:main_propagator}), where
   $$
   H_t := \int_0^t q_{s-}^a\,\mathrm{d}N_s^a - \int_0^t q_{s-}^b\,\mathrm{d}N_s^b, \quad \forall t \geq 0,
   $$
is an observable queue-weighted order-flow imbalance, while $J_t$ is a conditional anticipation term. We first calibrate the Hawkes and queue-reactive dynamics, then regress prices on the resulting propagator and queue-weighted signals, treating the unobserved anticipation term as a residual or a control. This two-step procedure avoids direct calibration of the full state-dependent price functional.
Empirically, the queue-state channel adds statistically significant out-of-sample predictive content for price changes beyond the aggressive-flow propagator.

\paragraph{Related literature.} Early empirical work established that trades have persistent effects on prices and linked these effects to information and order flow \cite{kyle1985continuous, hasbrouck1991measuring}. The propagator approach later provided a tractable framework in which prices respond to past signed trades through a decaying kernel \cite{bouchaud2003fluctuations, bouchaud2009markets, Hafsi2026MarketImpact}. Extensions include no-dynamic-arbitrage restrictions and optimal execution under transient impact \cite{gatheral2010no, huberman2004price, almgren2001optimal, obizhaeva2013optimal, gatheral2012transient, chevalier2025liquidityuncertainty, barzykin2026passive}.  At the aggregate level, empirical work has also documented the concave shape of impact and its slow decay after execution \cite{lillo2003master, almgren2005direct, toth2011anomalous, brokmann2015slow, mastromatteo2014agent, bucci2020crossover}. The paper is also related to the literature that models high-frequency order flow with Hawkes processes \cite{hawkes1971spectra, bowsher2007modelling, bacry2015hawkes}. In this setting, \cite{jaisson2015market} derived the propagator kernel from the Hawkes excitation kernel, while \cite{jaisson2015limit} studied the nearly unstable regime and its scaling limits. Hawkes-based calibrations have also been used to measure the degree of endogeneity in financial markets and to extend the framework to marked order flow \cite{filimonov2012quantifying, hardiman2013critical, rambaldi2017role}. The paper further connects to stochastic models of the limit order book. Queue-based and state-dependent models describe how insertions, cancellations, and trades depend on the current state of the book \cite{smith2003statistical, cont2010stochastic, cont2013price, abergel2013mathematical}. In particular, the queue-reactive framework of \cite{huang2015simulating} provides an empirically tractable description of best-queue dynamics and is a natural basis for modeling how liquidity affects price sensitivity. The passive-impact model of \cite{ouazzaniChahdiRosenbaumSzymanskiPassive} combines this queue-reactive view of the book with Hawkes-driven order flow. Related work directly estimates the impact of market, limit, and cancellation events \cite{eisler2012price,hautsch2012market,cont2014price}. Here passive events act through state-dependent trade sensitivity rather than additive signed flows. Section~\ref{sec:empirical-specification} discusses the corresponding order-flow-imbalance (OFI) benchmark. 
We show that the propagator structure emerges within this setting and develop a calibration methodology for the resulting model.

\paragraph{Outline.} Section \ref{sec:model-passive-liquidity} introduces the price-formation model and derives the decomposition of prices into a classical propagator component and a passive-liquidity correction. Section \ref{sec:calibration} describes the calibration of the order-flow and liquidity dynamics from high-frequency order-book data: the Hawkes kernel and the queue-reactive insertion and cancellation intensities. Section \ref{sec:empirical-specification} translates the theoretical decomposition into price regressions and estimates the queue-dependent price-sensitivity coefficients. Section \ref{sec:passive-impact-curves} defines passive market impact counterfactually and computes impact curves by Monte Carlo simulation.
\label{subsec:outline}

\section{A Price-Formation Model with Passive Liquidity}
\label{sec:model-passive-liquidity}

This section introduces the price-formation model used throughout the paper. We work on a complete filtered probability space $(\Omega,\Fc,(\Fc_t)_{t\ge0},\PP)$ satisfying the usual conditions and supporting all random variables considered below. The filtration represents the information available to the market, and event sizes are normalized to one throughout this section. 

\subsection{Linear Propagator Model}
\label{subsec:classical-propagator}

We first recall the constant-sensitivity benchmark. It is the special case in which every aggressive order has the same permanent price sensitivity, independently of the limit-order-book state. The construction follows the martingale price-formation argument of \cite{jaisson2015market}. Let $N^a$ and $N^b$ denote, respectively, buy market orders consuming the ask queue and sell market orders consuming the bid queue, and set $D:=N^a-N^b$. Under linear permanent impact, a terminal imbalance $D_T$ has a long-run contribution $\kappa D_T$, with $\kappa>0$ measured in price units per market order. If the price is the conditional expectation of long-run value, the \textit{efficient price} is
\begin{equation}
P_t = P_0 + \kappa\,\lim_{T\to\infty}\E\left[N_T^a-N_T^b\mid\Fc_t\right],
\quad \forall t \geq 0,
\label{eq:trad_price_def}
\end{equation}
where $N^a$ and $N^b$ count buy and sell market orders and $\kappa$ is permanent impact per trade, see \cite{bouchaud2003fluctuations,jaisson2015market}. The limit in \eqref{eq:trad_price_def} requires a law for future order flow. We interpret $\Fc_t$ as the current market information and the conditional expectation as the expected future order-flow imbalance under that law. The Hawkes model below supplies this conditional law. In the empirical part, stationarity is imposed only locally on short intraday windows rather than globally over the trading day.\\

To make \eqref{eq:trad_price_def} explicit, we model $N^a$ and $N^b$ as independent Hawkes processes with common baseline $\mu\ge0$ and excitation kernel $\phi\ge0$. Thus, for all $t \geq 0$ and $i\in\{a,b\}$, we write the intensity of $N^i$ as
\begin{equation}
\lambda_t^i = \mu + \int_0^{t-}\phi(t-u)\,\ud N_u^i.
\label{eq:trad_hawkes}
\end{equation}
We impose the standard subcriticality condition
\begin{equation}
\|\phi\|_{L^1}<1,
\label{eq:hawkes_stability}
\end{equation}
which guarantees non-explosion and finite mean intensity for the Hawkes process \cite{hawkes1971spectra,bacry2015hawkes}. The limit \eqref{eq:trad_price_def} can be computed explicitly, see \cite[Proposition 3.2]{jaisson2015market}, and yields the continuous-time propagator representation
\begin{equation}
P_t = P_0 + \kappa\int_0^t \xi(t-s) \,\ud(N_s^a-N_s^b),
\quad \forall t \geq 0,
\label{eq:trad_propagator}
\end{equation}
where
\begin{equation}
\xi(t)=1+\Big(1+\int_0^\infty \psi(u)\,\ud u\Big)\int_t^\infty \phi(u)\,\ud u,
\qquad
\psi=\sum_{n\ge1}\phi^{*n},
\quad \forall t \geq 0.
\label{eq:trad_xi}
\end{equation}
Writing $\eta:=\|\phi\|_{L^1}$, the stability condition \eqref{eq:hawkes_stability} gives $\int_0^\infty\psi(u)\,\ud u=\eta/(1-\eta)<\infty$. It follows that $\xi(t)\geq 1$, $\xi(t)\to 1$ as $t\to\infty$, and $\xi$ is absolutely continuous, with
$$
\xi'(t)=-(1-\eta)^{-1}\phi(t)
$$
for almost every $t\geq 0$. 
Since $\phi\ge0$, we have $\xi'\le0$ almost everywhere, so $\xi$ is non-increasing.
This identity expresses the infinitesimal martingale-balance condition: the deterministic decay of the impact of past trades exactly offsets the expected same-sided future order flow induced by Hawkes self-excitation. In \eqref{eq:trad_propagator}, a buy trade at time $s$ produces an immediate jump $+\kappa\xi(0)$, and a sell trade produces the symmetric jump $-\kappa\xi(0)$. The contribution of either trade at a later lag $u$ is $\pm\kappa\xi(u)$, so the permanent contribution is $\pm\kappa$. The transient excess $\kappa(\xi(u)-1)$ is entirely determined by the tail of the Hawkes excitation kernel. \\

The preceding formulas also clarify why the near-critical Hawkes regime is relevant for market impact. When $\eta$ is close to one, the multiplier $(1-\eta)^{-1}$ amplifies the transient part of $\xi$. If, in addition, the tail of $\phi$ is regularly varying, then the rescaled response kernels appearing in the nearly unstable limit have power-law singularities \cite{jaisson2015limit,jaisson2015market}. For a deterministic buy metaorder executed aggressively at rate $g$, the price displacement $\mathrm{MI}^{\mathrm{MO}}$ generated by this additional signed flow then satisfies the heuristic convolution \cite{jaisson2015market,ouazzaniChahdiRosenbaumSzymanskiPassive}
\begin{equation*}
\mathrm{MI}_t^{\mathrm{MO}} \simeq
\kappa\int_0^t\bigl(1+C_\alpha (t-s)^{-\alpha}\bigr)g(s)\,\ud s,
\quad \alpha\in(0,1),\ \forall t \geq 0,
\end{equation*}
with a positive constant $C_\alpha$ determined by the Hawkes parameters. This is the standard mechanism generating concave transient impact. The calibration below does not rely on a scaling-limit estimate. The same nearly unstable Hawkes asymptotics generate rough fractional limits for order-flow intensities \cite{jaisson2016rough,gatheral2018volatility}, and the resulting short-time singularity is analogous to the one appearing in rough-volatility asymptotics.\\

In numerical work, one often replaces a general kernel by a finite sum of exponentials,
\begin{equation*}
    \phi(t)=\sum_{j=1}^m c_j e^{-\beta_j t},\qquad c_j,\beta_j>0.
\end{equation*}
Setting $X_t^{j}=\int_0^{t-}e^{-\beta_j(t-s)}\,\ud N_s$ gives
\begin{equation*}
\lambda_t=\mu+\sum_{j=1}^m c_jX_t^j,
\qquad
\ud X_t^j=-\beta_jX_t^j\,\ud t+\ud N_t,
\quad \forall t \geq 0.
\end{equation*}
This representation offers two practical advantages. First, each decay rate $\beta_j$ identifies a characteristic timescale $\beta_j^{-1}$ of order-flow excitation. Second, the state $(X^1,\ldots,X^m)$ is finite-dimensional and Markovian, which makes simulation, likelihood evaluation, and filtering more tractable. It is the specification used in Section~\ref{subsec:hawkes-calibration}.  Heavy-tailed kernels can be approximated over finite horizons by finite exponential mixtures, and such Markovian lifts are standard for Volterra and rough-volatility models \cite{abi2019multifactor,cuchiero2019markovian,callegaro2021fast,beylkin2010approximation}. For this class, the computations remain explicit: the resolvent $\psi$ is a finite exponential-polynomial combination (Appendix~\ref{app:sum-of-exponential-kernels}), and Lemma~\ref{lem:xi} gives directly
$$
\xi(t)=1+\frac{1}{1-\sum_{j=1}^m c_j/\beta_j}
\sum_{j=1}^m\frac{c_j}{\beta_j}e^{-\beta_j t},
\quad \forall t \geq 0,
$$
provided $\sum_{j=1}^m c_j/\beta_j<1$. Thus the same fitted Hawkes kernel determines the propagator signal used later in the price regression. The next subsection keeps this aggressive-order structure fixed and replaces the constant sensitivity $\kappa$ by a queue-dependent function $\kappa(q)$.

\begin{rque}\label{rem:cross-excitation}
The independence of $N^a$ and $N^b$ is imposed only  to keep the propagator calculation one-dimensional. It is not empirically realistic. Figure~\ref{fig:xcorr-flow-front} shows strong cross-dependence: correlations are $0.76$ contemporaneously, $0.36$ at one second, and $0.15$ at one minute, and the signed flow $N^a-N^b$ has positive autocorrelation, although a cross-correlogram alone does not identify a Hawkes cross-kernel.

The derivation is not fundamentally tied to independence. Consider instead the symmetric bivariate Hawkes model
\[
\begin{aligned}
\lambda_t^a
&=\mu+\int_0^{t-}\phi_s(t-u)\,\ud N_u^a
      +\int_0^{t-}\phi_c(t-u)\,\ud N_u^b,\\
\lambda_t^b
&=\mu+\int_0^{t-}\phi_c(t-u)\,\ud N_u^a
      +\int_0^{t-}\phi_s(t-u)\,\ud N_u^b,
\end{aligned}
\]
where $\phi_s,\phi_c\ge0$ are integrable self- and cross-excitation kernels. Writing $\eta_s:=\int_0^\infty\phi_s(u)\,\ud u$ and $\eta_c:=\int_0^\infty\phi_c(u)\,\ud u$, the standard multivariate Hawkes stability condition is that the spectral radius of the integrated kernel matrix be smaller than one. In this symmetric nonnegative case its eigenvalues are $\eta_s+\eta_c$ and $\eta_s-\eta_c$, so the stability condition is $\eta_s+\eta_c<1$. For the signed mode $D=N^a-N^b$, the relevant kernel is $h:=\phi_s-\phi_c$. Indeed, $\lambda^a-\lambda^b$ is obtained by convolving past signed flow with $h$. Moreover, $\|h\|_{L^1}\le\eta_s+\eta_c<1$, so the signed resolvent
\[
\psi_h:=\sum_{n\ge1}h^{*n}
\]
converges in $L^1$. Repeating the martingale argument in this signed mode gives the same propagator representation as in \eqref{eq:trad_propagator},
\[
P_t=P_0+\kappa\int_0^t\xi_{\mathrm{bi}}(t-s)\,\ud(N_s^a-N_s^b),
\]
but with
\[
\xi_{\mathrm{bi}}(t)
=
1+\left(1+\int_0^\infty\psi_h(u)\,\ud u\right)
\int_t^\infty h(u)\,\ud u =
1+\frac{1}{1-\eta_h}
\int_t^\infty\bigl(\phi_s(u)-\phi_c(u)\bigr)\,\ud u,
\]
where $\eta_h:=\int_0^\infty h(u)\,\ud u=\eta_s-\eta_c$ satisfies $|\eta_h|<1$. Unlike the independent-case kernel in \eqref{eq:trad_xi}, $\xi_{\mathrm{bi}}$ need not be monotone because $\phi_s-\phi_c$ need not be nonnegative pointwise. Thus cross-excitation changes the kernel governing the persistence of signed flow from $\phi$ to the antisymmetric mode $\phi_s-\phi_c$, but it does not destroy the propagator representation \cite{jaisson2015market}.

The independence assumption does not affect the subsequent price-regression strategy. If the structural Hawkes restrictions are imposed, the relevant propagator kernel is derived from the corresponding fitted Hawkes kernel or kernels. In the reduced-form empirical specification of Section~\ref{sec:empirical-specification}, the propagator signal is instead calibrated directly through the exponential dictionary, without imposing independence or ignoring the cross-dependence in Figure~\ref{fig:xcorr-flow-front}. We do not estimate a full bivariate Hawkes model.
\end{rque}

\subsection{Incorporating the Limit Order Book State}
\label{subsec:queue-dependent-price-sensitivity}
The model retains Hawkes aggressive flow but lets price sensitivity depend on best-queue volume. This preserves the intuition of \cite{jaisson2015market,bouchaud2003fluctuations} and adds the queue-reactive channel needed for passive execution \cite{huang2015simulating,ouazzaniChahdiRosenbaumSzymanskiPassive}. Aggressive flow remains as in \eqref{eq:trad_hawkes}. Only passive book dynamics are added. \\

Let $q^a$ and $q^b$ denote the aggregated volumes at the best ask and best bid. For all $t\ge 0$,
\begin{equation}
q_t^a=q_0^a+L_t^a-C_t^a-N_t^a,
~~\text{and}~~
q_t^b=q_0^b+L_t^b-C_t^b-N_t^b.
\label{eq:queues_def}
\end{equation}
The limit-order and cancellation flows are inhomogeneous Poisson processes with queue-reactive intensities given, for all $t \geq 0$, by
\begin{align}
\lambda_t^{L,a}=\lambda^L(q_{t-}^a),
\qquad
\lambda_t^{L,b}=\lambda^L(q_{t-}^b),
\qquad
\lambda_t^{C,a}=\lambda^C(q_{t-}^a),
\qquad
\lambda_t^{C,b}=\lambda^C(q_{t-}^b).
\label{eq:limit_cancel_intensities}
\end{align}
A pathwise construction is obtained from independent Poisson random measures $\pi^{N,a}$, $\pi^{N,b}$, $\pi^{L,a}$, $\pi^{L,b}$, $\pi^{C,a}$, and $\pi^{C,b}$ on $\R_+\times\R_+$. For all $t\ge 0$ and $x\in\{a,b\}$,
\begin{equation}
\begin{aligned}
N_t^{x}&=\int_0^t\int_0^\infty \mathbbm{1}_{\{z\le \lambda_s^{x}\}}\,\pi^{N,x}(\ud s,\ud z),\\
L_t^{x}&=\int_0^t\int_0^\infty \mathbbm{1}_{\{z\le \lambda_s^{L,x}\}}\,\pi^{L,x}(\ud s,\ud z),\\
C_t^{x}&=\int_0^t\int_0^\infty \mathbbm{1}_{\{z\le \lambda_s^{C,x}\}}\,\pi^{C,x}(\ud s,\ud z).
\end{aligned}
\label{eq:prm_construction}
\end{equation}
Passive liquidity thus enters through the book state, not through a direct price jump at submission: posted liquidity changes the queue faced by later market orders and hence their price response. 

\begin{rque}
    This formulation abstracts from direct quote changes caused by passive events themselves, such as limit orders placed inside the spread or cancellations that remove the last quantity at the best quote. We isolate the channel through which passive events, conditional on quotes, affect the price via the book state faced by future market orders. This approximation is particularly natural for large-tick assets, where the spread is often one tick and many passive events modify queue sizes without immediately changing the mid-price. In our sample, the spread is one tick for $97\%$ of trade events. We measure it on trades because message-level time weighting is not retained. 
Direct quote-revision effects are treated empirically as part of the regression residual or as a robustness issue.
\end{rque}

The price is defined, for all $t \geq 0$, as the infinite-horizon conditional expectation of weighted future order-flow imbalance
\begin{equation}
P_t
=
P_0+\lim_{T\to\infty}\E\left[
\int_0^T \kappa(q_{s-}^a)\,\ud N_s^a-\int_0^T \kappa(q_{s-}^{b})\,\ud N_s^b
\ \Big|\ \Fc_t
\right].
\label{eq:price_def}
\end{equation}
The function $\kappa$ is a primitive of the price-formation model, alongside $(\mu,\phi)$ and $(\lambda^L,\lambda^C)$: the intensities determine the law of order flow and the book, whereas $\kappa$ determines the price contribution of each trade conditional on the book state. It is not derived from the intensities and cannot be identified from event data alone. When $\kappa$ is constant, \eqref{eq:price_def} reduces to the classical definition \eqref{eq:trad_price_def}, so a queue-dependent $\kappa$ is the minimal extension through which the book state enters price formation. We calibrate $(\mu,\phi,\lambda^L,\lambda^C)$ from events and $\kappa$ from prices.\\

Here $\kappa(q)$ is the price contribution assigned to a market order when the predictable queue size on the consumed side is $q$. The economically relevant case is a non-increasing $\kappa$, because a thicker queue should attenuate the price response to a trade of fixed normalized size. The boundedness imposed below is a sufficient condition for the infinite-horizon conditional expectation in \eqref{eq:price_def} to be well defined. Since an affine $\kappa$ is unbounded, the affine specification of Section~\ref{sec:empirical-specification} is interpreted only locally, on the observed queue range.

\begin{assump}\label{suficient_assump}
A sufficient set of assumptions, adapted from \cite{ouazzaniChahdiRosenbaumSzymanskiPassive}, is the following:
\begin{itemize}
\item $\lambda^L:\ZZ\to\R_+$ is non-increasing and $\lambda^C:\ZZ\to\R_+$ is non-decreasing on the integer queue lattice (or on the corresponding integer translate determined by the initial queue),
\item $\kappa:\R\to\R_+$ is bounded and non-increasing,
\item The mixing condition:
\end{itemize}
\begin{equation}
\mathfrak m
:=\inf_{k\in\N^ *}\inf_{q\in\ZZ}\Big\{\lambda^L(q)-\lambda^L(q+k)+\lambda^C(q+k)-\lambda^C(q)\Big\}>0.
\label{eq:mixing_condition}
\end{equation}

\end{assump}
 Under these assumptions, the price definition \eqref{eq:price_def} is finite almost surely for each fixed $t>0$ \cite{ouazzaniChahdiRosenbaumSzymanskiPassive}. Condition \eqref{eq:mixing_condition}, which is stronger than monotonicity of $\lambda^L-\lambda^C$ alone, imposes a uniform mean-reversion strength on the queue dynamics and controls the long-horizon contribution of future trades conditional on $\Fc_t$. For an affine drift $\lambda^L(q)-\lambda^C(q)=c_\lambda q+d_\lambda$ on the lattice, \eqref{eq:mixing_condition} gives $\mathfrak m=-c_\lambda$ when $c_\lambda<0$.\\

At this stage, \eqref{eq:price_def} is not yet operational because its conditional expectation depends on the entire future paths of both the Hawkes intensities and the queues. Splitting the integral at time $t$ and compensating the market-order processes gives, whenever the displayed improper integral is finite,
\begin{equation*}
P_t
=
P_0
+
\int_0^t \kappa(q_{s-}^{a})\,\ud N_s^a
-
\int_0^t \kappa(q_{s-}^{b})\,\ud N_s^b
+
\int_t^\infty
\E\!\left[
\kappa(q_{s-}^{a})\lambda_s^a
-
\kappa(q_{s-}^{b})\lambda_s^b
\ \middle|\ \Fc_t
\right]\ud s .
\end{equation*}
The first two integrals represent realized queue-weighted trades, whereas the final integral is a conditional anticipation term. This term is essential: when $\kappa$ is constant, the realized part contains only the permanent contribution of past trades, whereas the transient propagator component comes from the conditional expectation of future Hawkes descendants. With state-dependent $\kappa$, the same anticipation term also contains the future interaction between order-flow persistence and queue mean reversion.\\

The following decomposition establishes the precise link with the classical propagator. It separates the constant-sensitivity component, which is exactly the Hawkes propagator of \cite{jaisson2015market}, from the residual generated by fluctuations of $\kappa(q)$ around its stationary level. Once the preceding well-posedness conditions justify the infinite-horizon limits, this decomposition is algebraic.

\begin{theo}
\label{thm:general_propagator}
Assume that the conditions in Assumptions \ref{suficient_assump} hold, so that \eqref{eq:price_def} is finite for the times under consideration. Suppose, in addition, that $\phi(t)=\sum_{j=1}^m c_je^{-\beta_jt}$ is a finite sum of exponentials, so that $Z_t:=(q^a_t,q^b_t,X^{1,a}_t,\ldots,X^{m,a}_t,X^{1,b}_t,\ldots,X^{m,b}_t)$, with $X^{j,x}_t:=\int_0^{t-}e^{-\beta_j(t-s)}\,\ud N^x_s$, is a Markov process, and that $Z$ admits an invariant probability distribution $\pi$ whose ask and bid queue marginals coincide. Let $D:=N^a-N^b$ and define $P_q := \pi\circ(q^a)^{-1}$ as the marginal law of $q^a$ under $\pi$ (a time-weighted law). By symmetry, the same marginal law is used for $q^b$. Define also
$$
\psi:=\sum_{n\ge1}\phi^{*n},
\qquad
\bar{\kappa}:=\int_{\R}\kappa(q)\,P_q(\ud q),
\qquad
\xi(u):=1+\left(1+\int_0^\infty\psi(r)\,\ud r\right)\int_u^\infty\phi(r)\,\ud r,
\quad \forall u \geq 0.
$$
Then, for all $t \geq 0$,
\begin{equation}
\label{eq:general_propagator_decomp}
P_t
=
P_0+\bar{\kappa}\int_0^t \xi(t-s)\,\ud D_s+\mathcal{R}_t
\end{equation}
with
\begin{equation*}
    \mathcal{R}_t
=
\lim_{T\to\infty}
\E\left[
\int_0^T (\kappa(q_{s-}^a)-\bar\kappa)\,\ud N_s^a
-
\int_0^T (\kappa(q_{s-}^b)-\bar\kappa)\,\ud N_s^b
\ \middle|\ \Fc_t
\right].
\end{equation*}
Equivalently, compensating the future parts of $N^a$ and $N^b$ yields
\begin{equation*}
\begin{aligned}
\mathcal{R}_t={}&
\int_0^t (\kappa(q_{s-}^a)-\bar\kappa)\,\ud N_s^a
-
\int_0^t (\kappa(q_{s-}^b)-\bar\kappa)\,\ud N_s^b
+\int_t^\infty
\E\left[
(\kappa(q_{s-}^a)-\bar\kappa)\lambda_s^a
-
(\kappa(q_{s-}^b)-\bar\kappa)\lambda_s^b
\ \middle|\ \Fc_t
\right]\ud s.
\end{aligned}
\end{equation*}
\end{theo}
The proof of Theorem \ref{thm:general_propagator} is given in Appendix \ref{app:proofs-price-decomposition}. The theorem does not assert a closed form for $\mathcal R$. It identifies exactly the component of \eqref{eq:price_def} not captured by the constant-sensitivity propagator. This decomposition is useful because it isolates three distinct mechanisms: the standard Hawkes propagator signal
$$
G_t:=\int_0^t\xi(t-s)\,\ud D_s,
\quad \forall t \geq 0,
$$
which enters the price through $\bar\kappa\,G_t$,
the realized queue-dependent correction
$$
I_t^\kappa:=
\int_0^t(\kappa(q_{s-}^a)-\bar\kappa)\,\ud N_s^a
-
\int_0^t(\kappa(q_{s-}^b)-\bar\kappa)\,\ud N_s^b,
\quad \forall t \geq 0,
$$
and the anticipation of future queue-weighted order flow 
$$
J_t^\kappa:=
\int_t^\infty
\E\!\left[
(\kappa(q_{s-}^a)-\bar\kappa)\lambda_s^a
-
(\kappa(q_{s-}^b)-\bar\kappa)\lambda_s^b
\mid\Fc_t
\right]\ud s,
\quad \forall t \geq 0,
$$
so that $\mathcal R_t=I_t^\kappa+J_t^\kappa$. The propagator term averages the price response to signed flow over liquidity states, whereas $\mathcal R_t$ captures how identical signed flow has different impacts in different queue states.

\begin{rque}\label{rem:J-small}
The anticipation term $J^\kappa$ has no closed form in general and is treated as part of the regression error in Section~\ref{sec:empirical-specification}. This is justified at the sampling horizons used there, of a few seconds or more. For $\delta>0$ write $\Delta_\delta X_t:=X_{t+\delta}-X_t$. The price \eqref{eq:price_def} is a conditional expectation, hence a martingale. Under the stationary law $\pi$ of Theorem~\ref{thm:general_propagator}, and when $P$ is square integrable, $\operatorname{Var}(\Delta_\delta P)=\sigma_P^2\delta$, where $\sigma_P^2$ is the mean rate of its quadratic variation. By contrast, $J^\kappa$ is a stationary function of the current state, so $\operatorname{Var}(\Delta_\delta J^\kappa)\le4\operatorname{Var}_\pi(J^\kappa)$ for every $\delta$. Whenever $\operatorname{Var}_\pi(J^\kappa)<\infty$, its share in price increments is therefore at most of order $\delta^{-1}$, and satisfies
\[
\frac{\operatorname{Var}(\Delta_\delta J^\kappa)}{\operatorname{Var}(\Delta_\delta P)}\le\frac{4\operatorname{Var}_\pi(J^\kappa)}{\sigma_P^2\,\delta}.
\]
In the decoupled case of Proposition~\ref{prop:explicit_bound}, let $\rho:=-c_\lambda$ and $\kappa(q)=\kappa_0+\kappa_1q$. Then $J^\kappa_t=\kappa_1\frac{\mu}{\rho}(q^a_t-q^b_t)$, and the ratio becomes
\[
\frac{\operatorname{Var}(\Delta_\delta J^\kappa)}{\operatorname{Var}(\Delta_\delta P)}=\vartheta_J\,g(\rho\delta)\le\frac{1}{\rho\delta},
\qquad
g(x):=\frac{1-e^{-x}}{x}.
\]
Here $\vartheta_J\in[0,1]$ is the share of the quadratic variation of $P$ carried by the jumps of $J^\kappa$. Beyond the decoupled case, if $\phi(t)=c\,e^{-\beta t}$ and $\lambda^L$, $\lambda^C$ are affine in the queue, then $J^\kappa$ is an explicit quadratic function of the queue and intensity states. Since all intensities are then affine in these states, their stationary moments up to order four solve a finite linear system, which gives $\operatorname{Var}_\pi(J^\kappa)$, $\operatorname{Var}(\Delta_\delta J^\kappa)$ and $\sigma_P^2$ exactly.

We now use the calibration values of Sections~\ref{sec:calibration} to~\ref{sec:passive-impact-curves}: $\rho\approx1$\,s$^{-1}$, $\bar\lambda/\rho\approx16$ AES, where $\bar\lambda:=\mu/(1-\|\phi\|_{L^1})$ is the mean aggressive intensity, $\kappa_1=-8.1\times10^{-6}$ and $\bar\kappa=4.1\times10^{-3}$ at the mean best queue. For $\|\phi\|_{L^1}\le0.95$, half-lives $\ln2/\beta$ up to $5$\,s and cancellation-to-trade ratios from $5$ to $50$, the ratio $\operatorname{Var}(\Delta_\delta J^\kappa)/\operatorname{Var}(\Delta_\delta P)$, computed exactly from this linear system, is at most $5.5\%$ at one second, $0.87\%$ at ten seconds, $0.15\%$ at one minute and $0.055\%$ at the horizon $\delta^*=212$\,s of Section~\ref{subsec:propagator-benchmark}.
In absolute terms, the standard deviation of $\Delta_\delta J^\kappa$ never exceeds $0.67$ ticks over this range.
\end{rque}

\section{Calibration of Stationary Order Flow and Liquidity Dynamics}
\label{sec:calibration}

This section describes the empirical construction of the order-flow and liquidity inputs used in Sections \ref{sec:empirical-specification} to \ref{sec:passive-impact-curves}. The calibration follows the Hawkes-queue-reactive structure of \cite{huang2015simulating, wu2020single, ouazzaniChahdiRosenbaumSzymanskiPassive}. It concerns the aggressive Hawkes flow and the passive insertion and cancellation dynamics, not $\kappa$, whose coefficients are estimated from prices in Section~\ref{sec:empirical-specification}. This order is necessary because the price regression uses the propagator kernel $\xi$ and the queue paths as inputs. All hatted quantities in this section are estimated from the message data before the price regression is run.

\subsection{Data and Order-Book Reconstruction}
\label{subsec:data-order-book-reconstruction}

The empirical input is a time-stamped order-book message stream for E-mini S\&P 500 futures. For each trading day $d$, we reconstruct the best ask and best bid queues and the associated event processes
$$
(N^{a,d},N^{b,d},L^{a,d},L^{b,d},C^{a,d},C^{b,d},q^{a,d},q^{b,d}).
$$
At the best quotes, limit-order insertions ($L^{a/b}$) increase the corresponding queue, while cancellations ($C^{a/b}$) and aggressive executions ($N^{a/b}$) decrease it. If several messages share a timestamp, the feed ordering is retained, so that each event is associated with the queue immediately before it. Events away from the best quotes enter the reconstructed state only after a quote update makes them relevant to the best-queue path.

\paragraph{Sample.} The data are order-book messages from the CME for the E-mini S\&P 500 futures contract, whose price moves in steps (ticks) of $0.25$ index points. The sample covers 2025, plus the first days of January 2026 needed to complete the last test windows. After removing shortened sessions, days with FOMC announcements, and sessions with trading halts, 263 regular trading days remain, and 193 of them are used for testing after the calendar filter of Section~\ref{sec:empirical-specification}. Futures contracts expire every quarter, and in the weeks before an expiry the old and the next contract trade side by side at different prices (Figure~\ref{fig:roll-mixing}). Mixing them would create price jumps unrelated to trading, so in each session we keep only the contract with the most trades (the front-month stream). The switch to the next contract happens on the Monday of expiry week, and the same rule is used for the Treasury futures. Prices are not adjusted across contracts, since all regressions use price changes within a session. The same construction is applied to the E-mini Nasdaq-100 (NQ), E-mini Russell 2000 (RTY) and E-mini Dow (YM) futures over 2023 to 2025, to the E-mini S\&P 500 in 2023 and 2024, and to the Treasury futures ZN, ZB, ZF and TN in 2025, which are used for the out-of-sample comparisons of Section~\ref{subsec:fitting-results}. The calendar filter also drops FOMC weeks, which in 2025 include three of the four switch weeks. Between $10{:}00$ and $16{:}00$, the window used for the regressions, a session has about $330{,}000$ trades on average, split almost evenly between buys and sells, and the best bid and ask are one tick apart for $97\%$ of them. The average event size (AES) is $\widehat{\mathrm{AES}}=1.22$ contracts, the mean size of best-quote insertions (Appendix~\ref{app:calibration-details}). For the other instruments, the regressions use the mean trade size as AES. In ES, the queue of waiting orders at the three best prices is on average $101$ AES.

\paragraph{Time window.} Within the retained regular sessions, the Hawkes and queue-reactive calibrations of this section use the intraday window $10{:}00$ to $16{:}00$ local exchange time, while the price regressions of Section~\ref{sec:empirical-specification} and the sample statistics above use $10{:}00$ to $16{:}00$. Writing the retained window as $[T_d^0,T_d^1]$, we restart all counting processes at $T_d^0$ and use the observed queue sizes at that time as initial states. This convention only fixes the origin for the finite-sample likelihoods, exposures, and signals. Excluding irregular sessions reduces deterministic nonstationarity and event-classification artifacts. Our results do not extend to such stressed conditions.

\subsection{Hawkes Calibration of Aggressive Order Flow}
\label{subsec:hawkes-calibration}

\begin{figure}[htb!]
    \centering
    \includegraphics[width=0.7\textwidth]{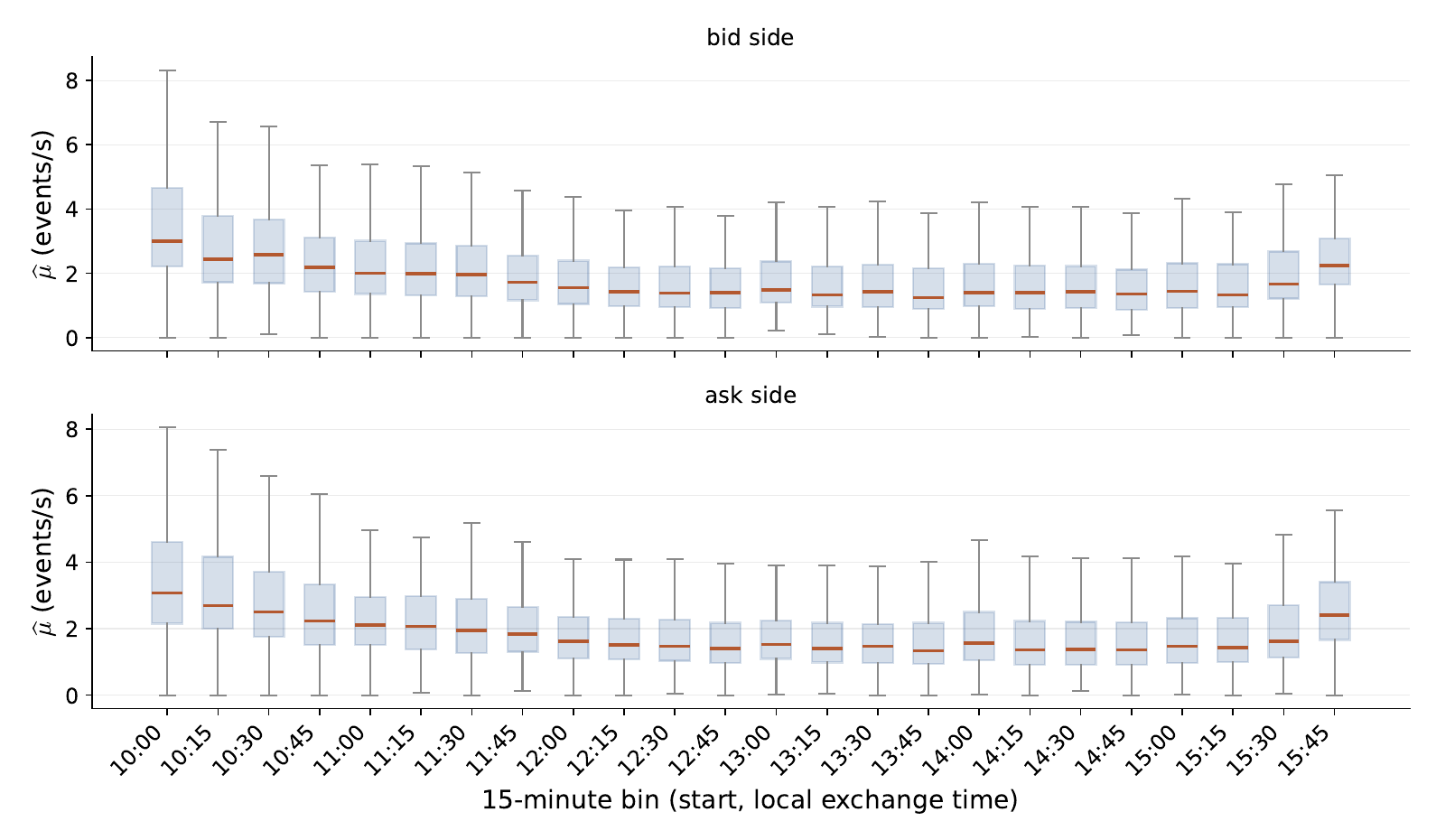}
    \caption{Cross-day distribution of the estimated baseline $\widehat\mu_j$ for each 15-minute bin (top: bid, bottom: ask).}
    \label{fig:mu_whiskers}
\end{figure}

The aggressive-order calibration estimates the excitation kernel $\phi$ in \eqref{eq:trad_hawkes} and assesses the stability of the temporal dependence in order flow. Market-order intensity is not stationary over the full trading day: it displays a systematic U- or J-shaped intraday pattern, with elevated activity in the morning, lower activity around midday, and a moderate increase in the afternoon. Such intraday patterns are well documented for volatility and also appear in volume, consistent with the variance-volume link \cite{wood1985transactions, baillie1991intraday, andersen2024intraday, karpoff.87, madhavan1997security, kyle2016market, muhle2026unified}. Figure~\ref{fig:mu_whiskers} shows this pattern in the fitted baselines. For price formation, the key object is temporal dependence in $\phi$, not the intensity level. We therefore assess the stability of $\phi$ locally rather than impose global intraday stationarity: within each short bin, order-flow statistics are treated as approximately stable, while variation across bins captures deterministic intraday seasonality. Each day is partitioned into $R=24$ intervals of 15 minutes,
$
I_{d,r}=[t_{d,r-1},t_{d,r})$, for $r=1,\ldots,R.
$ All order-flow and liquidity estimators below are computed on this grid. Relative to a single full-day fit, this local calibration makes intraday seasonality and day-to-day variation in the Hawkes and queue-reactive parameters visible, at the cost of greater sampling error within each bin.\\

Fix a day $d$ and a bin $r$, and write $h:=t_{d,r}-t_{d,r-1}$. All likelihoods in this subsection are conditional on the information available at the left endpoint $t_{d,r-1}$, or on a specified initialization of the Hawkes state at that endpoint, and provide the finite-sample counterpart for estimating the stationary Hawkes law in \eqref{eq:trad_hawkes}. On $I_{d,r}$ and for each side $x\in\{a,b\}$, aggressive order arrivals are fitted by a univariate Hawkes process with constant local baseline and exponential-mixture excitation,
$$
\lambda_t^{x,d,r}(\theta)
=
\mu^{x,d,r}
+
\int_{t_{d,r-1}}^{t-}
\phi^{x,d,r}(t-s)\,\ud N_s^{x,d},
\qquad
\phi^{x,d,r}(u)=\sum_{j=1}^{3} c_j^{x,d,r}e^{-\beta_j^{x,d,r}u}\mathbbm{1}_{\{u\ge0\}},
\quad \forall t\in I_{d,r}.
$$
The parameter vector is
$
\theta^{x,d,r}
=
(\mu^{x,d,r},c_1^{x,d,r},c_2^{x,d,r},c_3^{x,d,r},
\beta_1^{x,d,r},\beta_2^{x,d,r},\beta_3^{x,d,r}),
$
with $\mu^{x,d,r}>0$, $c_j^{x,d,r}\ge0$, $\beta_j^{x,d,r}>0$. The choice $m=3$ balances flexibility and parsimony and allows for fast, intermediate, and slow decay modes suitable for 15-minute bins. Day-level fits with a much richer 12-exponential kernel (and a piecewise-constant baseline) on every trade summary give branching ratios close to but mostly below criticality (mean $\widehat\eta=0.78$, 5th to 95th percentile $0.62$ to $0.95$).\\

In this model, the stability condition \eqref{eq:hawkes_stability} becomes $$
\eta^{x,d,r}:=\|\phi^{x,d,r}\|_{L^1}
=
\sum_{j=1}^{3}\frac{c_j^{x,d,r}}{\beta_j^{x,d,r}}<1.
$$
The stationary propagator interpretation is valid only for stable fitted kernels \cite{hawkes1971spectra,bacry2015hawkes,jaisson2015market}, so we either enforce stability or retain only stable fits. An unstable bin-level estimate is treated as a diagnostic of misspecification or finite-sample noise, not as a valid input to the stationary price decomposition. Of the $404$ day-level fits, $5$ have $\widehat\eta\ge1$, all below $1.05$, and are excluded as unstable. With the kernel truncated at $1$\,s, the mean is $0.67$ and no fit exceeds $0.79$. At the 15-minute bin level, the across-day median and upper quartile of $\widehat\eta$ exceed one only in the last quarter hour before the close ($15{:}45$ to $16{:}00$). Elsewhere the bin medians lie between $0.72$ and $0.80$, so bin-level exclusions occur in the tail. Estimation is by conditional maximum likelihood on each bin: the exact log-likelihood and compensator formulas for the exponential-mixture kernel, together with the initialization conventions, are standard and are collected in Appendix~\ref{app:calibration-details}.\\

Side-specific estimates provide a direct diagnostic of the symmetry assumption in \eqref{eq:trad_hawkes}. When a single propagator signal is required in Section \ref{sec:empirical-specification}, $\widehat\phi$ and $\widehat\xi$ denote the pre-specified pooled or side-averaged estimate used for that regression. The pooled kernel would maximize the joint likelihood across stable bins and both sides, with common $(c_j,\beta_j)$ and bin-specific baselines (parameters are not averaged across bins because the intensity depends on them nonlinearly). This fully restricted alternative has not been estimated on this sample. For any fitted stable kernel, we check
$$
\widehat\eta=\int_0^\infty \widehat\phi(u)\,\ud u
=\sum_{j=1}^{3}\frac{\widehat c_j}{\widehat\beta_j}<1,
$$
and Lemma \ref{lem:xi} gives
$$
\widehat\xi(t)
=
1+
\sum_{j=1}^{3}
\widehat w_j\,
e^{-\widehat\beta_j t},
\qquad
\widehat w_j:=\frac{\widehat c_j/\widehat\beta_j}{1-\widehat\eta},
\quad \forall t \geq 0.
$$
The $\widehat w_j$ are the \textit{propagator weights}: they measure the contribution of each fitted time scale to the transient response, and their stability across bins and days is examined below.
 Since $\widehat c_j$ and $\widehat\beta_j$ have units of inverse time, $\widehat\xi$ is dimensionless. Values of $\widehat\eta$ close to one and slowly decaying fitted kernels are interpreted only as evidence that order flow is endogenous.\\

We assess the Hawkes fit using the cross-day and intraday distributions of $\widehat\mu$ and of the induced propagator weights shown in Figures \ref{fig:mu_whiskers} and \ref{fig:hawkes_sumexp_weights}.

\paragraph{Empirical results.}
 Figure \ref{fig:mu_whiskers} reports the cross-day distribution of the fitted baseline, with the intraday pattern described above. The bid and ask estimates are close, which supports the symmetric specification. 

Figure \ref{fig:hawkes_sumexp_weights} (Appendix~\ref{app:additional-empirics}) reports the cross-day distribution of the fitted propagator weights $\widehat w_1,\widehat w_2,\widehat w_3$ defined above. Their stability across bins supports using a common kernel shape to construct the empirical propagator signal $\widehat G$ in Section~\ref{subsec:propagator-benchmark}, the sample analogue of $G_t=\int_0^t\xi(t-s)\,\ud D_s$.

\subsection{Queue-Reactive Calibration of Passive Liquidity}
\label{subsec:queue-reactive-calibration}

The passive side of the model is calibrated through the queue-reactive intensities $\lambda^L$ and $\lambda^C$ in \eqref{eq:limit_cancel_intensities}. Let
$\mathcal Q=\{0,1,\ldots,Q_{\max}\}
$
be the capped AES-normalized queue grid, with $Q_{\max}=1000$, where $q^{x,d}$ denotes the aggregated depth at the three best price levels on side $x$ of day $d$, in AES units. We calibrate on this aggregate rather than on the best queue because the best queue is depleted and refilled at every price change, so its dynamics are not a stationary birth-death chain on $\mathcal Q$. The aggregate smooths these resets, and it is the object to which the mixing condition \eqref{eq:mixing_condition} is applied in practice. Figure~\ref{fig:qrm-intensities-clean} (Appendix~\ref{app:additional-empirics}) shows the estimated intensities.
For $q\in\mathcal Q$, define the total exposure time
$$
\mathcal T(q)
:=
\sum_{d,r}\sum_{x\in\{a,b\}}
\int_{I_{d,r}}
\mathbbm{1}_{\{q_{s^-}^{x,d}=q\}}\,\ud s.
$$
For $E\in\{L,C\}$, the corresponding state-conditional event count is
$$
N^{E}(q)
:=
\sum_{d,r}\sum_{x\in\{a,b\}}
\int_{I_{d,r}}
\mathbbm{1}_{\{q_{s^-}^{x,d}=q\}}\,\ud E_s^{x,d}.
$$
On states with $\mathcal T(q)>0$, the maximum-likelihood estimator under the queue-reactive Poisson model is the exposure-normalized count
$$
\widehat\lambda^E(q)
=
\frac{N^E(q)}{\mathcal T(q)},
\quad \forall E\in\{L,C\}.
$$
This is the nonparametric estimator corresponding to a piecewise-constant intensity on the finite queue grid. Its derivation and the treatment of rarely visited states are in Appendix~\ref{app:calibration-details}, and exposure-based filtering is used only to regularize the plotted diagnostics. Pooling bid and ask exposures enforces the symmetry used in the theoretical model. We estimate the two sides separately only to diagnose departures from that symmetry. \\

 Market orders are treated differently. In the main model, $N^a$ and $N^b$ are Hawkes-driven and their intensities are not functions of the current queue size. Nevertheless, for diagnostic purposes we also compute the empirical state projection
$$
\widehat\lambda^M(q)
=
\frac{N^M(q)}{\mathcal T(q)},
\qquad
N^M(q)
:=
\sum_{d,r}\sum_{x\in\{a,b\}}
\int_{I_{d,r}}
\mathbbm{1}_{\{q_{s^-}^{x,d}=q\}}\,\ud N_s^{x,d}.
$$
This quantity is not a primitive of \eqref{eq:trad_hawkes}. It records the observed frequency of aggressive depletions conditional on queue state $q$. The full empirical depletion rate
$$
\widehat\lambda^D(q)
:=
\widehat\lambda^C(q)+\widehat\lambda^M(q)
$$
is therefore used only for order-book accounting and for comparison with standard queue-reactive diagnostics \cite{huang2015simulating}. Because \eqref{eq:queues_def} includes market-order depletions, the observed calendar-time queue drift involves $\lambda^L-\lambda^C-\lambda^M$ when market orders are projected on queue state. By contrast, the well-posedness condition \eqref{eq:mixing_condition} concerns only the cancellation-only drift $b(q):=\lambda^L(q)-\lambda^C(q)$ of the passive queue-reactive component, because aggressive flow is modeled separately by queue-independent Hawkes intensities. We report both: $\widehat\lambda^L-\widehat\lambda^D$ for queue accounting and $\widehat\lambda^L-\widehat\lambda^C$ for the theoretical mean-reversion condition.
\begin{rque}\label{rem:qr-mo}
The empirical projection $\widehat\lambda^M(q)$ in Figure~\ref{fig:qrm-intensities-clean} is visibly non-constant in $q$, consistent with the queue-reactive evidence of \cite{huang2015simulating,wu2020single}. We nevertheless keep market-order intensities queue-independent to preserve the exact propagator and a shared aggressive flow in the counterfactual. The resulting limitation is that the passive-impact experiments omit feedback from posted liquidity to market-order arrivals. Extending the robustness analysis to a queue-reactive market-order specification as in \cite{wu2020single}, with the market-order process coupled counterfactually, is left for future work. See also Section~\ref{sec:passive-impact-curves}.
\end{rque}

 We fit an affine drift $\lambda^L(q)-\lambda^C(q)\approx c_\lambda q+d_\lambda$ by weighted least squares,
$$
(\widehat c_\lambda,\widehat d_\lambda)
\in
\arg\min_{c,d}
\sum_{q\in\mathcal Q:\,\mathcal T(q)>0}
\mathcal T(q)
\left(
\widehat\lambda^L(q)-\widehat\lambda^C(q)-cq-d
\right)^2,
$$
and fit $\widehat\lambda^L-\widehat\lambda^D$ in the same way. These weights give more influence to states whose exposure-normalized intensity estimates are based on longer observation times. A fitted negative slope supports mean reversion but is only an empirical diagnostic: it does not prove the uniform lattice condition \eqref{eq:mixing_condition} outside the observed support, so the affine drift is interpreted only locally on the observed queue range.

\begin{figure}[htb!]
    \centering

    \begin{subfigure}{0.4\textwidth}
        \centering
        \includegraphics[width=\textwidth]{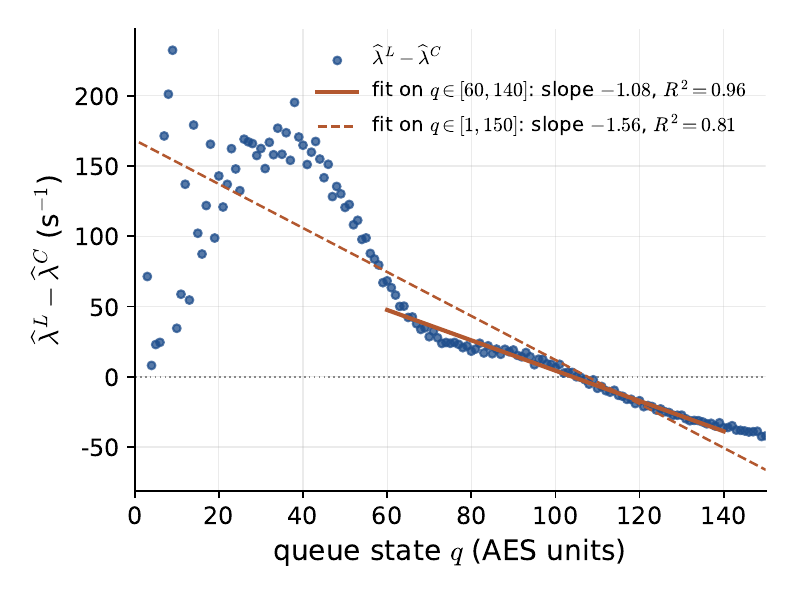}
        \caption{$\widehat\lambda^L-\widehat\lambda^C$, bid}
    \end{subfigure}
    \qquad
    \begin{subfigure}{0.4\textwidth}
        \centering
        \includegraphics[width=\textwidth]{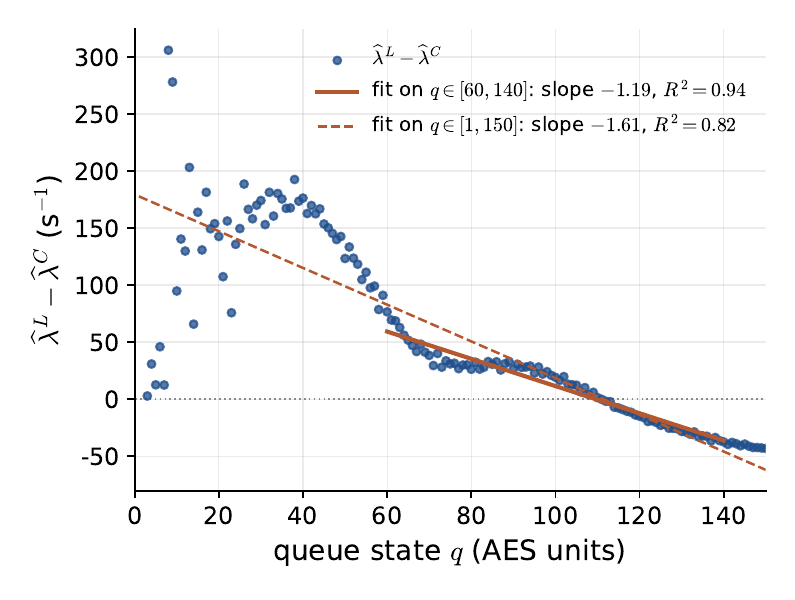}
        \caption{$\widehat\lambda^L-\widehat\lambda^C$, ask}
    \end{subfigure}

    \vspace{0.15cm}

    \begin{subfigure}{0.4\textwidth}
        \centering
        \includegraphics[width=\textwidth]{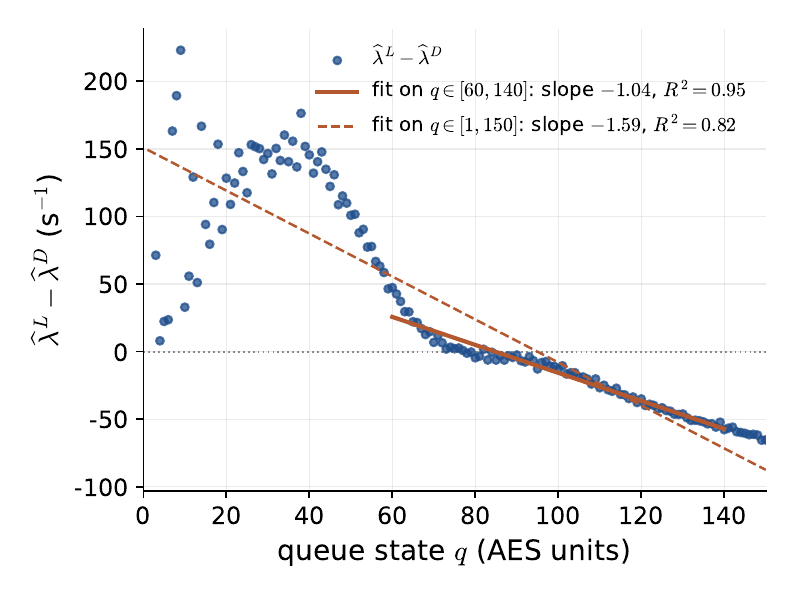}
        \caption{$\widehat\lambda^L-\widehat\lambda^D$, bid}
    \end{subfigure}
    \qquad
    \begin{subfigure}{0.4\textwidth}
        \centering
        \includegraphics[width=\textwidth]{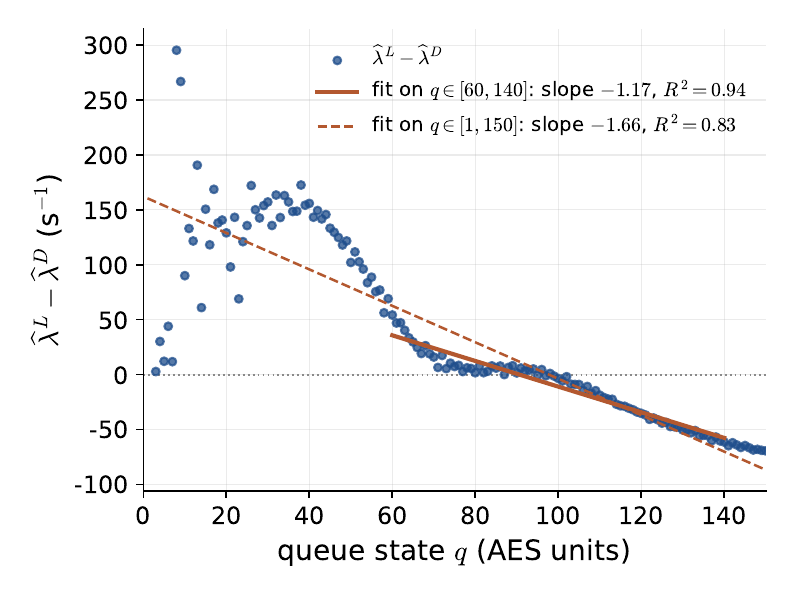}
        \caption{$\widehat\lambda^L-\widehat\lambda^D$, ask}
    \end{subfigure}

    \caption{Queue-reactive drift diagnostics for the aggregated three-level queue (ES 2025). Top: $\widehat\lambda^L-\widehat\lambda^C$. Bottom: $\widehat\lambda^L-\widehat\lambda^D$. Lines: affine fits on $60\le q\le140$ (solid) and $1\le q\le150$ (dashed).}
    \label{fig:qrm_grouped_drift_full_year}
\end{figure}

The same estimates provide an internal accounting check. On every state with positive exposure,
$$
\widehat\lambda^L(q)-\widehat\lambda^D(q)
=
\frac{N^L(q)-N^C(q)-N^M(q)}{\mathcal T(q)}.
$$
Failure of this identity indicates a mismatch between event classification and exposure computation. Figures \ref{fig:qrm-intensities-clean} and \ref{fig:qrm_grouped_drift_full_year} report the corresponding intensity and drift diagnostics, which validate the calibrated liquidity dynamics used to construct the queue process and the signal $H_t$.

\paragraph{Empirical results.}

 Figure~\ref{fig:qrm_grouped_drift_full_year} shows the affine drift fits. The cancellation-only drift $\widehat\lambda^L-\widehat\lambda^C$ is the one closer to the theoretical drift, and under the affine approximation a negative slope $c_\lambda$ gives the lattice mixing constant $\mathfrak m=-c_\lambda>0$ of Section~\ref{subsec:queue-dependent-price-sensitivity}.

\section{Empirical Specification of the Price Decomposition}
\label{sec:empirical-specification}

This section translates the continuous-time decomposition of Section \ref{sec:model-passive-liquidity} into the finite-sample price regressions used in the empirical analysis. It specifies the observable signals, the sampling scheme, and the interpretation of the affine coefficients. 

\subsection{Propagator Signal Benchmark Model}
\label{subsec:propagator-benchmark}

Let $P^{\mathrm{o}}$ denote the observed price proxy used on the left-hand side of the regression. Unless stated otherwise, the primary proxy is the \textit{weighted mid-price},
$$
P_t^{\mathrm{wmid}}
=
\frac{A_t q_t^b+B_t q_t^a}{q_t^a+q_t^b},
$$
where $A_t$ and $B_t$ are the best ask and best bid prices and the queues are the displayed best-queue sizes reconstructed in Section \ref{subsec:data-order-book-reconstruction}, in contracts. The usual mid-price and the last traded price are used only as robustness alternatives. We use the weighted mid-price as the baseline because it incorporates the same best-queue state variable as the passive-liquidity channel and attenuates mechanical bid-ask bounce.\\

Because the weighted mid-price is itself a function of the best queues, part of the estimated state dependence under this proxy is necessarily mechanical. Even with fixed quotes, a buy of size $v$ (in the unit of the queues) changes the proxy by $\Delta P^{\mathrm{wmid}}=(A-B)\,q^b\,v/\{(q^a+q^b)(q^a+q^b-v)\}$, with a larger move for a thinner ask. Accordingly, Section~\ref{subsec:fitting-results} gives equal prominence to the plain mid-price, which has no such mechanical queue dependence. With the plain mid-price, the coefficient keeps the same sign and magnitude ($t$-statistics $-5.9$ to $-11.4$ for $\delta$ from $30$ to $300$\,s, $-7.7$ at $\delta^*$), so the result is not a weighted-mid artifact. An explicit per-trade decomposition subtracting the mechanical increment is not reported here.\\

Fix a sampling horizon $\delta>0$. For each retained trading day $d$, define the regular grid
$$
t_i^d=T_d^0+i\delta,\quad i=0,\ldots,n_d(\delta),
$$
inside the cleaned intraday window $[T_d^0,T_d^1]$. For any c\`adl\`ag process $X$, write
$$
\Delta_i^\delta X:=X_{t_i^d}-X_{t_{i-1}^d}.
$$
All explanatory variables (including the stochastic integrals) are computed in event time, without binning, and then sampled through these increments. Only the final regression uses the regular grid.\\

The theoretical model in Section \ref{sec:model-passive-liquidity} normalizes aggressive orders to unit size. In the empirical implementation, aggressive order sizes are observed. We therefore define the AES-normalized marked aggressive-volume processes (normalization details in Appendix~\ref{app:calibration-details})
$$
V_t^{x,d}
:=
\sum_{T_d^0<\tau\le t,\ \tau\in N^{x,d}}
\frac{v_\tau^{x,d}}{\widehat{\mathrm{AES}}},
\quad \forall x\in\{a,b\},\ t\in[T_d^0,T_d^1],
$$
where $v_\tau^{x,d}$ is the raw executed size at the best quote. The empirical signed aggressive flow is
$$
D_t^{V,d}:=V_t^{a,d}-V_t^{b,d}.
$$
If all event sizes are set equal to one, $V^{x}=N^{x}$ for all $x\in\{a,b\}$ and this notation reduces to the unit-size convention.\\

The model-consistent propagator benchmark uses the stable fitted kernel $\widehat\xi(t)=1+\sum_{\ell=1}^{m}\widehat w_\ell e^{-\widehat\beta_\ell t}$ of Section~\ref{subsec:hawkes-calibration}, with propagator weights $\widehat w_\ell$. The corresponding empirical signal is
$$
\widehat G_t^d
:=
\int_{T_d^0}^t \widehat \xi(t-s)\,\ud D_s^{V,d}
=
D_t^{V,d}+\sum_{\ell=1}^{m}\widehat w_\ell \widehat X_t^{\ell,d},
\quad\text{ where }\quad
\widehat X_t^{\ell,d}
:=
\int_{T_d^0}^t e^{-\widehat\beta_\ell(t-s)}\,\ud D_s^{V,d},
\quad \forall t\in[T_d^0,T_d^1].
$$
For grid evaluation, the exponential factors satisfy the exact recursion
$$
\widehat X_{t_i^d}^{\ell,d}
=
e^{-\widehat\beta_\ell\delta}\widehat X_{t_{i-1}^d}^{\ell,d}
+
\int_{(t_{i-1}^d,t_i^d]}
e^{-\widehat\beta_\ell(t_i^d-s)}\,\ud D_s^{V,d}.
$$
The last term is the finite sum of observed signed marked trades in $(t_{i-1}^d,t_i^d]$, each weighted by its exact lag to $t_i^d$. The constrained benchmark regression is 
\begin{equation}
\Delta_i^\delta P^{\mathrm{o},d}
=
\kappa_0^\delta\,\Delta_i^\delta \widehat G^d
+\varepsilon_i^{\delta,d}.
\label{eq:empirical-benchmark-regression-revised}
\end{equation}
The coefficient $\kappa_0^\delta$ is measured in price units per AES-normalized aggressive-volume unit. Under the exact unit-size Hawkes model and a correctly specified price proxy, it coincides with the constant-sensitivity coefficient in \eqref{eq:trad_propagator}. Empirically, however, it is a projection coefficient at horizon $\delta$.\\

We also consider a reduced-form propagator dictionary. Fix positive decay rates
$
\mathcal B=\{\beta_1,\ldots,\beta_{n_\beta}\}
$
and define
$$
X_t^{\ell,d}:=\int_{T_d^0}^t e^{-\beta_\ell(t-s)}\,\ud D_s^{V,d},
\quad \forall t\in[T_d^0,T_d^1],\ \ell=1,\ldots,n_\beta.
$$
For any selected subset $S\subset\{1,\ldots,n_\beta\}$, the reduced-form benchmark is
\begin{equation}
\Delta_i^\delta P^{\mathrm{o},d}
=
\theta_0^\delta\,\Delta_i^\delta D^{V,d}
+\sum_{\ell\in S}\theta_\ell^\delta\,\Delta_i^\delta X^{\ell,d}
+\varepsilon_i^{\delta,d}.
\label{eq:empirical-reduced-form-propagator}
\end{equation}
This is a linear propagator approximation in the sense of \cite{bouchaud2003fluctuations,jaisson2015market}. It is deliberately less restrictive than the Hawkes-based specification: the coefficients $\theta_\ell^\delta$ are unrestricted projection coefficients and are not interpreted as the weights of a Hawkes resolvent. Heavy-tailed Hawkes models motivate a broad decay grid, but $S$ is selected solely for price prediction \cite{jaisson2015limit,jaisson2016rough,gatheral2018volatility}.\\

The sampling horizon $\delta$, the price proxy, and the exponential dictionary are finite-sample tuning choices. A very small $\delta$ provides many observations but increases exposure to bid-ask bounce, timestamp effects, price discreteness, and the immediate mechanical component of queue depletion. A larger $\delta$ attenuates some of these microstructure effects but averages over more variation in the unobserved anticipation term. We therefore compare a pre-specified grid of horizons and decay dictionaries on an in-sample/out-of-sample split. In the forward selection, each additional decay factor is admitted only according to its out-of-sample contribution to the chosen loss, not because a larger number of factors is theoretically required. Thus out-of-sample results assess predictive content, not a particular rough or near-critical regime.\\

\begin{figure}[htb!]
    \centering
    \begin{subfigure}{0.42\textwidth}
        \centering
        \includegraphics[width=\textwidth]{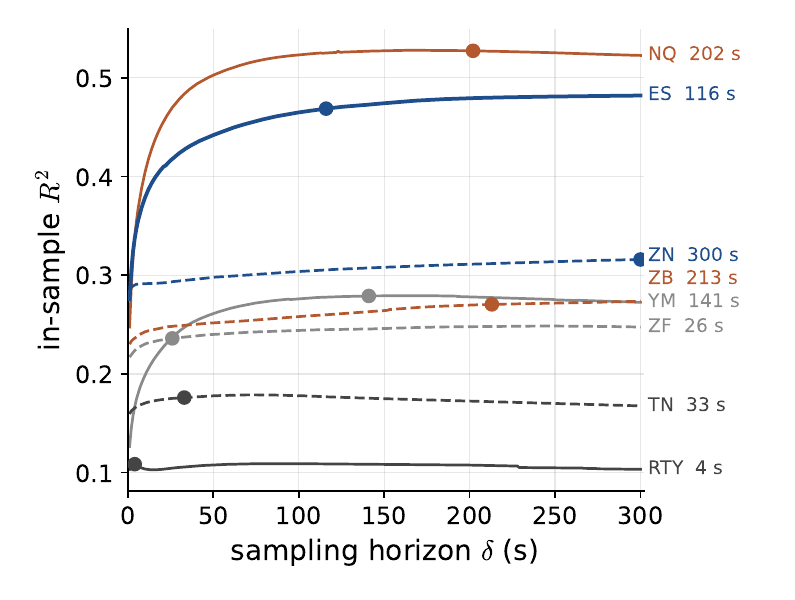}
        \caption{In-sample}
    \end{subfigure}
    \qquad
    \begin{subfigure}{0.42\textwidth}
        \centering
        \includegraphics[width=\textwidth]{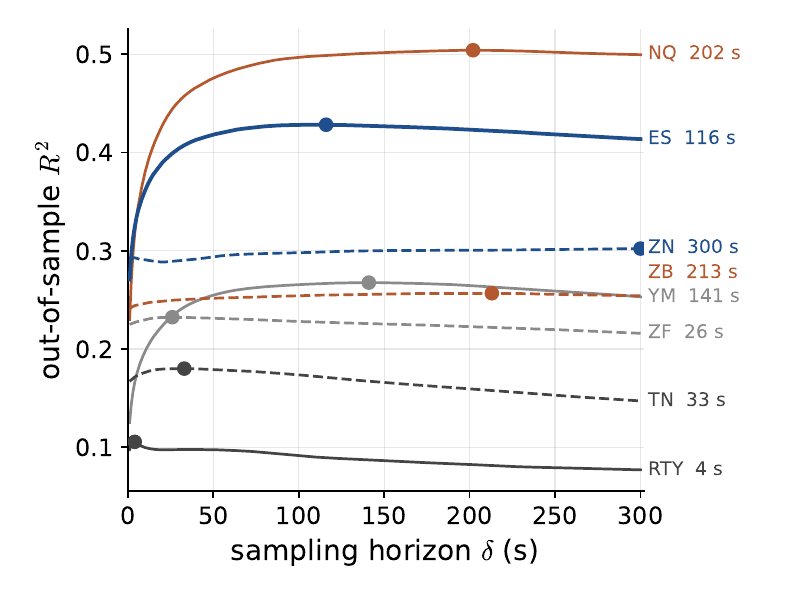}
        \caption{Out-of-sample}
    \end{subfigure}
    \caption{Horizon selection: average split-level $R^2$ of the best propagator dictionary at each horizon, in sample (left) and out of sample (right), 2025, weighted mid, one curve per instrument, 10:00 to 16:00, rolling 16-day windows with the ex-ante filter ($193$ to $194$ test days). Solid lines are equity indices and dashed lines are Treasuries. Dots mark the $\delta^*$ of each instrument.}
    \label{fig:propagator-delta-selection}
\end{figure}

In the implementation reported here, $\delta$ is selected on a grid ranging from one second to five minutes. Figure \ref{fig:propagator-delta-selection} reports the corresponding out-of-sample criterion. The minimum out-of-sample error, equivalently the maximum out-of-sample $R^2$, is attained in a flat region of a few minutes. On the 10:00 to 16:00 sample, the same procedure gives the following $\delta^*$ for 2023, 2024 and 2025: $258$, $235$ and $116$\,s for ES, $215$, $187$ and $202$\,s for NQ, $177$, $292$ and $141$\,s for YM, and $6$, $64$ and $4$\,s for RTY. For the Treasuries (2025 only) it gives $213$\,s for ZB, $300$\,s for ZN, $26$\,s for ZF and $33$\,s for TN. The horizon is a few minutes for all products except RTY, ZF and TN, where it is about one minute or less. The criterion is, however, flat over a wide neighborhood of $\delta^*$. Re-estimating the specifications of this section at other horizons between half a minute and five minutes leaves the signs and orders of magnitude of all fitted coefficients unchanged, and the affine model improves on the propagator benchmark out of sample at every horizon (Appendix~\ref{app:additional-empirics}, Table~\ref{tab:delta-robustness-clean}). The results are therefore not driven by the selected horizon, which also limits concerns about selecting $\delta^*$ on the reporting split. We use $\delta^*$ as the benchmark horizon in the affine specification below and interpret it as a tuning choice for the retained sample and price proxy, not as a structural time scale of the continuous-time model.

\begin{figure}[htb!]
    \centering
    \begin{subfigure}{0.42\textwidth}
        \centering
        \includegraphics[width=\textwidth]{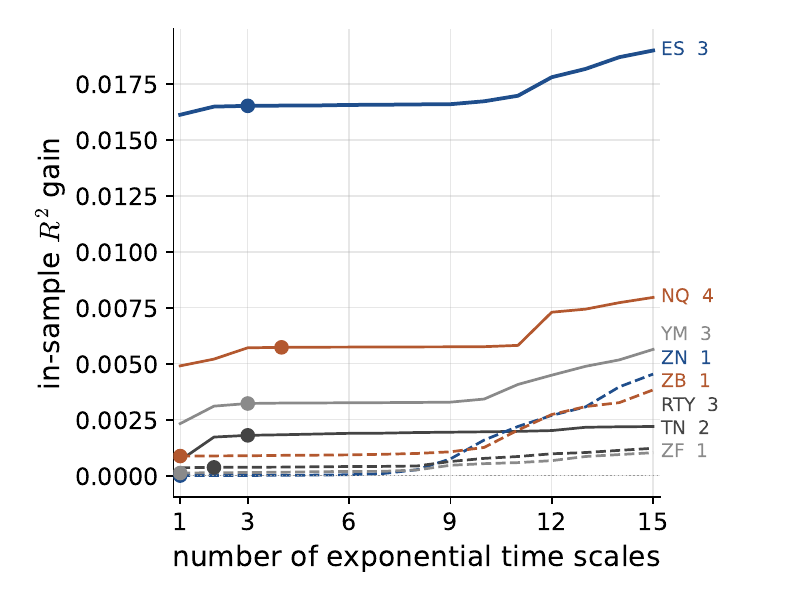}
        \caption{In-sample}
    \end{subfigure}
    \qquad
    \begin{subfigure}{0.42\textwidth}
        \centering
        \includegraphics[width=\textwidth]{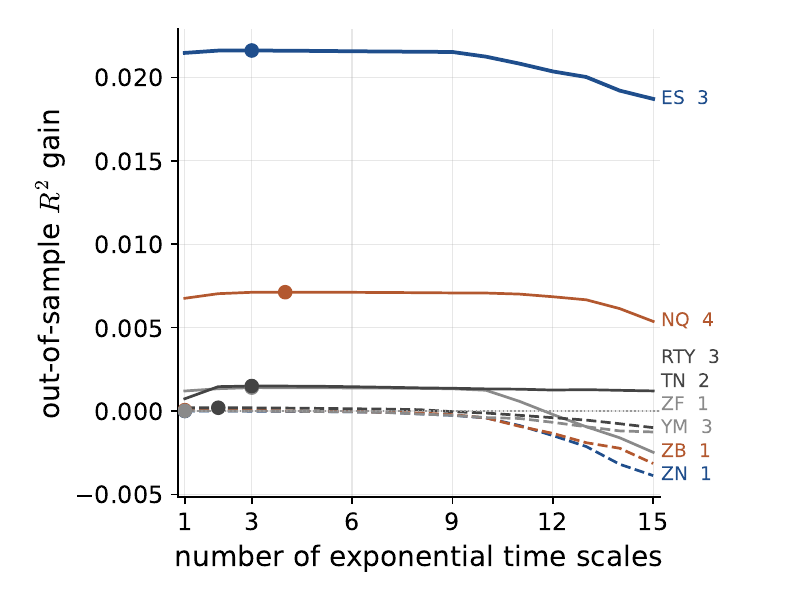}
        \caption{Out-of-sample}
    \end{subfigure}
    \caption{Forward selection over the exponential dictionary at each instrument's $\delta^{*}$ (weighted mid-price, 2025, 10:00 to 16:00). Gain in $R^2$ over the model without exponential factors, in sample (left) and out of sample (right). Solid lines are equity indices and dashed lines are Treasuries. Dots and labels give the number of time scales retained out of sample.}
    \label{fig:propagator-timescale-selection}
\end{figure}

Conditional on $\delta=\delta^*$, the forward-selection exercise over the exponential dictionary selects between one and four time scales according to the out-of-sample criterion (Figure~\ref{fig:propagator-timescale-selection}): $3$ for ES, $4$ for NQ, $3$ for YM and RTY, and $1$ or $2$ for the Treasuries. Further factors leave the criterion on a plateau before it degrades. The in-sample fit may continue to improve mechanically as additional factors are added, but the out-of-sample comparison does not justify interpreting a larger number of retained factors as evidence for a power-law kernel in this price-regression exercise. Since the criterion is nearly flat between one and about nine factors for every instrument, we use one common two-factor specification (decay rates $0.003$ and $0.1$\,s$^{-1}$) for all instruments as a parsimonious predictive representation of the propagator signal at the chosen horizon. Conversely, the deterioration of the out-of-sample criterion with more factors does not mean that the corresponding time scales are economically irrelevant: market-impact estimates are noisy and require very large samples for stable averages \cite{almgren2005direct,brokmann2015slow,bucci2020crossover}. Our estimates recover the relevant order of magnitude and test whether order-flow imbalance is a significant channel through which information enters prices. They are not exact market-impact measurements at each lag.

The same convention applies to the price proxy: robustness across the weighted mid-price, the usual mid-price, and the last traded price is reported separately because these proxies contain different microstructure noise at high frequency, and any disagreement is treated as a price-measure robustness issue rather than as evidence against the decomposition in Section \ref{sec:model-passive-liquidity}.\\

We now specialize the price equation to the affine specification used for calibration. Because an affine $\kappa$ is unbounded, this is not a direct corollary of the bounded theorem but a conditional affine representation in which the displayed improper conditional expectations are assumed to exist. Bounded or truncated affine forms satisfy the preceding conditions.

\begin{corro}
\label{cor:main_propagator} 
Assume the Hawkes and queue-reactive dynamics of Theorem \ref{thm:general_propagator}. Suppose that the affine price-sensitivity function is
$$
\kappa(q)=\kappa_0+\kappa_1 q,
$$
and assume that the improper conditional expectation
$$
J_t:=\lim_{T\to\infty}\int_t^T
\E\left[
q_{s-}^a\lambda_s^a-q_{s-}^b\lambda_s^b
\mid\Fc_t
\right]\ud s,
\quad \forall t \geq 0,
$$
exists almost surely, and that $\E_\pi[q^{x}\lambda^{x}]<\infty$ for all $x\in\{a,b\}$ (used in the dominated-convergence step of the proof). 
Then, with
$$
H_t:=\int_0^t q_{s-}^a\,\ud N_s^a-\int_0^t q_{s-}^b\,\ud N_s^b,
$$
one has, for all $t \geq 0$,
\begin{align}
P_t
&=
P_0+\kappa_0\int_0^t \xi(t-s)\,\ud D_s+\kappa_1(H_t+J_t).
\label{eq:main_propagator_decomp}
\end{align}
\end{corro}
The proof is in Appendix \ref{app:proofs-price-decomposition}. The coefficient $\kappa_0$ is the constant-sensitivity benchmark, while $\kappa_1$ has units of price per market order per queue unit. On any queue range where the affine specification is interpreted literally, the monotonicity condition on $\kappa$ corresponds to $\kappa_1\le0$. If $\kappa_1=0$, the queue-weighted terms vanish and \eqref{eq:main_propagator_decomp} reduces to the classical propagator representation, so the passive-impact component is entirely captured by the terms involving $\kappa_1$.
\begin{rque}\label{rem:centered}
Writing $\bar q:=\int q\,P_q(\ud q)$ and $\kappa(q)=\bar\kappa+\kappa_1(q-\bar q)$ with $\bar\kappa=\kappa_0+\kappa_1\bar q$ gives the equivalent centered decomposition
$P_t=P_0+\bar\kappa\,G_t+\kappa_1(\widetilde H_t+\widetilde J_t)$, where
$\widetilde H_t:=\int_0^t (q_{s-}^a-\bar q)\,\ud N_s^a-\int_0^t (q_{s-}^b-\bar q)\,\ud N_s^b$ for all $t \geq 0$
and $\widetilde J_t$ is the corresponding centered anticipation term. This form matches Theorem~\ref{thm:general_propagator} term by term: the propagator coefficient is $\bar\kappa$ and $\kappa_1(\widetilde H+\widetilde J)=\mathcal R$. Empirically, the centered form is also better conditioned because it removes from $H$ the component $\bar q\,D_t$ that is collinear with the propagator signal. Section~\ref{subsec:affine-queue-sensitivity} uses the centered signal as a collinearity diagnostic.
\end{rque}
\begin{rque}
The calibration of Section~\ref{sec:calibration} and the price regressions serve distinct roles. Section \ref{sec:calibration} estimates the primitives of the order-flow and queue-reactive dynamics (the Hawkes kernel entering $\xi$ and the state-dependent intensities $\lambda^L$ and $\lambda^C$), which provide a model-consistent propagator signal when the Hawkes restrictions are imposed. The regressions instead project observed price changes on the signals implied by Theorem~\ref{thm:general_propagator} and therefore allow for finite-sample misspecification, volume marks, price-proxy noise, and the omitted anticipation term $J$. Two empirical consequences follow. First, the Hawkes-based kernel $\xi$ is a disciplined benchmark, but not the only admissible finite-sample signal: the fixed-decay reduced-form dictionary is a robustness device whose coefficients are Hawkes weights only under the restrictions of Lemma~\ref{lem:xi}. Second, because regressions use horizon-$\delta$ increments, the relevant time scales are those predictive at that horizon, which need not coincide exactly with the time scales that dominate the point-process likelihood at the event scale.

\end{rque}

\subsection{Affine Specification of Queue-Dependent Price Sensitivity}
\label{subsec:affine-queue-sensitivity}

We now translate Corollary \ref{cor:main_propagator} into the queue-dependent empirical component. We use the affine specification $\kappa(q)=\kappa_0+\kappa_1 q$, where $q$ denotes the predictable best-queue size on the consumed side, measured empirically as the AES-normalized queue introduced in Section \ref{subsec:data-order-book-reconstruction} and Appendix~\ref{app:calibration-details}. We restrict the affine form to observed queues, as the theory assumes bounded $\kappa$. In \eqref{eq:main_propagator_decomp}, each trade is weighted by the queue available immediately before it, and the anticipation term $J$ cannot be observed from a single book path and has no closed-form estimator here.\\

The marked empirical counterpart of $H$ is
$$
H_t^{V,d}
:=
\int_{T_d^0}^t q_{s^-}^{a,d}\,\ud V_s^{a,d}
-
\int_{T_d^0}^t q_{s^-}^{b,d}\,\ud V_s^{b,d}
=
\sum_{T_d^0<\tau\le t,\ \tau\in N^{a,d}}q_{\tau-}^{a,d}
\frac{v_\tau^{a,d}}{\widehat{\mathrm{AES}}}
-
\sum_{T_d^0<\tau\le t,\ \tau\in N^{b,d}}q_{\tau-}^{b,d}
\frac{v_\tau^{b,d}}{\widehat{\mathrm{AES}}},
\quad \forall t\in[T_d^0,T_d^1].
$$
Both $q$ and $v/\widehat{\mathrm{AES}}$ are dimensionless in AES units, so the coefficient on $H^V$ is measured in price units per aggressive-volume unit per queue unit. If event sizes are ignored, $V^{a/b}=N^{a/b}$ and $H^V$ reduces to the observable term in Corollary \ref{cor:main_propagator}.

For a fixed horizon $\delta$, the model-implied increment identity is
$$
\Delta_i^\delta P^d
=
\kappa_0\Delta_i^\delta G^d
+\kappa_1\Delta_i^\delta H^d
+\kappa_1\Delta_i^\delta J^d
$$
in the unit-size model, and its marked empirical analogue replaces $G,H$ by the volume-based signals. Because $J$ is unobserved, the baseline affine price regression is
\begin{equation}
\Delta_i^\delta P^{\mathrm{o},d}
=
\kappa_0^\delta\,\Delta_i^\delta \widehat G^d
+\kappa_1^\delta\,\Delta_i^\delta H^{V,d}
+u_i^{\delta,d}.
\label{eq:empirical-affine-regression-revised}
\end{equation}
In the reduced-form propagator version, $\Delta_i^\delta \widehat G^d$ is replaced by the selected dictionary regressors in \eqref{eq:empirical-reduced-form-propagator}, yielding
\begin{equation}
\Delta_i^\delta P^{\mathrm{o},d}
=
\theta_0^\delta\,\Delta_i^\delta D^{V,d}
+\sum_{\ell\in S}\theta_\ell^\delta\,\Delta_i^\delta X^{\ell,d}
+\gamma^\delta\,\Delta_i^\delta H^{V,d}
+u_i^{\delta,d}.
\label{eq:empirical-affine-reduced-form}
\end{equation}
The constrained specification \eqref{eq:empirical-affine-regression-revised} is closest to the theory because the propagator kernel is inherited from the Hawkes calibration. The reduced form aids prediction, but $(\theta_\ell^\delta)_{\ell\in S}$ and $\gamma^\delta$ remain projection coefficients absent the structural restrictions. Thus \eqref{eq:empirical-affine-reduced-form} tests incremental prediction by queue state, while \eqref{eq:empirical-affine-regression-revised} tests the Hawkes-implied structural specification. 

The regression residual contains several distinct terms:
$$
u_i^{\delta,d}
=
\kappa_1^\delta\Delta_i^\delta J^d
+
\Delta_i^\delta(P^{\mathrm{o},d}-P^d)
+
\text{kernel-estimation, volume-marking, and sampling errors}.
$$
Least squares therefore identifies the structural affine coefficient only under an additional orthogonality condition between this residual and the displayed regressors. Otherwise, the $\Delta H^V$ coefficient is only its best linear projection conditional on the propagator. The sign restriction remains economically meaningful: if $\kappa$ is non-increasing on the observed queue range, then the affine coefficient should satisfy $\kappa_1\le0$. A negative coefficient means that, for a fixed aggressive volume, a thicker queue on the consumed side attenuates the price response: it lowers the upward price contribution of buy market orders and reduces the magnitude of the downward contribution of sell market orders, because the signed queue-weighted term enters with the opposite sign.

\paragraph{Controlling for the anticipation term.} $J$ need not be orthogonal to the regressors because it depends on the same queues and intensities driving $\widehat G$ and $H^V$. Proposition~\ref{prop:explicit_bound} suggests an observable control: in the decoupled case $\phi=0$ with affine drift, $J_t=\frac{\mu}{-c_\lambda}(q_t^a-q_t^b)$ exactly. We therefore also estimate
$$
\Delta_i^\delta P^{\mathrm{o},d}
=
\kappa_0^\delta\,\Delta_i^\delta \widehat G^d
+\kappa_1^\delta\,\Delta_i^\delta H^{V,d}
+\kappa_J^\delta\,\Delta_i^\delta\big(q^{a,d}-q^{b,d}\big)
+u_i^{\delta,d},
$$
where under the Poisson benchmark $\kappa_J=\kappa_1\mu/(-c_\lambda)$. The stability of $\widehat\kappa_1$ when this control is added is our primary diagnostic for omitted-$J$ bias. The estimate is unchanged at $-8.1\times10^{-6}$, and the out-of-sample $R^2$ moves from $0.437$ to $0.438$ (Table~\ref{tab:main-results-clean}). When $\phi\neq0$, $J$ also loads on the Hawkes state, part of which is already spanned by the exponential factors $X^{\ell}$ of the reduced form.

The empirical contribution of passive liquidity is assessed by comparing \eqref{eq:empirical-affine-regression-revised} or \eqref{eq:empirical-affine-reduced-form} with the corresponding constant-sensitivity benchmark on the same horizon, price proxy, and train/test split. This comparison isolates the incremental information in the queue-weighted imbalance $H^V$ from effects attributable solely to the propagator dictionary or price measure.

The same affine specification also gives an empirical estimate of the average price sensitivity $\bar{\kappa}$ appearing in Theorem \ref{thm:general_propagator}. Since $\bar\kappa=\kappa_0+\kappa_1\int q\,P_q(\ud q)$, we set
$$
\widehat{\bar{\kappa}}^\delta
=
\widehat{\kappa}_0^\delta
+\widehat{\kappa}_1^\delta
\sum_q q\,\widehat P_q(q),
$$
where $\widehat P_q$ is the empirical distribution of the AES-normalized consumed-side best queues immediately before each trade, the regressor of the $\kappa_1$ fit, evaluated only on the observed queue support. This trade-weighted average matches the regression, while $P_q$ in Theorem~\ref{thm:general_propagator} is time-weighted.

We accompany the affine estimates with two specification checks. First, $\widehat\kappa_0+\widehat\kappa_1 q$ remains positive far beyond the observed range, crossing zero only at $-\widehat\kappa_0/\widehat\kappa_1\approx$$528$ AES units, against $85$ AES at the $99$th percentile of the queue found by trades. Second, we re-estimate the regressions using the bounded alternatives $\kappa(q)=\kappa_0+\kappa_1\log(1+q)$ and $\kappa(q)=\kappa_0+\kappa_1/(1+q)$, both consistent with decreasing sensitivity. Their out-of-sample $R^2$ is $0.444$ and $0.426$, against $0.437$ for the affine form (Table~\ref{tab:main-results-clean}).

\subsection{More General Models}

The affine best-queue specification is the most parsimonious empirical version of Corollary \ref{cor:main_propagator}, but the scalar queue state can also be replaced by observable book-state vectors. These empirical extensions of \eqref{eq:empirical-affine-regression-revised} do not yield a new closed form for $J$.

Let $q_{j,t}^{a,d}$ and $q_{j,t}^{b,d}$ denote the AES-normalized displayed volumes at depth $j$ on the ask and bid sides, with $j=1,\ldots,\ell$. With the same predictable convention, the consumed-side depth vector is $(q_{1,t-}^{a,d},\ldots,q_{\ell,t-}^{a,d})$ for a buy market order and $(q_{1,t-}^{b,d},\ldots,q_{\ell,t-}^{b,d})$ for a sell market order.\\

\textbf{Aggregated consumed-side depth.}
A first extension keeps one liquidity coefficient but replaces the best queue by the total displayed depth over the first $\ell$ levels:
$$
\kappa(q_1,\ldots,q_\ell)
=
\kappa_0+\kappa_1\sum_{j=1}^{\ell}q_j.
$$
The associated observable signal is
$$
H_t^{V,\mathrm{agg},d}
:=
\int_{T_d^0}^t\left(\sum_{j=1}^{\ell}q_{j,s^-}^{a,d}\right)\ud V_s^{a,d}
-
\int_{T_d^0}^t\left(\sum_{j=1}^{\ell}q_{j,s^-}^{b,d}\right)\ud V_s^{b,d},
\quad \forall t\in[T_d^0,T_d^1].
$$
This specification allows liquidity near the best quote to contribute to price resistance even when it is not concentrated at the first level. It is less local than the best-queue model, but it preserves a single queue-sensitivity coefficient, and on the observed support the monotonicity interpretation remains $\kappa_1\le0$.\\

\textbf{Depth-specific consumed-side coefficients.}
A more flexible specification lets each depth have its own coefficient:
$$
\kappa(q_1,\ldots,q_\ell)
=
\kappa_0+\sum_{j=1}^{\ell}\kappa_j q_j.
$$
Define
$$
H_{j,t}^{V,d}
:=
\int_{T_d^0}^t q_{j,s^-}^{a,d}\,\ud V_s^{a,d}
-
\int_{T_d^0}^t q_{j,s^-}^{b,d}\,\ud V_s^{b,d},
\quad \forall t\in[T_d^0,T_d^1],\ j=1,\ldots,\ell.
$$
The empirical regression then replaces the single queue signal by
$$
\sum_{j=1}^{\ell}\kappa_j^\delta\,\Delta_i^\delta H_j^{V,d}.
$$
This model can distinguish the predictive content of liquidity at different depths, but it increases estimation variance and exposure to collinearity because nearby displayed depths are strongly correlated, so individual signs should not be over-interpreted without stability checks. Monotonicity applies to the fitted function on observed support, not to each unconstrained coefficient.\\

\textbf{Opposite-side information.}
One may also include the state of the non-consumed side (the bid for a buy market order, the ask for a sell market order). A linear specification is
$$
\kappa(q_1,\ldots,q_\ell,q_1^{\mathrm{opp}},\ldots,q_\ell^{\mathrm{opp}})
=
\kappa_0+\sum_{j=1}^{\ell}\kappa_j q_j
+\sum_{j=1}^{\ell}\zeta_j q_j^{\mathrm{opp}}.
$$
The additional empirical signals are
$$
K_{j,t}^{V,d}
:=
\int_{T_d^0}^t q_{j,s^-}^{b,d}\,\ud V_s^{a,d}
-
\int_{T_d^0}^t q_{j,s^-}^{a,d}\,\ud V_s^{b,d},
\quad \forall t\in[T_d^0,T_d^1],\ j=1,\ldots,\ell.
$$
These terms are related to order-book imbalance: consumed-side liquidity measures the resistance of the queue being hit, while opposite-side liquidity records the support behind the trade direction. A positive $\zeta_j$ matches the usual imbalance intuition but is not implied by Theorem~\ref{thm:general_propagator}. Because the regressors are highly correlated, we treat this mainly as an out-of-sample robustness check. For each extension, $\bar\kappa$ is the fitted $\kappa$ averaged over the empirical state distribution, a finite-sample analogue of the stationary average in Theorem \ref{thm:general_propagator}, not a separate structural parameter.\\

\subsection{Estimation Results and Stability Diagnostics}
\label{subsec:fitting-results}

Table~\ref{tab:main-results-clean} reports the main estimates at $\delta=\delta^*$ for the weighted mid-price proxy, with day-by-day (Fama MacBeth) standard errors. Each rolling fit uses 16 days for estimation and the next day for testing, with 193 filtered test days. $R^2_{\mathrm{OOS}}$ is the mean daily out-of-sample $R^2$ and $R^2_{\mathrm{IS}}$ the mean training-window fit. In the implementation, the propagator block is the reduced-form dictionary with the selected decay rates $\{0.003,0.1\}$, so the affine specification \eqref{eq:empirical-affine-regression-revised} and the reduced-form specification \eqref{eq:empirical-affine-reduced-form} coincide. Units are index points per AES volume unit and, for $\kappa_1$, additionally per AES queue unit.
\begin{table}[htb!]

\centering\footnotesize
\resizebox{\ifdim\width>\textwidth\textwidth\else\width\fi}{!}{%
\begin{tabular}{lccccc}
\toprule
Specification & $\widehat\kappa_0$ & $\widehat\kappa_1$ (s.e.) & $R^2_{\mathrm{IS}}$ & $R^2_{\mathrm{OOS}}$ & $\widehat{\bar\kappa}$ \\
\midrule
Propagator only & $3.8\times10^{-3}$ & & 0.482 & 0.422 & \\
Affine & $4.3\times10^{-3}$ & $-8.1\times10^{-6}\ (1.7\times10^{-6})$ & 0.500 & 0.437 & $4.1\times10^{-3}$ \\
Affine + J-control & $4.3\times10^{-3}$ & $-8.1\times10^{-6}\ (1.7\times10^{-6})$ & 0.500 & 0.438 & \\
Logarithmic & & & 0.513 & 0.444 & \\
Inverse & & & 0.500 & 0.426 & \\
\bottomrule
\end{tabular}}
\caption{Headline estimates at $\delta^*=212$\,s (weighted mid-price, 193 test days). Standard errors are day-by-day (Fama MacBeth). $\widehat{\bar\kappa}$ is evaluated at the mean consumed-side best queue, the queue on which $\kappa_1$ is fitted, of $20$ AES.}
\label{tab:main-results-clean}
\end{table}

\paragraph{Full book-state dictionary.}
Beyond the affine and logarithmic forms, we let the price sensitivity depend on the
full displayed state through a dictionary of $102$ volume-weighted signals: four
transforms (linear, logarithmic, square root and inverse) of the ten displayed
queues on the consumed side and on the resting side, the per-level and cumulative
depth imbalances, the center of mass of each side, the spread, and four indicators
of the level-one queue bin. Each signal $W_k$ enters as
$\sum_{t<s\le t+\delta}\varepsilon_s v_s W_k(\text{state}_{s^-})$, so that the
specification remains a linear model for the increment of the weighted mid with
the propagator block unchanged. With one hundred and two correlated regressors and
sixteen training days, ordinary least squares overfits: on the 2025 ES sample it
reaches a mean out-of-sample $R^2$ of $0.425$ with two test days below $-1$,
against $0.597$ for the partial ridge described next ($0.603$ with the forecast
truncation). We therefore estimate the
dictionary by a partial ridge in which the propagator block ($\kappa$ and the two
kernel terms) is left unpenalized and the dictionary coefficients receive a common
penalty, selected inside each training window on a thirteen-day fit and three-day
validation split over a logarithmic grid, so that no test information is used.
Finally, the one-day-ahead forecast is truncated at three training standard
deviations of the target. For the $102$-signal ridge this guard is nearly
inactive ($0.597$ without and $0.603$ with truncation, no test day below $-1$). It
matters for the extended power-series dictionaries of Section~\ref{sec:passive-impact-curves},
where single windows can extrapolate to an unseen book state.
Table~\ref{tab:ladder_es_2025} reports the resulting ladder with and without the
truncation, and Table~\ref{tab:clip_sensitivity_es_2025} the sensitivity to the
truncation level.

\begin{table}[htb!]

\centering\footnotesize
\caption{Daily out-of-sample $R^2$ by model (ES 2025, 193 test days), without and with the $3\sigma$ forecast truncation: number of parameters $p$, mean, median, fifth percentile, minimum, and days with $R^2<-1$.}
\label{tab:ladder_es_2025}
\resizebox{\ifdim\width>\textwidth\textwidth\else\width\fi}{!}{%
\begin{tabular}{l r r r r r r r}
\toprule
model & $p$ & mean & median & $p_{5}$ & min & days $R^2<-1$ & $n$ days \\
\midrule
\multicolumn{8}{l}{\emph{unclipped}} \\
propagator & 3 & 0.422 & 0.487 & 0.020 & -0.788 & 0 & 193 \\
affine & 4 & 0.437 & 0.500 & 0.056 & -0.940 & 0 & 193 \\
log & 4 & 0.446 & 0.495 & 0.014 & -0.930 & 0 & 193 \\
pair & 5 & 0.556 & 0.585 & 0.284 & -0.353 & 0 & 193 \\
full dictionary ridge & 105 & 0.597 & 0.624 & 0.342 & 0.010 & 0 & 193 \\
\midrule
\multicolumn{8}{l}{\emph{forecast clipped at 3 sigma of the training increments}} \\
propagator & 3 & 0.458 & 0.501 & 0.148 & -0.469 & 0 & 193 \\
affine & 4 & 0.469 & 0.515 & 0.148 & -0.434 & 0 & 193 \\
log & 4 & 0.473 & 0.512 & 0.164 & -0.469 & 0 & 193 \\
pair & 5 & 0.563 & 0.591 & 0.289 & -0.289 & 0 & 193 \\
full dictionary ridge & 105 & 0.603 & 0.626 & 0.352 & 0.146 & 0 & 193 \\
\bottomrule
\end{tabular}}
\end{table}

\begin{table}[htb!]

\centering\footnotesize
\caption{Mean daily out-of-sample $R^2$ and days below $-1$ for different forecast truncation levels (ES 2025).}
\label{tab:clip_sensitivity_es_2025}
\resizebox{\ifdim\width>\textwidth\textwidth\else\width\fi}{!}{%
\begin{tabular}{l r r r r r}
\toprule
model & no clip & 2 & 2.5 & 3 & 4 \\
\midrule
\multicolumn{6}{l}{\emph{mean daily $R^2$}} \\
propagator & 0.422 & 0.483 & 0.470 & 0.458 & 0.442 \\
affine & 0.437 & 0.492 & 0.480 & 0.469 & 0.455 \\
log & 0.446 & 0.495 & 0.484 & 0.473 & 0.460 \\
pair & 0.556 & 0.572 & 0.568 & 0.563 & 0.558 \\
full dictionary ridge & 0.597 & 0.604 & 0.604 & 0.603 & 0.600 \\
\bottomrule
\end{tabular}}
\end{table}

Fixed once on the 2025 ES sample and applied without re-tuning (the ridge penalty
is still selected inside each rolling window), the full dictionary improves on the
propagator in all twelve instrument-year cells of the four equity-index futures over
2023 to 2025, by $0.06$ to $0.19$ of mean daily out-of-sample $R^2$ with Newey West
$t$ statistics between $7$ and $13$ and gains on $85$ to $95\%$ of the test days,
and in the four Treasury futures of 2025 as well, by $0.06$ to $0.09$ with $t$
between $7$ and $12$ (Table~\ref{tab:cells}). The
paired daily tests of the ladder specifications in every cell are reported in
Tables~\ref{tab:significance} and~\ref{tab:significance-2025}.

\begin{table}[htb!]

\centering\footnotesize
\caption{Mean daily out-of-sample $R^2$ by instrument and year (specification fixed on ES 2025), lift of the full dictionary over the propagator, its Newey West $t$-statistic, and the share of days with a positive gain.}
\label{tab:cells}
\resizebox{\ifdim\width>\textwidth\textwidth\else\width\fi}{!}{%
\begin{tabular}{l l r r r r r r r}
\toprule
symbol & year & propagator & log & pair & champion & lift & NW $t$ & share positive \\
\midrule
ES & 2023 & 0.593 & 0.595 & 0.628 & 0.662 & 0.069 & 7.92 & 0.899 \\
ES & 2024 & 0.530 & 0.531 & 0.593 & 0.635 & 0.106 & 13.05 & 0.898 \\
ES & 2025 & 0.458 & 0.473 & 0.563 & 0.603 & 0.145 & 10.99 & 0.943 \\
NQ & 2023 & 0.638 & 0.638 & 0.652 & 0.697 & 0.058 & 12.64 & 0.879 \\
NQ & 2024 & 0.661 & 0.672 & 0.683 & 0.725 & 0.064 & 10.96 & 0.903 \\
NQ & 2025 & 0.520 & 0.540 & 0.537 & 0.612 & 0.092 & 8.14 & 0.855 \\
RTY & 2023 & 0.134 & 0.155 & 0.163 & 0.325 & 0.190 & 13.12 & 0.950 \\
RTY & 2024 & 0.184 & 0.197 & 0.262 & 0.359 & 0.176 & 11.66 & 0.929 \\
RTY & 2025 & 0.085 & 0.094 & 0.111 & 0.198 & 0.113 & 10.65 & 0.870 \\
YM & 2023 & 0.381 & 0.387 & 0.397 & 0.462 & 0.081 & 7.16 & 0.884 \\
YM & 2024 & 0.362 & 0.370 & 0.435 & 0.474 & 0.111 & 8.82 & 0.867 \\
YM & 2025 & 0.269 & 0.280 & 0.285 & 0.390 & 0.121 & 8.73 & 0.860 \\
\midrule
TN & 2025 & 0.159 & 0.169 & 0.176 & 0.222 & 0.064 & 7.83 & 0.814 \\
ZB & 2025 & 0.261 & 0.265 & 0.270 & 0.324 & 0.063 & 7.33 & 0.871 \\
ZF & 2025 & 0.222 & 0.222 & 0.227 & 0.293 & 0.071 & 9.24 & 0.851 \\
ZN & 2025 & 0.312 & 0.317 & 0.336 & 0.401 & 0.090 & 11.88 & 0.933 \\
\bottomrule
\end{tabular}}
\end{table}

Because the preceding specifications are estimated on a finite sample, we assess their stability across calendar time and intraday segments, to test whether the fitted projection coefficients are driven by particular subsamples. We hold the price proxy, $\delta^*$, dictionary $S$, and split convention fixed, varying only the estimation sample.

Calendar-time stability is assessed with twenty-day rolling windows: if $d_1,\ldots,d_n$ are the retained trading days, we re-estimate the affine regression on each window $\mathcal W_m:=\{d_m,\ldots,d_{m+19}\}$, $m=1,\ldots,n-19$. The relevant objects are the signs, orders of magnitude, and relative variation of the propagator and queue-sensitivity coefficients. A consistently negative queue coefficient supports interpreting $H^V$ as a persistent liquidity-related signal, whereas large movements would reflect local conditions, sampling noise, or projection misspecification.

For intraday stability, we re-estimate the regression in each 15-minute bin, pooling days. Similar coefficients across bins indicate that the price response to signed and queue-weighted order flow is approximately stable over the trading day, despite nonstationarity in the price level, whereas material intraday variation makes full-sample coefficients time-of-day averages.

Exact equality of rolling or intraday estimates is not expected because observed price increments also contain the unobserved anticipation term $J$, price-proxy errors, and other high-frequency effects. Stability is therefore assessed through signs, relative magnitudes, and predictive performance rather than pointwise constancy of the fitted coefficients.

\paragraph{Stability of the queue slope.}
Figure~\ref{fig:stability} tracks the affine coefficients over the year and over
the trading day. In rolling windows of twenty retained sessions the pooled
estimate of $\kappa_1$ is negative in every window, between $-5\times10^{-6}$ and
$-2\times10^{-5}$ index points per AES unit of volume and per AES unit of queue
for most of the year, and reaches $-2.5\times10^{-5}$ in the windows ending in
mid May 2025, when $\kappa_0$ also rises from about $3.5\times10^{-3}$ to
$1.0\times10^{-2}$: the volatility episode of the spring raised the price
sensitivity to trades and steepened its dependence on the queue at the same time.
Split at mid-year, the day-by-day mean slope is $-1.8\times10^{-5}$ ($t=-5.8$) in
the first half and $-9.5\times10^{-6}$ ($t=-7.1$) in the second, a difference of
$-8.9\times10^{-6}$ that is itself significant ($t=-2.6$), while the out-of-sample
gain of the affine term over the propagator is positive in both halves ($0.015$
and $0.015$). Across the twenty-four fifteen-minute bins of the session, the pooled
slope is negative in all of them and its day-by-day mean in twenty, so the intraday evidence is one of a stable sign
without detectable time-of-day structure. The in-sample gain of the affine term is
positive in every bin but small (at most about six percentage points of $R^2$).
Per-bin out-of-sample gains are mostly negative and imprecise (twenty-two of twenty-four
bins below zero, none with $t<-2$), which reflects the small size of a
fifteen-minute test block rather than a time-of-day effect.

\section{Passive Impact Curves and Execution Interpretation}
\label{sec:passive-impact-curves}
\subsection{Counterfactual Definition of Passive Impact}

The price-formation model also yields a counterfactual definition of passive market impact. It compares the price generated by the observed order book with that generated by an alternative book containing an additional passive schedule, while holding the primitive market randomness fixed. Shared randomness isolates the passive intervention from independent simulation noise. This is the pathwise viewpoint used in \cite{jaisson2015market,ouazzaniChahdiRosenbaumSzymanskiPassive} and is consistent with the queue-reactive construction of Section \ref{subsec:queue-dependent-price-sensitivity}.

Consider a passive sell schedule, represented by a cumulative process $L^o$ posting additional liquidity on the ask side. The passive buy case is symmetric, with the bid side replacing the ask side and signs reversed. At valuation time $t$, the market is not assumed to anticipate future child orders. We therefore use the truncated schedule
$$
L_s^{o,t}:=L^o_{s\wedge t},\quad \forall s \geq 0.
$$
The counterfactual ask queue is
$$
\widetilde q_s^{a,t}
=
q_0^a+\widetilde L_s^{a,t}-\widetilde C_s^{a,t}-N_s^a+L_s^{o,t},
\quad \forall s \geq 0,
$$
where $\widetilde L^{a,t}$ and $\widetilde C^{a,t}$ are generated from the same Poisson random measures $\pi^{L,a}$ and $\pi^{C,a}$ as the factual queue, but with thresholds evaluated at $\widetilde q_{s^-}^{a,t}$:
$$
\widetilde L_s^{a,t}
=
\int_0^s\int_0^\infty
\mathbbm{1}_{\{z\le \lambda^L(\widetilde q_{u-}^{a,t})\}}
\,\pi^{L,a}(\ud u,\ud z),
\qquad
\widetilde C_s^{a,t}
=
\int_0^s\int_0^\infty
\mathbbm{1}_{\{z\le \lambda^C(\widetilde q_{u-}^{a,t})\}}
\,\pi^{C,a}(\ud u,\ud z).
$$
Because $N^a$ follows a Hawkes process with queue-independent intensity, the factual and counterfactual worlds share the same $N^a$. Remark~\ref{rem:qr-mo} discusses the queue-reactive extension, in which aggressive flow must also be coupled counterfactually.

The counterfactual price of the passive sell schedule is
$$
\widetilde P_t^{\,\mathrm{sell}}
=
P_0+
\lim_{T\to\infty}
\E\left[
\int_0^T\kappa(\widetilde q_{s-}^{a,t})\,\ud N_s^a
-
\int_0^T\kappa(q_{s-}^{b})\,\ud N_s^b
\ \middle|\ \Fc_t
\right],
\quad \forall t \geq 0,
$$
whenever the displayed limit is finite. The passive impact is
$$
\mathrm{MI}_t^{\mathrm{sell}}
:=
\widetilde P_t^{\,\mathrm{sell}}-P_t,
\quad \forall t \geq 0.
$$
Subtracting the factual price gives
$$
\mathrm{MI}_t^{\mathrm{sell}}
=
\lim_{T\to\infty}
\E\left[
\int_0^T
\left\{\kappa(\widetilde q_{s-}^{a,t})-\kappa(q_{s-}^a)\right\}
\,\ud N_s^a
\ \middle|\ \Fc_t
\right].
$$
Under the monotone shared-noise coupling, the added ask-side liquidity gives $\widetilde q_s^{a,t}\ge q_s^a$ pathwise, so if $\kappa$ is non-increasing, then $\kappa(\widetilde q_{s-}^{a,t})-\kappa(q_{s-}^a)\le0$. Conditional monotone convergence therefore justifies the infinite-horizon representation
\begin{equation}
\mathrm{MI}_t^{\mathrm{sell}}
=
\E\left[
\int_0^\infty
\left\{\kappa(\widetilde q_{s-}^{a,t})-\kappa(q_{s-}^a)\right\}
\,\ud N_s^a
\ \middle|\ \Fc_t
\right]
\approx
\int_0^t
\left\{\kappa(\widetilde q_{s-}^{a,t})-\kappa(q_{s-}^a)\right\}
\,\ud N_s^a
,
\label{eq:impact_infinite}
\end{equation}
provided the factual and counterfactual price limits are finite. Under this sign convention, a passive sell schedule therefore has non-positive price impact and a passive buy schedule has the symmetric non-negative impact. This does not make passive execution costless: passive posting changes future price formation by modifying the queue state faced by subsequent aggressive orders. The approximation in \eqref{eq:impact_infinite} drops the anticipated future difference $\int_t^\infty\E[\{\kappa(\widetilde q_{s-}^{a,t})-\kappa(q_{s-}^a)\}\lambda_s^a\mid\Fc_t]\,\ud s$ and becomes exact as $t\to\infty$. Under affine $\kappa$, we evaluate the conditional expectation analytically as the realized displacement plus a resolvent-tail term, including the future component.
\begin{rque}
Because costless posting and cancellation can move prices, passive/aggressive round trips may enable manipulation \cite{huberman2004price,gatheral2010no}. We do not analyze the admissibility of such strategies here. Deriving conditions on $(\kappa,\phi,\lambda^L,\lambda^C)$ that exclude quasi-arbitrage from passive posting remains an open question.
\end{rque}

\subsection{Impact Profile in Time}

We next describe numerical experiments that recover market-impact time profiles under the calibrated Hawkes-queue-reactive model, computing counterfactual price paths from the calibrated primitives without imposing a parametric impact curve. All volumes and queue sizes are expressed in AES units, and the price sensitivity is evaluated with the affine specification $\kappa(q)=\kappa_0+\kappa_1 q$ on the observed queue range. The factual and counterfactual queues $q$ and $\widetilde q$ are simulated under the shared-noise conditional procedure of \cite{conditionalSimulationPaper}, which we use as a numerical input without reproducing its algorithm or justification.

The first experiment provides the standard aggressive benchmark. We consider a buy market-order strategy that executes a total AES-normalized volume $V$ over the execution interval $[0,T]$, split into $n_{\mathrm{MO}}$ equal child market orders placed at $\tau_i^{\mathrm{MO}}:=iT/n_{\mathrm{MO}}$, $i=1,\ldots,n_{\mathrm{MO}}$. The cumulative intervention volume, truncated at valuation time $t$, is
$$
N_s^{o,t}
=
\sum_{i=1}^{n_{\mathrm{MO}}}
\frac{V}{n_{\mathrm{MO}}}
\mathbbm{1}_{\{\tau_i^{\mathrm{MO}}\le s\wedge t\}},
\quad \forall s \geq 0.
$$
We choose $n_{\mathrm{MO}}$ sufficiently large that this step function approximates constant-rate execution at rate $V/T$. In a unit-event simulator, the same schedule is implemented by the corresponding AES-normalized sequence of unit or marked market-order arrivals.

Each of the $N_{\mathrm{MC}}=1000$ replicas simulates background Hawkes flow and factual queue dynamics from the calibrated model. The counterfactual ask queue $\widetilde q^{a,t,k}$ is then obtained by inserting the child market orders into the same latent noise environment. The associated aggressive market-impact path $\mathrm{MI}^{\mathrm{MO},k}_t$, $0\le t\le T_{\mathrm{plot}}$, is the difference between the counterfactual price under the buy market-order schedule and the factual price under the same background noise. The counterfactual uses the same price functional, replacing the ask queue by $\widetilde q^{a,t,k}$ and adding the buy intervention, and $T_{\mathrm{plot}}\ge T$ is chosen so that both the execution phase and the post-execution relaxation phase are visible.

The reported average curve is
$$
\overline{\mathrm{MI}}^{\mathrm{MO}}_t
=
\frac{1}{N_{\mathrm{MC}}}\sum_{k=1}^{N_{\mathrm{MC}}}
\mathrm{MI}^{\mathrm{MO},k}_t,
\quad \forall t\in[0,T_{\mathrm{plot}}].
$$
This reproduces the standard market-impact time-profile experiment associated with Bouchaud-type propagator curves, recording the build-up during execution, the peak or end-of-execution impact, and subsequent relaxation. No shape restriction is imposed ex ante.

\paragraph{Passive posted-liquidity impact.}
The second experiment applies the same construction to a passive posted-liquidity schedule, with the side convention of \eqref{eq:impact_infinite}: a passive sell schedule posting additional liquidity on the ask side (the passive buy case is symmetric).

The passive strategy posts a total AES-normalized volume $V$ over $[0,T]$, split into $n_{\mathrm{LO}}$ equal child limit orders placed at $\tau_i^{\mathrm{LO}}:=iT/n_{\mathrm{LO}}$, $i=1,\ldots,n_{\mathrm{LO}}$. The non-anticipative posted-volume process used at valuation time $t$ is
$$
L_s^{o,t}
=
\sum_{i=1}^{n_{\mathrm{LO}}}
\frac{V}{n_{\mathrm{LO}}}
\mathbbm{1}_{\{\tau_i^{\mathrm{LO}}\le s\wedge t\}},
\quad \forall s \geq 0.
$$
Again, $n_{\mathrm{LO}}$ is chosen sufficiently large to approximate constant-rate posting. For each of $N_{\mathrm{MC}}=1000$ replicas, factual and counterfactual queues share the same Poisson noise, with the latter including $L^{o,t}$. The passive market-impact path $\mathrm{MI}^{\mathrm{LO},k}_t$, $0\le t\le T_{\mathrm{plot}}$, is then the simulated version of \eqref{eq:impact_infinite} under the calibrated Hawkes kernel, calibrated queue-reactive insertion and cancellation intensities, and the affine queue-dependent sensitivity, and the mean curve $\overline{\mathrm{MI}}^{\mathrm{LO}}_t$ is defined as in the aggressive experiment.

The volume input in this experiment is the posted volume $V$, not the realized executed volume. Thus the experiment measures the price effect of displayed liquidity supplied to the book, not impact per executed order. It introduces no execution probability, fill model, or correction for conditional fill selection.

\paragraph{Liquidity regimes.}
To document how the price response varies with the prevailing state of the book, we
sort the sessions of the year into terciles of the time-weighted mean of the
consumed-side best queue, which define a low-liquidity regime (thin queues), a
medium regime and a high-liquidity regime (thick queues). Within each regime we
re-estimate the affine specification, which yields a regime-specific propagator
$G_r(\tau)$, the expected displacement of the weighted mid $\tau$ seconds after a
unit aggressive trade at the regime mean queue $\bar q_r$. Table~\ref{tab:regimes}
reports the estimates, and Figure~\ref{fig:three-propagators}, shown in the
introduction, plots the three curves. Thinner books exhibit a larger immediate
impact, about 2.3 times as large in the low-liquidity as in the high-liquidity regime,
and retain a larger share of it permanently.

\begin{table}[htb!]

\centering\footnotesize
\caption{Pooled affine fits by liquidity regime (ES 2025, AES units) with Fama MacBeth standard errors. $G_r(0)=\kappa_0^{r}+\kappa_1^{r}\bar q_r+w_1^{r}+w_2^{r}$ is the immediate response.}
\label{tab:regimes}
\resizebox{\ifdim\width>\textwidth\textwidth\else\width\fi}{!}{%
\begin{tabular}{l r r r r r r r}
\toprule
regime & $\bar q_r$ & $\kappa_0^{r}$ & $\kappa_1^{r}$ & $w_1^{r}$ & $w_2^{r}$ & $G_r(0)$ & $n$ days \\
\midrule
low & 7.7 & $6.3\times10^{-3}$ ($4.7\times10^{-4}$) & $-1.1\times10^{-5}$ ($4.8\times10^{-6}$) & $1.8\times10^{-3}$ ($2.9\times10^{-4}$) & $1.3\times10^{-3}$ ($3.0\times10^{-4}$) & $9.3\times10^{-3}$ & 61 \\
medium & 18.8 & $3.5\times10^{-3}$ ($1.6\times10^{-4}$) & $-4.5\times10^{-6}$ ($1.6\times10^{-6}$) & $1.1\times10^{-3}$ ($1.3\times10^{-4}$) & $1.5\times10^{-3}$ ($1.4\times10^{-4}$) & $5.9\times10^{-3}$ & 62 \\
high & 31.4 & $2.0\times10^{-3}$ ($1.2\times10^{-4}$) & $-4.7\times10^{-6}$ ($1.5\times10^{-6}$) & $1.3\times10^{-3}$ ($1.4\times10^{-4}$) & $9.0\times10^{-4}$ ($1.1\times10^{-4}$) & $4.0\times10^{-3}$ & 70 \\
\midrule
all sessions & 19.9 & $4.3\times10^{-3}$ ($2.0\times10^{-4}$) & $-8.1\times10^{-6}$ ($1.7\times10^{-6}$) & $1.5\times10^{-3}$ ($1.1\times10^{-4}$) & $1.1\times10^{-3}$ ($1.2\times10^{-4}$) & $6.7\times10^{-3}$ & 193 \\
\bottomrule
\end{tabular}}
\end{table}

\subsection{Impact Profile in Participation Rate}

We now use the preceding time-profile experiments for a comparative-static analysis in volume. Let $0<V_1<\cdots<V_J$ be a pre-specified grid of total AES-normalized strategy volumes. For each $V_j$, both experiments use the same calibrated inputs, affine sensitivity, and shared-noise simulation \cite{conditionalSimulationPaper}, with $N_{\mathrm{MC}}=1000$ Monte Carlo replicas.

The volume grid can equivalently be expressed in terms of participation rates,
$$
\nu_j
=
\frac{V_j}{Q_T^{\mathrm{bench}}},
\quad j=1,\ldots,J,
$$
where $Q_T^{\mathrm{bench}}$ is an empirical benchmark for market volume over the execution interval $[0,T]$. For example, $Q_T^{\mathrm{bench}}$ can be the average AES-normalized aggressive volume over comparable intervals of length $T$. The precise benchmark is fixed before running the experiment and reported with the results. This normalization does not change the simulated impact paths: it only rescales the horizontal axis from total volume $V_j$ to participation rate $\nu_j$.

For each total volume $V_j$, we compute the average pathwise peak absolute impact over the plotted horizon,
$$
I_{\max}^{\mathrm{MO}}(V_j)
=
\frac{1}{N_{\mathrm{MC}}}
\sum_{k=1}^{N_{\mathrm{MC}}}
\sup_{0\le t\le T_{\mathrm{plot}}}
\left|
\mathrm{MI}^{\mathrm{MO},k}_t
\right|,
\quad j=1,\ldots,J,
$$
and $I_{\max}^{\mathrm{LO}}(V_j)$ is defined in the same way from $\mathrm{MI}^{\mathrm{LO},k}$. Taking absolute values makes the statistic comparable across side conventions, particularly when comparing aggressive buy impact with the ask-side passive sell convention in \eqref{eq:impact_infinite}.

 The peak of the mean curve,
$$
\sup_{0\le t\le T_{\mathrm{plot}}}
\left|
\overline{\mathrm{MI}}^{\mathrm{MO}}_t(V_j)
\right|,
\qquad
\sup_{0\le t\le T_{\mathrm{plot}}}
\left|
\overline{\mathrm{MI}}^{\mathrm{LO}}_t(V_j)
\right|,
$$
is the \textit{primary} statistic, because pathwise $I_{\max}$ is upward biased by simulation and market noise ($\E\,\sup_t|\cdot|\ \ge\ \sup_t|\E[\,\cdot\,]|$), especially when true impact is small relative to noise. The pathwise peak $I_{\max}$ serves as a robustness check. We also report end-of-execution impact, $\overline{\mathrm{MI}}^{\mathrm{MO}}_T(V_j)$ and $\overline{\mathrm{MI}}^{\mathrm{LO}}_T(V_j)$, and post-execution impact at a fixed lag after $T$ within the plotted horizon. These alternatives distinguish the maximum path displacement from the end-of-execution and post-execution relaxation components.

 Both peak statistics are plotted against $V_j$ or $\nu_j$, and the main figures use the peak of the mean curve. 
The purpose is to assess whether the simulated curves are compatible with square-root-type market-impact laws, which is an empirical implication of the calibrated model at the selected execution horizon, not a maintained assumption. A standard diagnostic is a log-log regression of the peak statistic on volume, written here for $I_{\max}$ (the peak-of-mean version is analogous),
$$
\log I_{\max}^{\mathrm{MO}}(V_j)
=
a_{\mathrm{MO}}+b_{\mathrm{MO}}\log V_j+\varepsilon_j^{\mathrm{MO}},
\qquad
\log I_{\max}^{\mathrm{LO}}(V_j)
=
a_{\mathrm{LO}}+b_{\mathrm{LO}}\log V_j+\varepsilon_j^{\mathrm{LO}},
$$
or the same specification with $V_j$ replaced by $\nu_j$. The estimated slopes can then be compared with the square-root benchmark $b=1/2$, and the fitted curves can be plotted against $C\sqrt{V_j}$ or $C\sqrt{\nu_j}$ benchmarks.
 Because $\nu_j$ is a rescaled volume, the $\nu_j$-version is not a separate participation-rate test. The latent-liquidity models of \cite{toth2011anomalous,mastromatteo2014agent} motivate concave or square-root dependence on metaorder size rather than, by themselves, a generic $\sqrt{\nu}$ law. \cite{benzaquen2018multitimescale} shows theoretically that multi-timescale liquidity can recover a square-root size law over moderate participation-rate regimes and analyzes trading-rate dependence. \cite{durin2023twosquareroot} distinguishes square-root laws in cumulative volume at fixed participation and in participation at fixed volume, with near-linearity at low rates.

\begin{figure}[htb!]

  \centering
  \includegraphics[width=0.45\textwidth]{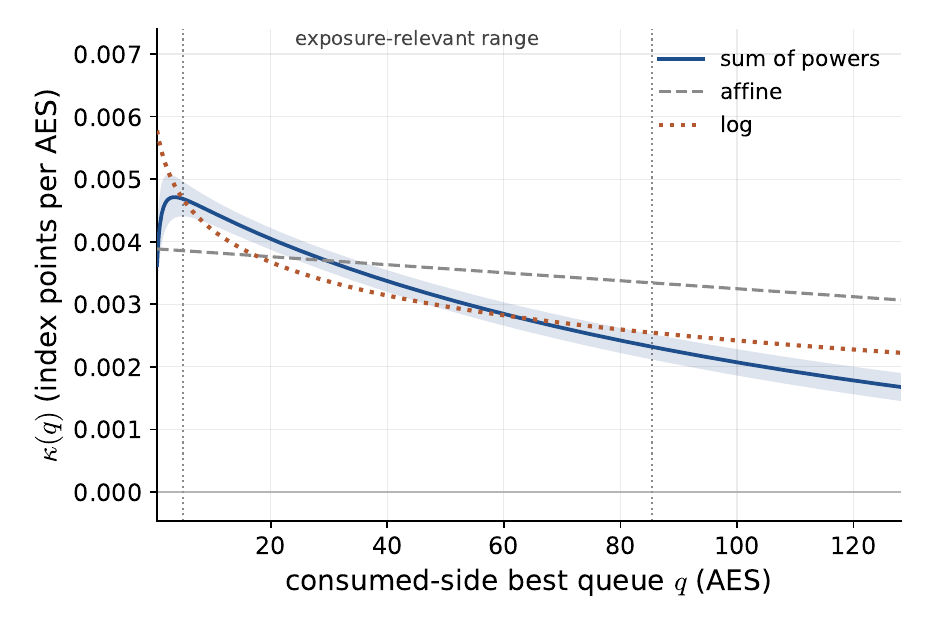}
  \caption{Sensitivity $\kappa(q)$ against the best queue (AES units): sum-of-powers fit with a two-standard-error band, and affine and logarithmic fits. Dotted lines: median and $99$th percentile of the queue found by trades.}
  \label{fig:kappa-powers}
\end{figure}

\paragraph{A nonlinear sensitivity for the participation-rate experiment.}
The affine form is a local description on the exposure-relevant range of queue
sizes. For the participation-rate experiment, which perturbs the stationary queue
by amounts that grow with the posted volume, we also estimate a sum-of-powers
specification
\[
\kappa(q)=c_0+\sum_{p\in\mathcal P}c_p\,q^{p}+c_{\log}\log(1+q),
\qquad
\mathcal P=\{\tfrac14,\tfrac12,\tfrac34,1,\tfrac32,2\},
\]
by the same rolling partial ridge, with the corresponding volume-weighted signals
added to the propagator block. The
specification reaches a mean out-of-sample $R^2$ of $0.479$ against $0.469$ for
the affine form and $0.473$ for the logarithmic form. Its paired daily gain over
the propagator is $0.021$ (Newey West $t=4.8$), the gain over the affine form, $0.010$, is significant ($t=2.9$), and the gain over the logarithmic form, $0.006$, is not ($t=1.6$). Figure~\ref{fig:kappa-powers} plots the fitted curve with a band of two
standard errors across the rolling windows, together with the ridge affine and
logarithmic fits. The curve is decreasing and convex on the exposure-relevant
range, from the median to the $99$th percentile of the queue found by trades
($5$ to $85$ AES): it lies close to the logarithmic form for small queues, crosses
the affine form near $30$ AES, and beyond that decays faster than both the affine and the logarithmic forms ($4.1\times10^{-3}$ index points per AES
at the trade-weighted mean queue of $20$ AES against $3.8\times10^{-3}$ affine and
$3.7\times10^{-3}$ logarithmic, and $2.3\times10^{-3}$, $3.3\times10^{-3}$,
$2.5\times10^{-3}$ at the $99$th percentile). Its local slope at the mean queue,
$-3.9\times10^{-5}$ per AES, is steeper than the affine slope
($-6.4\times10^{-6}$, the ridge shrinks it towards zero) and about as steep as the
logarithmic one ($-3.8\times10^{-5}$).

\paragraph{Stationary shift and participation rate.}
Following the asymptotic argument of \cite{ouazzaniChahdiRosenbaumSzymanskiPassive},
a passive schedule posting at a constant rate $\gamma$ (expressed as a fraction of
the aggressive rate $\bar\lambda$) shifts the stationary queue from
$q^\star$ to $\bar q^\star(\gamma)=q^\star+\gamma \bar\lambda/(-c_\lambda)$, where $c_\lambda$ $<0$ is the
slope of the calibrated affine drift $b=\lambda^L-\lambda^C$ and $q^\star$ is set to
the observed exposure-weighted mean queue ($101$ AES units) with the intercept of
$b$ adjusted so that $b(q^\star)=\bar\lambda$, and the impact per unit of
aggressive volume is $\kappa(\bar q^\star(\gamma))-\kappa(q^\star)$.
A power-law fit gives an exponent of
$0.99$ for the sum-of-powers form, against $0.99$ for the logarithmic form and $1$
by construction for the affine form. \cite{ouazzaniChahdiRosenbaumSzymanskiPassive} report $0.54$ under a
square-root sensitivity: with a mean reversion of about one per second the
stationary queue moves by only $\bar\lambda/(-c_\lambda)\approx16$ AES units at full
participation, over which every fitted sensitivity is close to linear.

\begin{figure}[htb!]

  \centering
  \includegraphics[width=\textwidth]{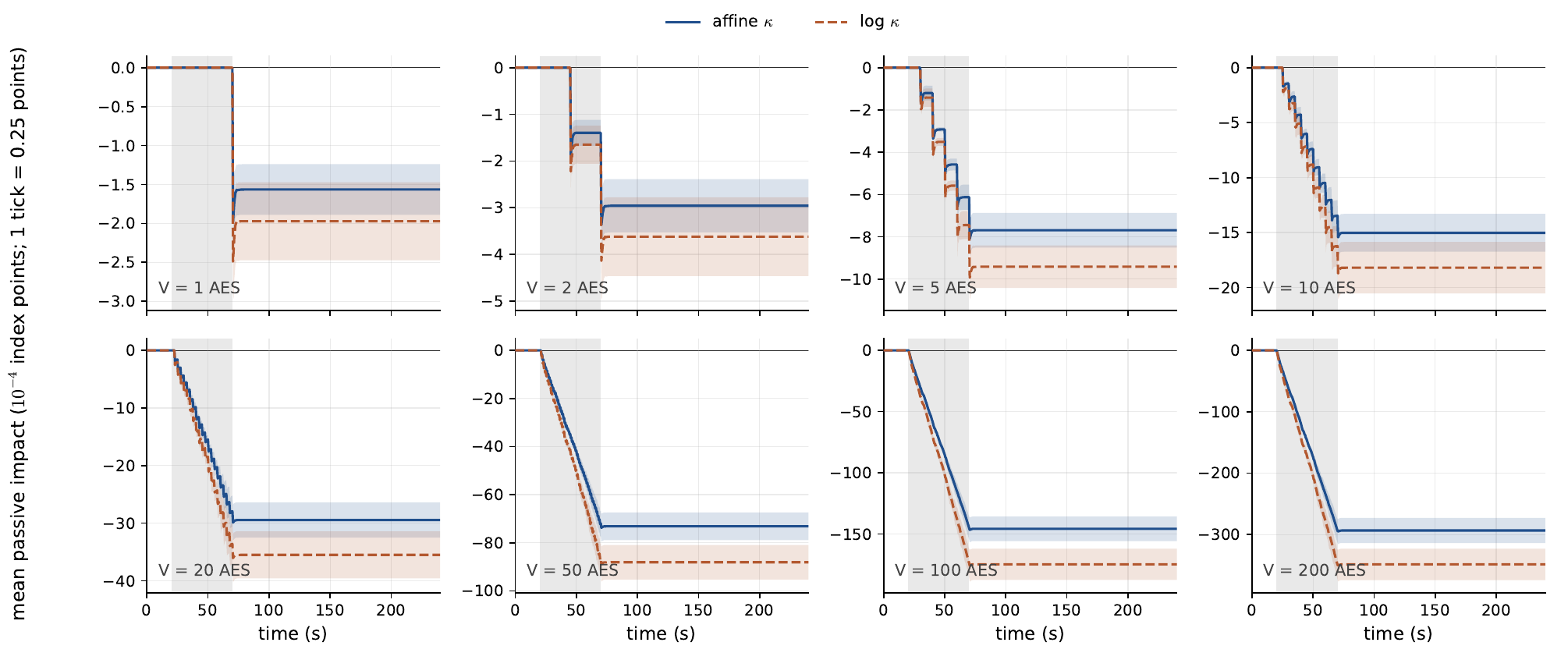}
  \caption{Simulated passive impact of a sell schedule of volume $V$ (AES units) posted over $50$ s (shaded), for the affine (solid) and logarithmic (dashed) sensitivities, with a band of one standard deviation across market histories.}
  \label{fig:cf-trajectories}
\end{figure}

\begin{figure}[htb!]

  \centering
  \includegraphics[width=0.48\textwidth]{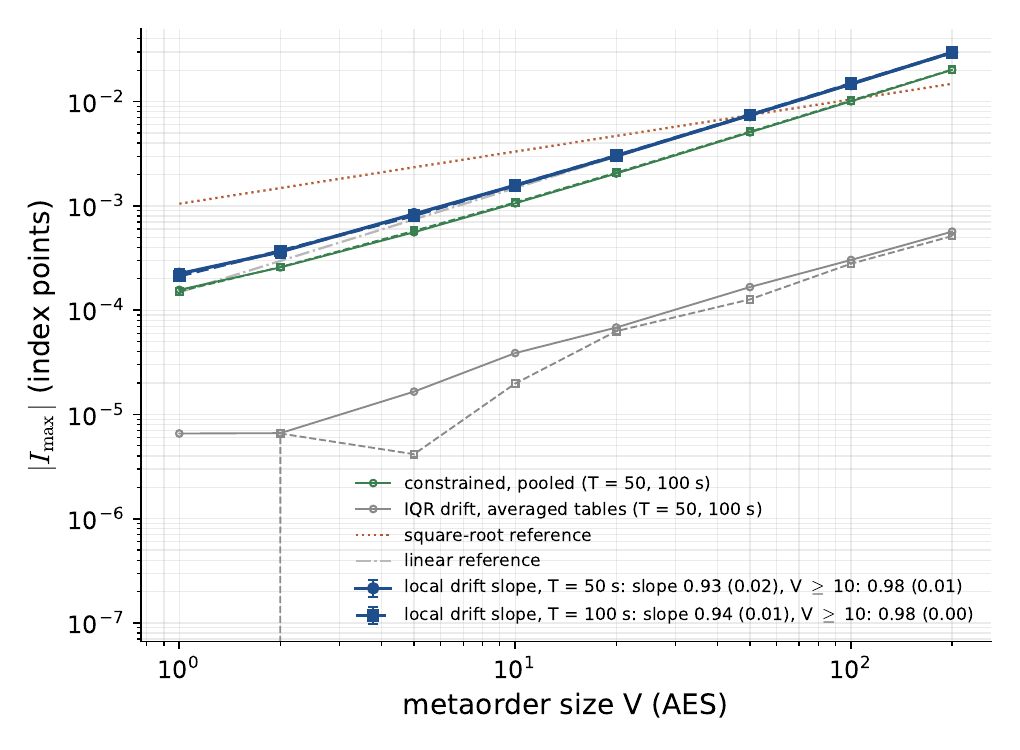}
  \includegraphics[width=0.48\textwidth]{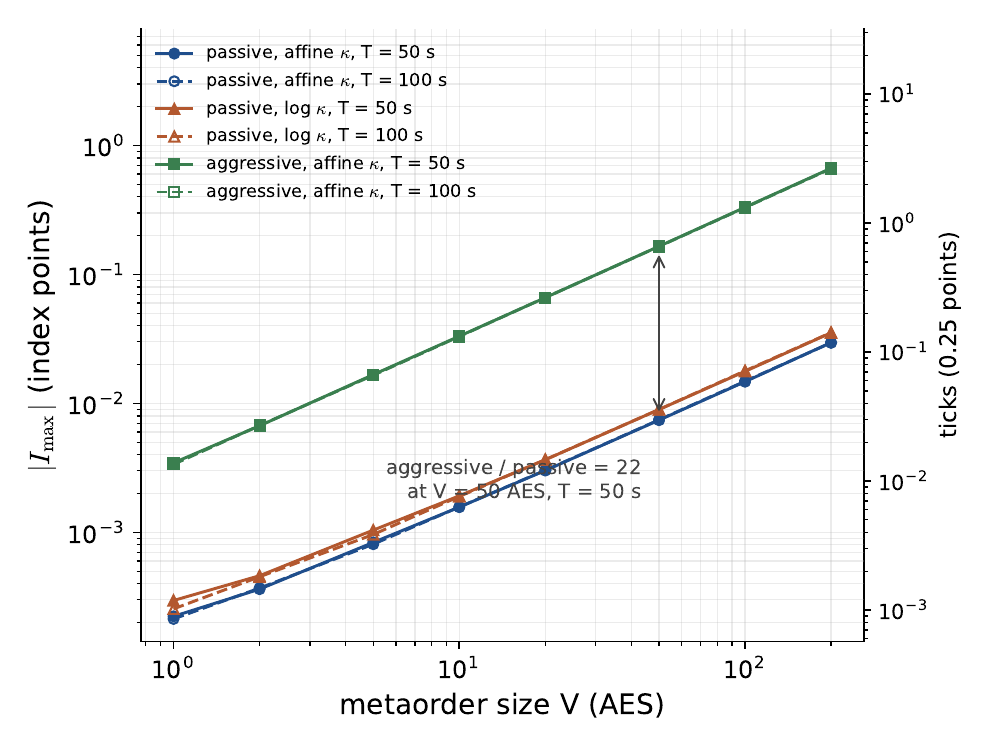}
  \caption{Left: peak of the mean passive impact against posted volume for three calibrations of the queue mean reversion, with square-root and linear references. Right: the same peak for passive posting and for the same volume executed with market orders.}
  \label{fig:cf-imax}
\end{figure}

\paragraph{Simulated passive impact.}
With the queue dynamics calibrated on the front-month stream
(Section~\ref{subsec:queue-reactive-calibration}), a passive sell schedule of $V$
AES units posted as unit limit orders equally spaced over $T=50$ s displaces the
weighted mid at its peak by $1.6\times10^{-3}$ index points for $V=10$, $7.4\times10^{-3}$ for
$V=50$ (about three percent of a tick), $1.5\times10^{-2}$ for $V=100$ and
$3.0\times10^{-2}$ for $V=200$ (12 percent of a tick) under the affine
sensitivity (mean over three independent market histories of $1000$ replicas
each, across-world dispersion of 7 to 10 percent). The displacement is
essentially permanent: the level at the end of the horizon is within 1 to 4
percent of the peak (4 percent at $V=10$, 1 percent at $V=200$).
Doubling the window to $T=100$ s changes the peak by less than 1\%, so
the impact depends on the posted volume rather than on the participation rate.
The log-log slope of the peak in $V$ is $0.98$ for $V \geq 10$ AES (standard error
$0.004$ on the ten world-mean points, $0.93$ over all sizes since the smallest
orders are absorbed before the peak), against $1$ for the aggressive benchmark. The same
volume traded aggressively displaces the price $22$ times more
(Figures~\ref{fig:cf-trajectories} and~\ref{fig:cf-imax}). The logarithmic and sum-of-powers sensitivities give peaks
1.2 and 1.7 times the affine one at $V=50$, in proportion to
their steeper local slope at the stationary queue. The affine sensitivity used in the simulations is the mean of the
rolling-window least-squares fits (slope $-9.3\times10^{-6}$ per AES unit of
queue), steeper than the pooled estimate of Table~\ref{tab:main-results-clean}
($-8.1\times10^{-6}$), and under the affine form the peaks scale in proportion. The calibration of the queue mean-reversion rate is the dominant uncertainty:
with the rate fitted on the whole state range ($1.6$ per second) the peak at
$V=100$ is $1.0\times10^{-2}$ index points instead of $1.5\times10^{-2}$, and
with the interquartile-filtered global fit of the drift on the same stream ($77$ per second), it falls
to $3.0\times10^{-4}$: a book that absorbs a posted
unit in a hundredth of a second leaves almost no trace on the price.

\paragraph{Regimes.}
Repeating the experiment on the same calibration with the regime-specific affine
sensitivities and stationary queues of Section~\ref{sec:passive-impact-curves}
(Figure~\ref{fig:cf-regimes}) gives peaks at $V=50$ of $8.5\times10^{-3}$ (low
liquidity), $3.6\times10^{-3}$ (medium) and $3.8\times10^{-3}$ (high liquidity) index points.
With a shared mean-reversion rate and an affine sensitivity the ratio of these
peaks is the ratio of the regime slopes $\kappa_1^{r}$ by construction, so this
experiment translates the regime estimates into price units rather than adding
independent evidence: passive impact, like aggressive impact, is about 2.2
times larger when the book is thin.

\subsection{Marginal Impact and the Order-Book Imbalance}
\label{subsec:imbalance}
The queue-dependent sensitivity can also be read as a function of the level-one
imbalance seen from the consumed side,
$\iota=(q^{c}_1-q^{o}_1)/(q^{c}_1+q^{o}_1)\in[-1,1]$, negative when the queue being
consumed is the thin one ($q^{c}_1$ and $q^{o}_1$ are the level-one queues on the consumed and opposite sides). We estimate
\[
\kappa(\iota)=\vartheta_0+\vartheta_1 \iota+\vartheta_2 \iota^{2}+\vartheta_3 \iota^{3}+\vartheta_5 \iota^{5}+\vartheta_{a}\,\lvert \iota\rvert+\vartheta_{r}\,\mathrm{sgn}(\iota)\sqrt{\lvert \iota\rvert}
\]
with the same rolling partial ridge, on its own and jointly with the affine queue
term. The imbalance-only sensitivity, left panel of
Figure~\ref{fig:kappa-imbalance}, is between $3.5$ and $4.1\times10^{-3}$ index
points per AES for $\iota$ between $-0.8$ and $0$, falling to $2.4\times10^{-3}$ as
the consumed queue empties, peaks at $4.1\times10^{-3}$ for a mild surplus on the
consumed side ($\iota\approx0.3$) and declines to $2.2\times10^{-3}$ when the
consumed queue holds the whole displayed depth. Added to the affine queue term, the
imbalance terms raise the mean out-of-sample $R^2$ from $0.469$ to $0.474$, a gain
over the affine form that is positive on $58\%$ of the test days. The gain of the joint specification over the propagator is $0.016$,
positive on $69\%$ of the days ($t=4.8$).

The right panel overlays two model-free curves on the same imbalance bins, in
ticks and in the consumed-side convention. The first is the displacement of the
mid-price $P^{\mathrm{mid}}$ per unit of traded volume after a trade,
$\mathbb E[\varepsilon_t(P^{\mathrm{mid}}_{t+\delta}-P^{\mathrm{mid}}_{t^-})\,|\,\iota_{t^-}]/\mathbb E[v_t\,|\,\iota_{t^-}]$
at $\delta=212$ s. The second is a finite-horizon version of the micro-price
adjustment of \cite{Stoikov2018}, $\mathbb E[P^{\mathrm{mid}}_{t+\delta}-P^{\mathrm{mid}}_t\,|\,\iota_t]$ evaluated
on the one second grid without conditioning on a trade, which measures the
mechanical drift of the mid towards the heavier side. Both decrease with $\iota$,
but the micro-price drift changes sign, from $+0.60$ ticks when the consumed side
is the thin one ($\iota=-0.8$) to $-0.60$ ticks when it is the thick one, whereas
the trade-conditional displacement remains non-negative: $0.31$ ticks per AES at
$\iota=-0.8$, $0.12$ at balance, and indistinguishable from zero once the consumed
side holds more than about seventy percent of the level-one depth. When a trade
consumes the thin queue it therefore rides a drift that would have occurred
anyway, and when it consumes the thick queue it offsets the mechanical drift in
the other direction. The sensitivity $\kappa(\iota)$ is not a restatement of the
micro-price: it is the incremental response of the price to a unit of traded
volume given the state the trade encounters, and it remains positive and of the
same order of magnitude at every imbalance.

\section{Conclusion}
\label{sec:conclusion}

We develop and calibrate a price-formation model in which sensitivity to aggressive orders depends on the state of the best queues. Theorem~\ref{thm:general_propagator} writes the price as the average-sensitivity Hawkes propagator plus a queue-state correction. With affine sensitivity, the correction is the observable queue-weighted imbalance plus an anticipation term, yielding a linear calibration. We estimate the Hawkes kernel and queue-reactive intensities using E-mini S\&P 500 futures data, construct the implied propagator signal, and estimate queue-sensitivity coefficients at an out-of-sample-selected horizon. The queue-sensitivity slope is negative, confirming that thicker queues attenuate trade impact. With the $3\sigma$ forecast truncation, the queue term raises the mean daily out-of-sample $R^2$ from $0.458$ to $0.469$ ($0.473$ for log sensitivity and $0.603$ for the full book state, Table~\ref{tab:ladder_es_2025}). Fixed on 2025 ES data, the full book-state specification improves on the propagator in all sixteen instrument-year cells, namely the twelve equity-index cells over 2023 to 2025 and the four Treasury futures of 2025 (Table~\ref{tab:cells}). 

The simulated passive-impact curves build during posting and relax toward a nonzero permanent component. On the front-month calibration of Section~\ref{sec:calibration}, a passive schedule of $100$ AES units posted over $50$~s displaces the weighted mid by $1.5\times10^{-2}$ index points, about 6 percent of a tick, permanently and in proportion to the posted volume. This is 22 times smaller than the impact of the same volume executed with market orders, and about 2.2 times larger in thin-book than in thick-book sessions. Finally, the marginal impact of a market order remains positive at every level-one imbalance while the micro-price drift changes sign (Section~\ref{subsec:imbalance}).

Extensions include cross-exciting Hawkes flow, queue-reactive market orders, richer sensitivity functions, and fill-adjusted execution costs.

\section*{Acknowledgments}

We are sincerely grateful to Jesse Davis, Youssef Ouazzani Chahdi and Mathieu Rosenbaum for many insightful and stimulating discussions, from which this work has greatly benefited.
Yadh Hafsi acknowledges support from the Chair AIDA-F with Bank of America at École Polytechnique and from the Institut Europlace de Finance (IEF).
Grégoire Szymanski gratefully acknowledges support from the ILB Chair \textit{Artificial Intelligence and Quantitative Methods for Finance} at University Paris Dauphine-PSL.

{
\sloppy

\printbibliography
}

\appendix

\section{Sum-of-Exponential Kernels}
\label{app:sum-of-exponential-kernels}
\begin{lemma}[Explicit solution of a linear Volterra convolution equation]
Let $\phi:[0,\infty)\to\mathbb{C}$ be given by
$$
\phi(t)=\sum_{i=1}^n c_i e^{-\beta_i t},\quad \forall t \geq 0,
$$
with $c_i,\beta_i\in\mathbb{C}$. Let $\psi$ be the unique locally integrable solution of the Volterra equation $\psi*\phi+\phi=\psi$ on $[0,\infty)$ It is continuous and given by $\psi=\sum_{n\ge1}\phi^{*n}$. Define
$$
Q(s):=\prod_{i=1}^n (s+\beta_i),
\qquad
P(s):=\sum_{i=1}^n c_i\prod_{j\neq i}(s+\beta_j),
\qquad
D(s):=Q(s)-P(s),
$$
and $\sigma:=\max_{i}\bigl(-\Re(\beta_i)\bigr)+\sum_{i=1}^n|c_i|$.
Then, for all $s\in\mathbb{C}$ with $\Re(s)>\sigma$, the Laplace transforms of $\phi$ and $\psi$ converge absolutely, $1-\Phi(s)\neq0$, and
$$
\Psi(s):=\mathcal{L}\{\psi\}(s)=\frac{\Phi(s)}{1-\Phi(s)}=\frac{P(s)}{D(s)},
\qquad
\Phi(s):=\mathcal{L}\{\phi\}(s)=\sum_{i=1}^n\frac{c_i}{s+\beta_i}.
$$
Moreover, if the zeros $r_1,\dots,r_{n_r}$ of $D$ are all simple (so $D(r_k)=0$ and
$D'(r_k)\neq 0$), then
$$
\psi(t)=\sum_{k=1}^{n_r} A_k e^{r_k t},
\qquad
A_k=\frac{P(r_k)}{D'(r_k)},\quad \forall t \geq 0.
$$
If $D$ has zeros with multiplicity $>1$, then $\psi$ is a finite sum of terms
$t^\ell e^{rt}$ corresponding to the partial fraction expansion of $P(s)/D(s)$.
\end{lemma}

\begin{proof}
Write $C:=\sum_{i=1}^n|c_i|$ and $a:=\max_i(-\Re(\beta_i))$, so that $|\phi(t)|\le Ce^{at}$ for all $t\ge0$ and $\sigma=a+C$. By induction, $|\phi^{*n}(t)|\le C^n t^{n-1}e^{at}/(n-1)!$ for all $n\ge1$ and $t\ge0$. Hence $\sum_{n\ge1}\phi^{*n}$ converges uniformly on compact sets to a continuous function $\psi$ with $|\psi(t)|\le Ce^{\sigma t}$, and $\psi*\phi=\sum_{n\ge2}\phi^{*n}=\psi-\phi$. If $\tilde\psi$ is another locally integrable solution, then $\delta:=\psi-\tilde\psi$ satisfies $\delta=\delta*\phi$, so $e^{-at}|\delta(t)|\le C\int_0^t e^{-au}|\delta(u)|\,\ud u$, and Gronwall's lemma gives $\delta=0$ almost everywhere.

Now fix $s\in\mathbb{C}$ with $\Re(s)>\sigma$. By the bounds above, the Laplace integrals of $\phi$, $\psi$ and $\psi*\phi$ converge absolutely at $s$, so Fubini's theorem gives $\mathcal{L}\{\psi*\phi\}(s)=\Psi(s)\Phi(s)$, and the Volterra equation yields $\Psi\Phi+\Phi=\Psi$. Since $\phi(t)=\sum_{i=1}^n c_i e^{-\beta_i t}$ and $\Re(s+\beta_i)\ge\Re(s)-a>C$ for all $i$, we have
$
\Phi(s)=\sum_{i=1}^n \frac{c_i}{s+\beta_i}
$
and $|\Phi(s)|\le C/(\Re(s)-a)<1$. Hence $1-\Phi(s)\neq0$ and $\Psi=\Phi/(1-\Phi)$.
Writing $\Phi=P/Q$, with
$
Q(s)=\prod_{i=1}^n (s+\beta_i)$ and $
P(s)=\sum_{i=1}^n c_i\prod_{j\neq i}(s+\beta_j),
$
we obtain
$$
\Psi(s)=\frac{P(s)/Q(s)}{1-P(s)/Q(s)}=\frac{P(s)}{Q(s)-P(s)}=\frac{P(s)}{D(s)},
\qquad \Re(s)>\sigma.
$$
Since $Q(s)\neq0$ and $1-\Phi(s)\neq0$ for $\Re(s)>\sigma$, the polynomial $D=Q(1-\Phi)$ has no zero in this half-plane, that is, $\Re(r_k)\le\sigma$ for all $k$. Moreover, $D$ is monic of degree $n$ and $P$ has degree at most $n-1$.
If the zeros $r_k$ of $D$ are simple, then $P(s)/D(s)$ has a partial fraction
decomposition
$$
\frac{P(s)}{D(s)}=\sum_{k=1}^{n_r} \frac{A_k}{s-r_k},
\qquad
A_k=\operatorname{Res}_{s=r_k}\frac{P(s)}{D(s)}=\frac{P(r_k)}{D'(r_k)}.
$$
The continuous function $f(t):=\sum_{k=1}^{n_r} A_k e^{r_k t}$ satisfies $\mathcal{L}\{f\}(s)=\sum_{k=1}^{n_r}A_k/(s-r_k)=P(s)/D(s)=\Psi(s)$ for $\Re(s)>\sigma$. Since $\psi$ and $f$ are continuous on $[0,\infty)$, of exponential order, and have the same Laplace transform on this half-plane, Lerch's theorem gives $\psi(t)=\sum_{k=1}^{n_r} A_k e^{r_k t}$ for all $t \geq 0$.
If $D$ has repeated zeros, the partial fractions contain terms $(s-r)^{-(\ell+1)}$,
which are the Laplace transforms of $t^\ell e^{rt}/\ell!$ for $\Re(s)>\Re(r)$, and the same argument yields the stated
exponential-polynomial form. This proves the lemma.
\end{proof}

\begin{lemma}[Explicit form of $\xi$]
\label{lem:xi}
Assume the setting of the previous lemma and, in addition, that $c_i\ge0$ and $\beta_i>0$ are real for all $i$ and
$
\eta:=\int_0^\infty \phi(u)\,\ud u=\sum_{i=1}^n \frac{c_i}{\beta_i}<1
$.
Define, for all $t \geq 0$,
$
\xi(t)=1+\Bigl(1+\int_0^\infty \psi(u)\,\ud u\Bigr)\int_t^\infty \phi(s)\,\ud s
$.
Then
$$
\xi(t)
=
1+\frac{1}{1-\eta}\int_t^\infty \phi(s)\,\ud s
=
1+\frac{1}{1-\sum_{j=1}^n c_j/\beta_j}
\sum_{i=1}^n \frac{c_i}{\beta_i}e^{-\beta_i t}.
$$
\end{lemma}

\begin{proof}
Since $c_i\ge0$ and $\beta_i>0$, $\phi$ is nonnegative and integrable with $\int_0^\infty\phi(u)\,\ud u=\eta$. By the previous lemma, $\psi=\sum_{n\ge1}\phi^{*n}$, where each $\phi^{*n}$ is nonnegative and, by Tonelli's theorem and induction, $\int_0^\infty\phi^{*n}(u)\,\ud u=\eta^n$. Tonelli's theorem again gives
$
\int_0^\infty \psi(u)\,\ud u
=
\sum_{n\ge1}\eta^n
=
\eta/(1-\eta)
$.
Hence
$
1+\int_0^\infty \psi(u)\,\ud u= (1-\eta)^{-1}
$.
Also, since $\phi(t)=\sum_{i=1}^n c_i e^{-\beta_i t}$ and $\beta_i>0$,
$$
\int_t^\infty \phi(s)\,\ud s
=
\sum_{i=1}^n c_i\int_t^\infty e^{-\beta_i s}\,\ud s
=
\sum_{i=1}^n \frac{c_i}{\beta_i}e^{-\beta_i t}.
$$
Substituting these two identities into the definition of $\xi$ yields the stated formula.
\end{proof}

To determine when transient impact falls to a proportion $\alpha$ of its initial value, we need to solve for $t$ such that
$\xi(t) - 1 = \alpha(\xi(0) - 1)$,
that is $\sum_{i=1}^n \frac{c_i}{\beta_i}e^{-\beta_i t}
  =
  \alpha \sum_{i=1}^n \frac{c_i}{\beta_i}$.

\section{Proofs of the Price Decomposition}
\label{app:proofs-price-decomposition}

\begin{proof}[Proof of Theorem~\ref{thm:general_propagator}]
Fix $t\in[0,\infty)$. For $T\in(0,\infty)$ and $s\in[0,T]$, write
$
\kappa(q_{s-}^{a/b})
=
\bar\kappa+(\kappa(q_{s-}^{a/b})-\bar\kappa)
$.
Inserting this decomposition into \eqref{eq:price_def} gives
$$
P_t = P_0 +
\lim_{T\to\infty}
\Big(
\bar\kappa\,\E[D_T\mid\Fc_t]
+\E\left[
\int_0^T\left(\kappa(q_{s-}^a)-\bar\kappa\right)\ud N_s^a
-
\int_0^T\left(\kappa(q_{s-}^{b})-\bar\kappa\right)\ud N_s^b
\ \middle|\ \Fc_t
\right]
\Big).
$$
Under $\|\phi\|_{L^1}<1$, the Hawkes resolvent argument gives, for all $t \geq 0$,
$$
\lim_{T\to\infty}\E[D_T\mid\Fc_t]
=
\int_0^t\xi(t-s)\,\ud D_s,
$$
as in the classical propagator derivation \cite{jaisson2015market}, which yields \eqref{eq:general_propagator_decomp}. For the compensator form, fix $x\in\{a,b\}$ and define $M_u^{x}:=N_u^{x}-\int_0^u\lambda_r^{x}\,\ud r$ for all $u \geq 0$. Then $(M_u^{x})_{u\ge0}$ is an $\Fc$-martingale. Hence, for all $T > t$, we have
$$
\E\left[
\int_0^TY_s^{x}\,\ud N_s^{x}
\ \middle|\ \Fc_t
\right]
=
\int_0^tY_s^{x}\,\ud N_s^{x}
+
\E\left[
\int_t^TY_s^{x}\lambda_s^{x}\,\ud s
\ \middle|\ \Fc_t
\right]
$$
where we write $
Y_s^{x}:=\kappa(q_{s-}^{x})-\bar\kappa
$
for all $s\in[0,T]$. By boundedness of $\kappa$ and the preceding well-posedness assumptions, the limit in \eqref{eq:price_def} is finite in this model class \cite{ouazzaniChahdiRosenbaumSzymanskiPassive}. Letting $T\to\infty$ in the finite-horizon identity yields the compensator representation of $\mathcal{R}_t$.
\end{proof}

\begin{proof}[Proof of Corollary~\ref{cor:main_propagator}]
Fix $t\in[0,\infty)$. For $T\in(0,\infty)$ and $s\in[0,T]$, inserting
$
\kappa(q_{s-}^{x})=\kappa_0+\kappa_1 q_{s-}^{x}
$ with $x\in\{a,b\}$
into \eqref{eq:price_def} gives
$$
\E\left[
\int_0^T\kappa(q_{s-}^a)\,\ud N_s^a
-\int_0^T\kappa(q_{s-}^{b})\,\ud N_s^b
\ \middle|\ \Fc_t
\right]
=
\kappa_0\,\E[D_T\mid\Fc_t]
+
\kappa_1\,\E\left[
\int_0^T q_{s-}^a\,\ud N_s^a
-\int_0^T q_{s-}^b\,\ud N_s^b
\ \middle|\ \Fc_t
\right].
$$
The first term converges, as $T\to\infty$, to
$
\kappa_0\int_0^t\xi(t-s)\,\ud D_s
$
by the same Hawkes resolvent argument as above. For the second term, fix $x\in\{a,b\}$. Since
$
M_u^{x}:=N_u^{x}-\int_0^u\lambda_r^{x}\,\ud r
$
is an $\Fc$-martingale, the compensator identity gives, for all $T > t$,
$$
\E\left[
\int_0^T q_{s-}^{x}\,\ud N_s^{x}
\ \middle|\ \Fc_t
\right]
=
\int_0^t q_{s-}^{x}\,\ud N_s^{x}
+
\E\left[
\int_t^T q_{s-}^{x}\lambda_s^{x}\,\ud s
\ \middle|\ \Fc_t
\right].
$$
Applying this identity for $x=a$ and $x=b$, subtracting the two expressions, and letting $T\to\infty$ by dominated convergence under the stated integrability condition $\E_\pi[q^{x}\lambda^{x}]<\infty$, for all $x\in\{a,b\}$, gives \eqref{eq:main_propagator_decomp}.
\end{proof}

\section{Additional Results}
Proposition~\ref{prop:explicit_bound} gives the closed form of the anticipation term in the decoupled case and motivates the $J$-control regression of Section~\ref{subsec:affine-queue-sensitivity}. 

\begin{prop}\label{prop:explicit_bound}
Suppose that
$
\lambda^L(q)-\lambda^C(q)=c_\lambda q+d_\lambda
$ with $c_\lambda<0$. Assume also that $\phi=0$. Then, for all $t \geq 0$,
$$
\int_t^\infty
\E\left[q_s^a\lambda_s^a-q_s^b\lambda_s^b\mid\Fc_t\right]\ud s
=
\frac{\mu}{-c_\lambda}\left(q_t^a-q_t^b\right),
$$
and therefore no bound depending only on the model parameters can hold on an unbounded queue state space.
\end{prop}

\begin{proof}
Because $\phi=0$, the market-order processes are independent Poisson processes with constant intensity $\mu$. Now fix $t\ge0$ and $x\in\{a,b\}$. Since
$$
q_s^{x}
=
q_t^{x}
+(L_s^{x}-L_t^{x})
-(C_s^{x}-C_t^{x})
-(N_s^{x}-N_t^{x}),
\quad \forall s \geq t,
$$
and since the predictable intensities of $L^{x}$, $C^{x}$, $N^{x}$ are $\lambda^L(q^{x})$, $\lambda^C(q^{x})$, and $\mu$, respectively, the compensated-process identity gives
$$
\E[q_s^{x}\mid\Fc_t]
=
q_t^{x}
+
\int_t^s
\E\left[\lambda^L(q_u^{x})-\lambda^C(q_u^{x})-\mu\mid\Fc_t\right]\ud u.
$$
Using the affine drift assumption,
$
\lambda^L(q)-\lambda^C(q)=c_\lambda q+d_\lambda
$,
we obtain, for
$
e_x(s):=\E[q_s^{x}\mid\Fc_t]
$
the linear ODE
$$
e_x'(s)=c_\lambda e_x(s)+d_\lambda-\mu,
\quad \forall s \geq t,
\quad e_x(t)=q_t^{x}.
$$
Subtracting the equations for $x=a$ and $x=b$ eliminates the constant term $d_\lambda-\mu$. Hence
$$
\frac{\ud}{\ud s}\left(e_a(s)-e_b(s)\right)
=
c_\lambda\left(e_a(s)-e_b(s)\right),
\quad
e_a(t)-e_b(t)=q_t^a-q_t^b.
$$
Therefore
$$
e_a(s)-e_b(s)=e^{c_\lambda(s-t)}\left(q_t^a-q_t^b\right).
$$
Since $c_\lambda<0$ and $\lambda_s^a=\lambda_s^b=\mu$, we get
$$
\begin{aligned}
\int_t^\infty
\E\left[q_s^a\lambda_s^a-q_s^b\lambda_s^b\mid\Fc_t\right]\ud s
&=
\mu\int_t^\infty\E[q_s^a-q_s^b\mid\Fc_t]\,\ud s=
\mu(q_t^a-q_t^b)\int_t^\infty e^{c_\lambda(s-t)}\,\ud s =
\frac{\mu}{-c_\lambda}(q_t^a-q_t^b).
\end{aligned}
$$
The expression therefore depends on the current queue imbalance. On an unbounded state space, $|q_t^a-q_t^b|$ cannot be bounded by model parameters alone. Thus the claimed parameter-only bound cannot hold in general.
\end{proof}

\section{Calibration Details}
\label{app:calibration-details}
\paragraph{Queue-size normalization.} 
Although the theoretical model normalizes event sizes to one, empirical queue sizes are converted to average-event-size (AES) units before queue-reactive calibration and construction of the queue-weighted signal $H$. If $v_m$ is the raw displayed size of a best-quote limit-order insertion, we set
$$
\widehat{\mathrm{AES}}
=
\frac{1}{n_L}\sum_{k=1}^{n_L}v_k,
$$
where the sum runs over best-quote insertion messages in the cleaned sample used for calibration. Raw queue volumes are then mapped to the dimensionless state
$$
q_t^{x}
=
\left\lceil
\frac{q_t^{x,\mathrm{raw}}}{\widehat{\mathrm{AES}}}
\right\rceil,
\quad \forall x\in\{a,b\}.
$$
The Hawkes likelihood is fitted to aggressive event times, while the queue-reactive estimators use AES-normalized queue states and event counts. This is the empirical counterpart of the unit-size convention in \eqref{eq:queues_def}. The ceiling maps every strictly positive displayed queue to a positive integer state, preserving compatibility with the birth-death lattice used by the queue-reactive model. Empty or locked/crossed quote configurations that remain after cleaning are excluded from exposure calculations rather than forced into the positive-state model. Since $q^{x,\mathrm{raw}}$ and $\widehat{\mathrm{AES}}$ are both measured in contracts, $q^{x}$ is dimensionless and measured in units of a typical best-quote insertion, so the coefficient on $H$ has units of price per market order per AES unit.

\paragraph{Hawkes likelihood on intraday bins.}
Let $\tau_1<\cdots<\tau_n$ be the aggressive event times of side $x$ in $I_{d,r}$, and write $u_i=\tau_i-t_{d,r-1}\in[0,h)$. For $j=1,2,3$, introduce the local exponential state
$$
X_{u}^{j,x,d,r}
=
X_{0}^{j,x,d,r}e^{-\beta_j^{x,d,r}u}
+
\sum_{i:\,u_i<u}e^{-\beta_j^{x,d,r}(u-u_i)},
\quad \forall u\in[0,h],
$$
where $X_{0}^{j,x,d,r}$ is either obtained by carrying the pre-bin history or set to zero in the reset likelihood. The predictable intensity at event time $u_i$ is therefore
$$
\lambda_{u_i}^{x,d,r}(\theta)
=
\mu^{x,d,r}
+
\sum_{j=1}^{3}c_j^{x,d,r}
\left(
X_{0}^{j,x,d,r}e^{-\beta_j^{x,d,r}u_i}
+
\sum_{k<i}e^{-\beta_j^{x,d,r}(u_i-u_k)}
\right).
$$
The conditional log-likelihood on the bin is
$$
\ell_{d,r}^{x}(\theta)
=
\sum_{i=1}^{n}\log \lambda_{u_i}^{x,d,r}(\theta)
-
\int_{0}^{h}\lambda_u^{x,d,r}(\theta)\,\ud u,
$$
with exact compensator
$$
\begin{aligned}
\int_{0}^{h}\lambda_u^{x,d,r}(\theta)\,\ud u
&=
\mu^{x,d,r}h
+
\sum_{j=1}^{3}
\frac{c_j^{x,d,r}X_{0}^{j,x,d,r}}{\beta_j^{x,d,r}}
\left(1-e^{-\beta_j^{x,d,r}h}\right)
+
\sum_{i=1}^{n}\sum_{j=1}^{3}
\frac{c_j^{x,d,r}}{\beta_j^{x,d,r}}
\left(1-e^{-\beta_j^{x,d,r}(h-u_i)}\right).
\end{aligned}
$$
The formula in the reset likelihood is recovered by taking $X_0^{j,x,d,r}=0$. Using this explicit expression avoids time-discretization error in the Hawkes likelihood.

\paragraph{Queue-reactive likelihood.}
Conditional on the observed queue path and under the queue-reactive Poisson specification, the contribution of event type $E$ to the log-likelihood is, up to an additive constant,
$$
\ell_E(\lambda^E)
=
\sum_{q\in\mathcal Q}
\left\{
N^E(q)\log \lambda^E(q)-\lambda^E(q)\mathcal T(q)
\right\}.
$$
Maximizing this expression state by state gives the estimator of Section~\ref{subsec:queue-reactive-calibration}. States with $\mathcal T(q)=0$ carry no likelihood information and are excluded from the nonparametric estimate. For rarely visited states, the estimator has high variance. Accordingly, plotted diagnostics are restricted to states with sufficient exposure or displayed after applying the stated filtering rule. This regularizes only the displayed finite-sample estimates, not the theoretical monotonicity assumptions.

\section{Additional Empirical Results}
\label{app:additional-empirics}
This appendix collects the tables and figures of the front-month sample referred to in Sections~\ref{sec:calibration} to~\ref{sec:passive-impact-curves}.

\begin{figure}[htb!]
    \centering

    \includegraphics[width=0.72\textwidth]{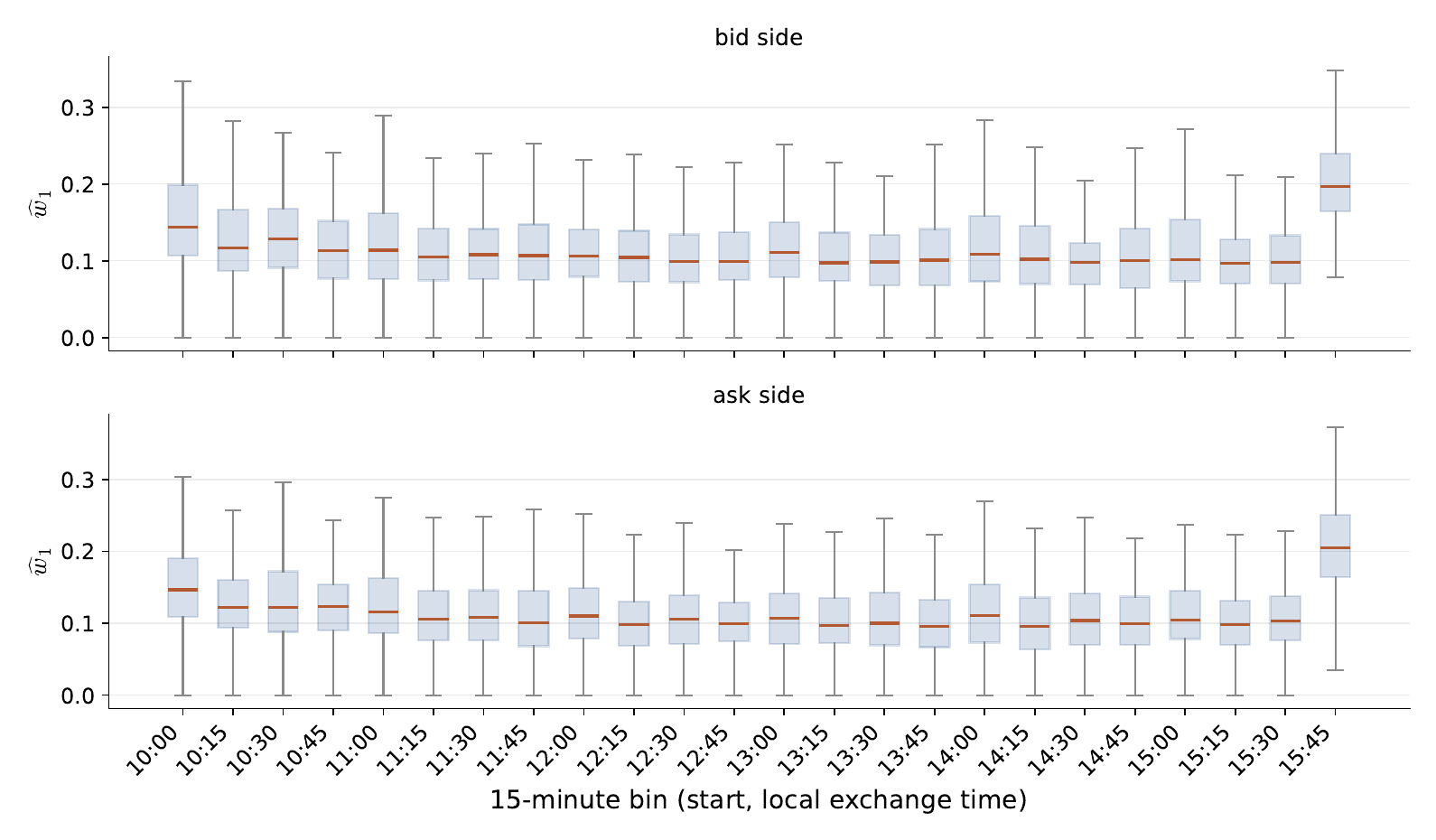}

    \vspace{0.2cm}

    \includegraphics[width=0.72\textwidth]{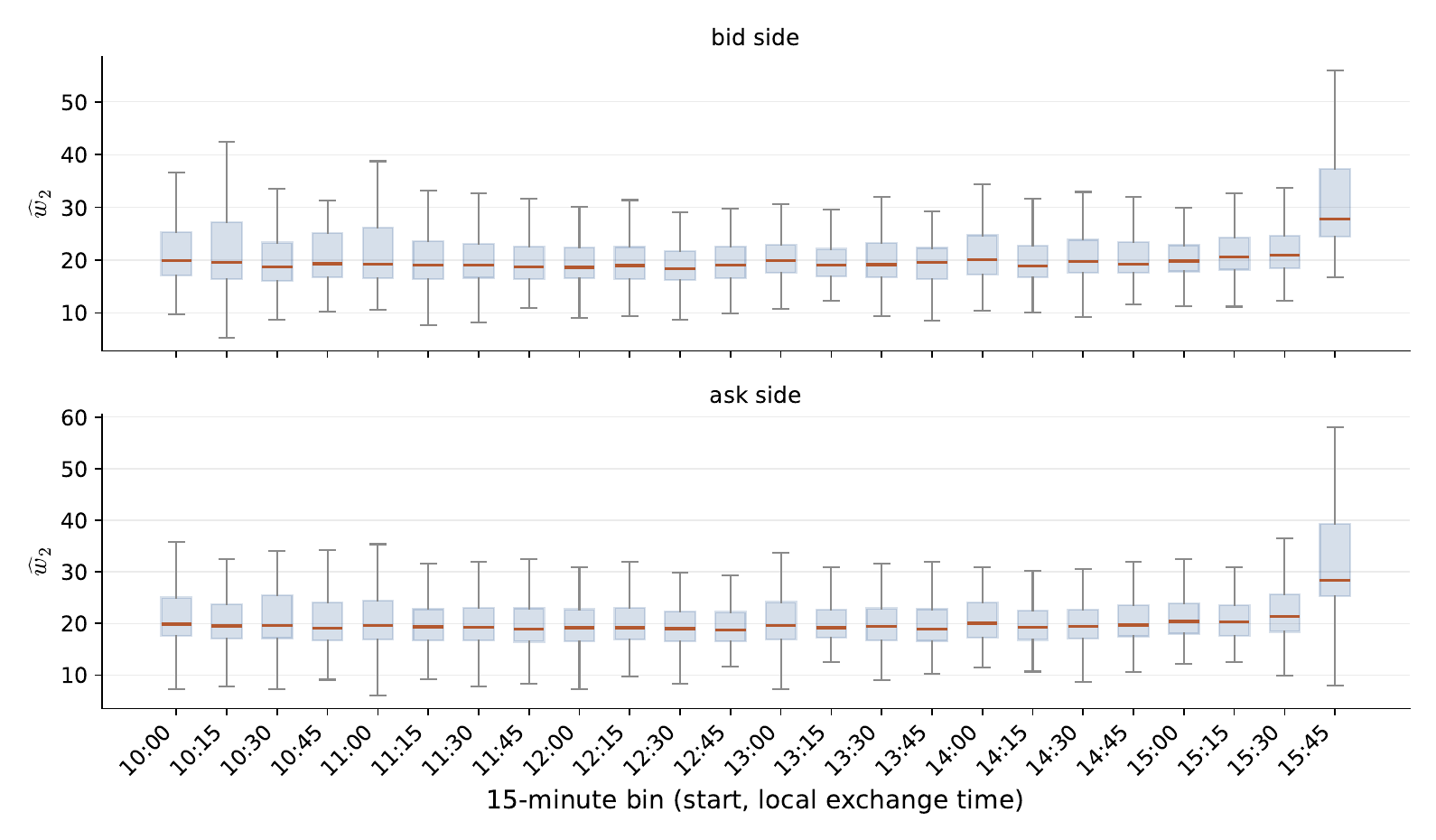}

    \vspace{0.2cm}

    \includegraphics[width=0.72\textwidth]{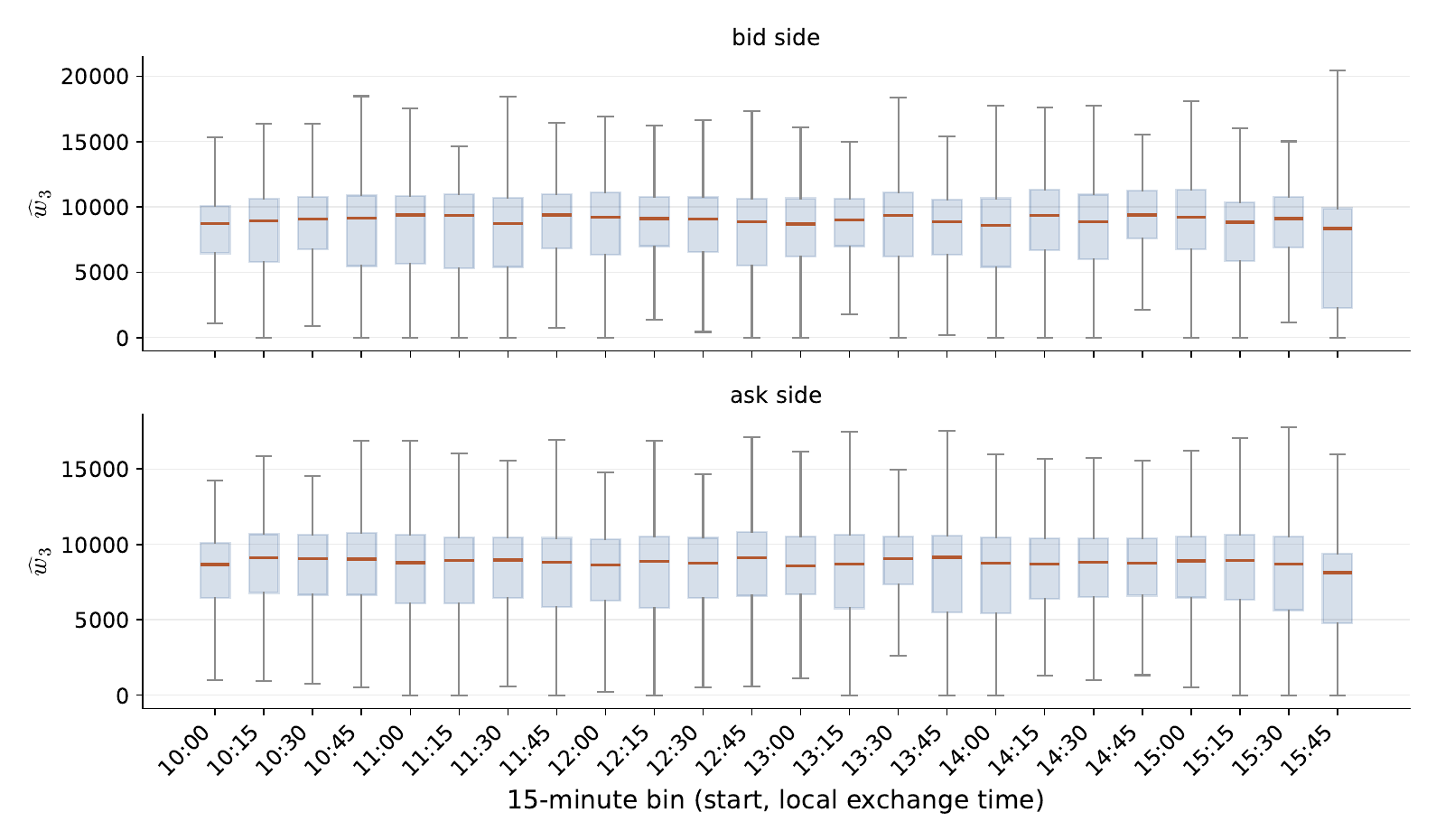}

    \caption{Cross-day distribution of the fitted propagator weights $\widehat w_1$, $\widehat w_2$ and $\widehat w_3$.}
    \label{fig:hawkes_sumexp_weights}
\end{figure}

\begin{table}[htb!]

\centering\footnotesize
\resizebox{\ifdim\width>\textwidth\textwidth\else\width\fi}{!}{%
\begin{tabular}{c c c c c c c}
\toprule
$\delta$ (s) & $\widehat\kappa_1$ (pooled) & Fama MacBeth $t$ & $R^2_{\mathrm{OOS}}$ propagator & $R^2_{\mathrm{OOS}}$ affine & $R^2_{\mathrm{OOS}}$ log & gain \\
\midrule
30 & $-1.01\times10^{-5}$ & $-11.4$ & 0.403 & 0.423 & 0.447 & $+0.021$ \\
60 & $-9.67\times10^{-6}$ & $-10.9$ & 0.420 & 0.438 & 0.456 & $+0.018$ \\
212 & $-8.07\times10^{-6}$ & $-7.8$ & 0.422 & 0.437 & 0.444 & $+0.015$ \\
300 & $-7.20\times10^{-6}$ & $-6.0$ & 0.414 & 0.425 & 0.431 & $+0.012$ \\
\bottomrule
\end{tabular}}
\caption{Horizon robustness of the affine specification on the front-month sample (weighted mid, ES 2025, 193 test days). $t$ is for the day-by-day mean of $\widehat\kappa_1$. Gain is affine minus propagator out-of-sample $R^2$.}
\label{tab:delta-robustness-clean}
\end{table}

\begin{figure}[htb!]

    \centering
    \includegraphics[width=0.5\textwidth]{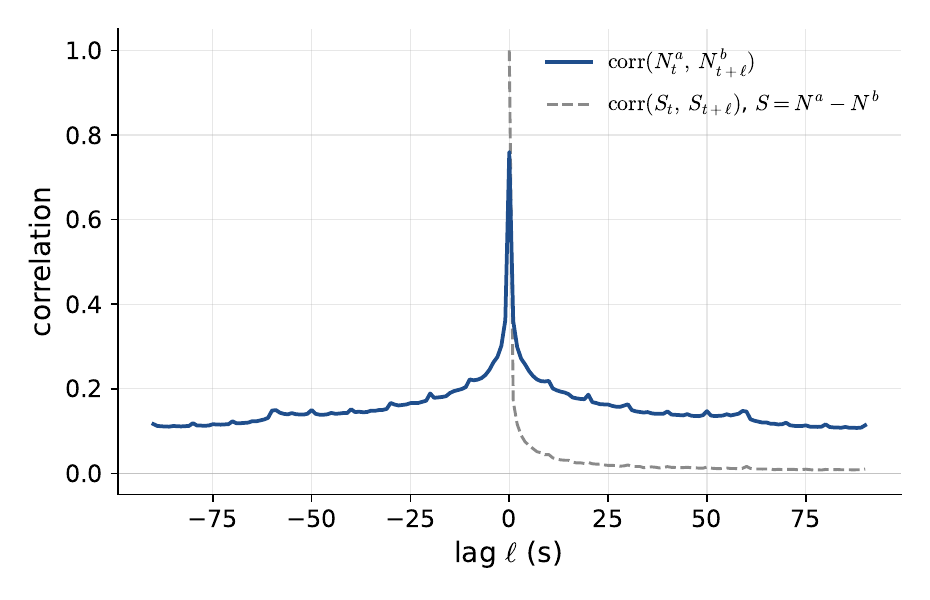}
    \caption{Cross-correlogram of buy and sell aggressive order flow on the front-month stream (one-second counts, ES 2025).}
    \label{fig:xcorr-flow-front}
\end{figure}

\begin{table}[htb!]

\centering\footnotesize
\caption{Paired daily tests of the gains over the propagator (differences in out-of-sample $R^2$ at the $3\sigma$ truncation): mean, Fama MacBeth and Newey West $t$, share of positive days, binomial and Wilcoxon $p$ values, bootstrap $95\%$ interval, and test days. Equity-index cells, 2023 and 2024.}
\label{tab:significance}
\resizebox{\ifdim\width>\textwidth\textwidth\else\width\fi}{!}{%
\begin{tabular}{l l r r r r r r c r}
\toprule
cell & model & mean $d$ & FM $t$ & NW $t$ & share $d>0$ & binomial $p$ & Wilcoxon $p$ & bootstrap 95\% CI & $n$ \\
\midrule
ES 2023 & affine & 0.002 & 2.73 & 2.62 & 0.598 & 0.007 & $<$0.001 & [0.001, 0.004] & 199 \\
 & log & 0.002 & 1.31 & 1.61 & 0.513 & 0.777 & 0.353 & [-0.000, 0.005] & 199 \\
 & pair & 0.035 & 6.86 & 5.03 & 0.844 & $<$0.001 & $<$0.001 & [0.024, 0.050] & 199 \\
 & full dictionary ridge & 0.069 & 10.22 & 7.92 & 0.899 & $<$0.001 & $<$0.001 & [0.052, 0.088] & 199 \\
\midrule
ES 2024 & affine & 0.006 & 4.97 & 4.45 & 0.735 & $<$0.001 & $<$0.001 & [0.003, 0.009] & 196 \\
 & log & 0.001 & 1.38 & 1.13 & 0.531 & 0.432 & 0.195 & [-0.001, 0.005] & 196 \\
 & pair & 0.063 & 12.69 & 9.97 & 0.878 & $<$0.001 & $<$0.001 & [0.052, 0.076] & 196 \\
 & full dictionary ridge & 0.106 & 15.24 & 13.05 & 0.898 & $<$0.001 & $<$0.001 & [0.092, 0.117] & 196 \\
\midrule
NQ 2023 & affine & 0.003 & 2.76 & 2.30 & 0.633 & $<$0.001 & $<$0.001 & [-0.000, 0.005] & 199 \\
 & log & -0.000 & -0.05 & -0.04 & 0.482 & 0.671 & 0.760 & [-0.002, 0.001] & 199 \\
 & pair & 0.014 & 6.48 & 5.59 & 0.734 & $<$0.001 & $<$0.001 & [0.008, 0.019] & 199 \\
 & full dictionary ridge & 0.058 & 12.70 & 12.64 & 0.879 & $<$0.001 & $<$0.001 & [0.051, 0.065] & 199 \\
\midrule
NQ 2024 & affine & 0.009 & 4.81 & 3.96 & 0.776 & $<$0.001 & $<$0.001 & [0.003, 0.014] & 196 \\
 & log & 0.011 & 6.65 & 4.63 & 0.745 & $<$0.001 & $<$0.001 & [0.006, 0.016] & 196 \\
 & pair & 0.022 & 9.51 & 6.33 & 0.867 & $<$0.001 & $<$0.001 & [0.014, 0.031] & 196 \\
 & full dictionary ridge & 0.064 & 12.22 & 10.96 & 0.903 & $<$0.001 & $<$0.001 & [0.053, 0.078] & 196 \\
\midrule
RTY 2023 & affine & 0.015 & 7.35 & 6.22 & 0.819 & $<$0.001 & $<$0.001 & [0.010, 0.020] & 199 \\
 & log & 0.021 & 7.40 & 6.24 & 0.769 & $<$0.001 & $<$0.001 & [0.013, 0.029] & 199 \\
 & pair & 0.028 & 8.92 & 7.13 & 0.809 & $<$0.001 & $<$0.001 & [0.019, 0.038] & 199 \\
 & full dictionary ridge & 0.190 & 18.83 & 13.12 & 0.950 & $<$0.001 & $<$0.001 & [0.160, 0.216] & 199 \\
\midrule
RTY 2024 & affine & 0.011 & 5.87 & 4.53 & 0.735 & $<$0.001 & $<$0.001 & [0.005, 0.017] & 196 \\
 & log & 0.013 & 5.03 & 3.84 & 0.730 & $<$0.001 & $<$0.001 & [0.004, 0.022] & 196 \\
 & pair & 0.078 & 12.90 & 8.78 & 0.888 & $<$0.001 & $<$0.001 & [0.050, 0.104] & 196 \\
 & full dictionary ridge & 0.176 & 17.72 & 11.66 & 0.929 & $<$0.001 & $<$0.001 & [0.136, 0.212] & 196 \\
\midrule
YM 2023 & affine & 0.012 & 4.13 & 3.86 & 0.719 & $<$0.001 & $<$0.001 & [0.006, 0.018] & 199 \\
 & log & 0.006 & 2.70 & 2.59 & 0.623 & 0.001 & 0.001 & [0.003, 0.010] & 199 \\
 & pair & 0.016 & 4.47 & 4.15 & 0.673 & $<$0.001 & $<$0.001 & [0.009, 0.023] & 199 \\
 & full dictionary ridge & 0.081 & 8.28 & 7.16 & 0.884 & $<$0.001 & $<$0.001 & [0.067, 0.098] & 199 \\
\midrule
YM 2024 & affine & 0.009 & 4.19 & 3.42 & 0.709 & $<$0.001 & $<$0.001 & [0.003, 0.015] & 196 \\
 & log & 0.008 & 4.30 & 3.38 & 0.658 & $<$0.001 & $<$0.001 & [0.003, 0.013] & 196 \\
 & pair & 0.073 & 13.84 & 8.20 & 0.898 & $<$0.001 & $<$0.001 & [0.048, 0.100] & 196 \\
 & full dictionary ridge & 0.111 & 14.00 & 8.82 & 0.867 & $<$0.001 & $<$0.001 & [0.076, 0.149] & 196 \\
\bottomrule
\end{tabular}}
\end{table}

\begin{table}[htb!]
\centering\footnotesize
\caption{Paired daily tests of the gains over the propagator, as in the previous table. 2025 cells, equity-index and Treasury futures.}
\label{tab:significance-2025}
\resizebox{\ifdim\width>\textwidth\textwidth\else\width\fi}{!}{%
\begin{tabular}{l l r r r r r r c r}
\toprule
cell & model & mean $d$ & FM $t$ & NW $t$ & share $d>0$ & binomial $p$ & Wilcoxon $p$ & bootstrap 95\% CI & $n$ \\
\midrule
ES 2025 & affine & 0.011 & 3.98 & 4.07 & 0.699 & $<$0.001 & $<$0.001 & [0.006, 0.016] & 193 \\
 & log & 0.015 & 4.69 & 4.31 & 0.741 & $<$0.001 & $<$0.001 & [0.007, 0.022] & 193 \\
 & pair & 0.105 & 14.26 & 10.29 & 0.917 & $<$0.001 & $<$0.001 & [0.081, 0.133] & 193 \\
 & full dictionary ridge & 0.145 & 14.70 & 10.99 & 0.943 & $<$0.001 & $<$0.001 & [0.115, 0.178] & 193 \\
\midrule
NQ 2025 & affine & 0.008 & 2.18 & 2.05 & 0.684 & $<$0.001 & $<$0.001 & [0.000, 0.015] & 193 \\
 & log & 0.019 & 5.44 & 4.88 & 0.689 & $<$0.001 & $<$0.001 & [0.012, 0.026] & 193 \\
 & pair & 0.017 & 5.04 & 4.49 & 0.679 & $<$0.001 & $<$0.001 & [0.010, 0.023] & 193 \\
 & full dictionary ridge & 0.092 & 11.23 & 8.14 & 0.855 & $<$0.001 & $<$0.001 & [0.069, 0.119] & 193 \\
\midrule
RTY 2025 & affine & 0.007 & 3.31 & 3.13 & 0.689 & $<$0.001 & $<$0.001 & [0.003, 0.010] & 193 \\
 & log & 0.009 & 3.51 & 3.30 & 0.648 & $<$0.001 & $<$0.001 & [0.003, 0.014] & 193 \\
 & pair & 0.026 & 5.43 & 3.73 & 0.699 & $<$0.001 & $<$0.001 & [0.009, 0.046] & 193 \\
 & full dictionary ridge & 0.113 & 14.88 & 10.65 & 0.870 & $<$0.001 & $<$0.001 & [0.086, 0.141] & 193 \\
\midrule
TN 2025 & affine & 0.010 & 4.66 & 3.93 & 0.711 & $<$0.001 & $<$0.001 & [0.005, 0.015] & 194 \\
 & log & 0.010 & 4.48 & 3.70 & 0.691 & $<$0.001 & $<$0.001 & [0.004, 0.018] & 194 \\
 & pair & 0.018 & 5.70 & 4.29 & 0.706 & $<$0.001 & $<$0.001 & [0.008, 0.029] & 194 \\
 & full dictionary ridge & 0.064 & 8.38 & 7.83 & 0.814 & $<$0.001 & $<$0.001 & [0.050, 0.079] & 194 \\
\midrule
YM 2025 & affine & 0.011 & 3.60 & 3.98 & 0.751 & $<$0.001 & $<$0.001 & [0.008, 0.013] & 193 \\
 & log & 0.011 & 4.06 & 3.24 & 0.668 & $<$0.001 & $<$0.001 & [0.005, 0.018] & 193 \\
 & pair & 0.016 & 4.43 & 4.51 & 0.762 & $<$0.001 & $<$0.001 & [0.010, 0.023] & 193 \\
 & full dictionary ridge & 0.121 & 11.55 & 8.73 & 0.860 & $<$0.001 & $<$0.001 & [0.090, 0.152] & 193 \\
\midrule
ZB 2025 & affine & 0.007 & 2.78 & 2.24 & 0.675 & $<$0.001 & $<$0.001 & [0.002, 0.015] & 194 \\
 & log & 0.004 & 1.85 & 1.39 & 0.608 & 0.003 & 0.002 & [-0.001, 0.011] & 194 \\
 & pair & 0.009 & 3.18 & 2.43 & 0.655 & $<$0.001 & $<$0.001 & [0.002, 0.018] & 194 \\
 & full dictionary ridge & 0.063 & 8.99 & 7.33 & 0.871 & $<$0.001 & $<$0.001 & [0.048, 0.076] & 194 \\
\midrule
ZF 2025 & affine & 0.004 & 1.82 & 2.38 & 0.655 & $<$0.001 & $<$0.001 & [0.002, 0.007] & 194 \\
 & log & 0.000 & 0.15 & 0.16 & 0.598 & 0.008 & 0.006 & [-0.003, 0.004] & 194 \\
 & pair & 0.005 & 2.10 & 2.20 & 0.634 & $<$0.001 & $<$0.001 & [0.001, 0.010] & 194 \\
 & full dictionary ridge & 0.071 & 10.32 & 9.24 & 0.851 & $<$0.001 & $<$0.001 & [0.059, 0.084] & 194 \\
\midrule
ZN 2025 & affine & 0.013 & 5.46 & 4.39 & 0.727 & $<$0.001 & $<$0.001 & [0.006, 0.020] & 194 \\
 & log & 0.005 & 2.30 & 2.41 & 0.655 & $<$0.001 & $<$0.001 & [0.001, 0.009] & 194 \\
 & pair & 0.024 & 5.97 & 4.45 & 0.768 & $<$0.001 & $<$0.001 & [0.013, 0.037] & 194 \\
 & full dictionary ridge & 0.090 & 13.99 & 11.88 & 0.933 & $<$0.001 & $<$0.001 & [0.073, 0.103] & 194 \\
\bottomrule
\end{tabular}}
\end{table}

\begin{figure}[htb!]

  \centering
  \includegraphics[width=0.5\textwidth]{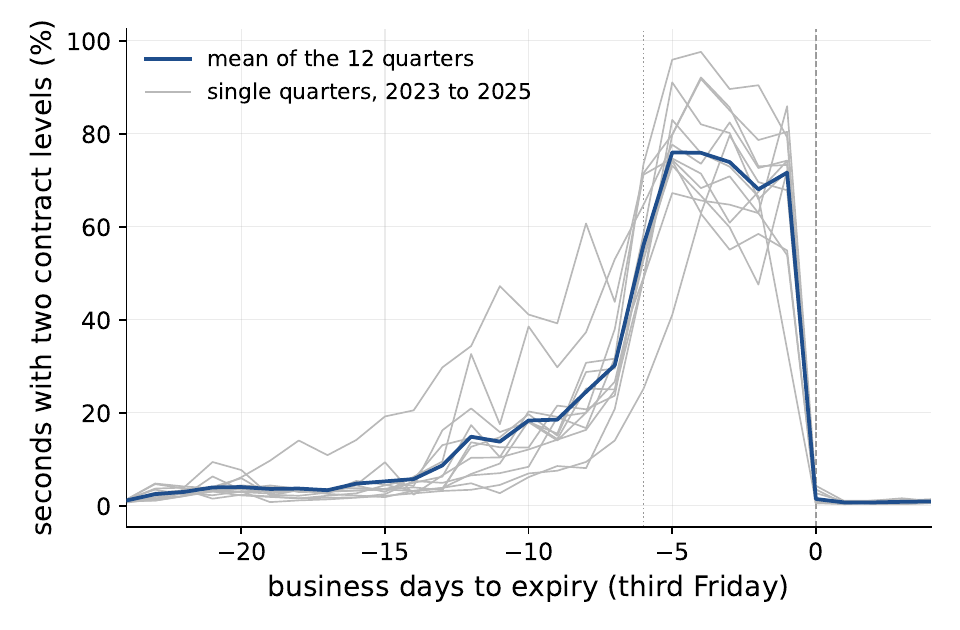}
  \caption{Share of one-second intervals with two contract price levels in the raw ES stream against business days to the quarterly expiry, for twelve quarters (grey) and their mean (blue). Dashed: expiry. Dotted: conventional volume roll date.}
  \label{fig:roll-mixing}
\end{figure}

\begin{figure}[htb!]

  \centering
  \includegraphics[width=0.72\textwidth]{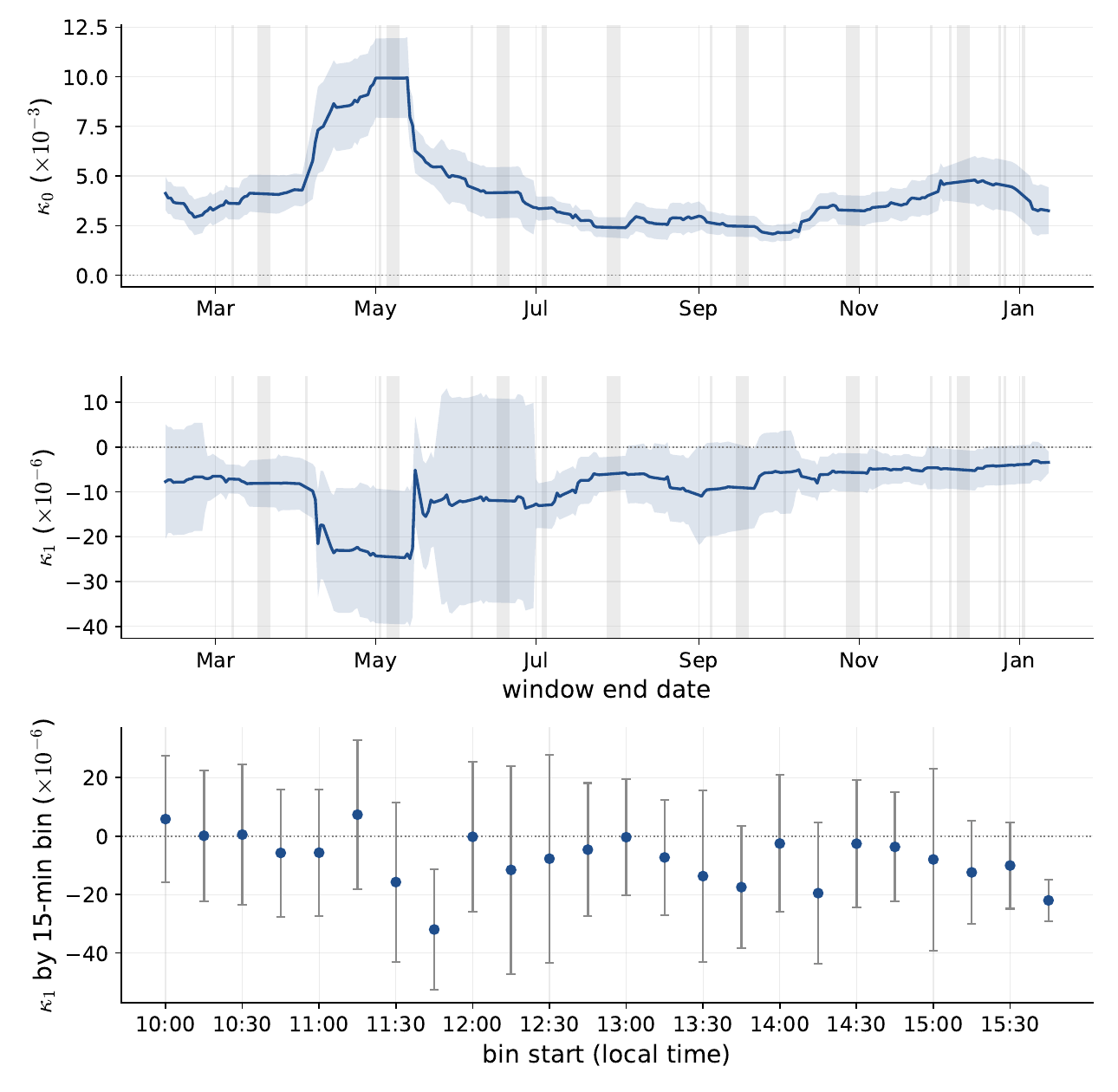}
  \caption{Stability of the affine coefficients, ES 2025. Top: $\kappa_0$ and $\kappa_1$ over rolling windows of twenty sessions with $95\%$ bands. Bottom: mean $\kappa_1$ by fifteen-minute bin with $95\%$ error bars.}
  \label{fig:stability}
\end{figure}

\begin{figure}[htb!]

  \centering
  \includegraphics[width=0.42\textwidth]{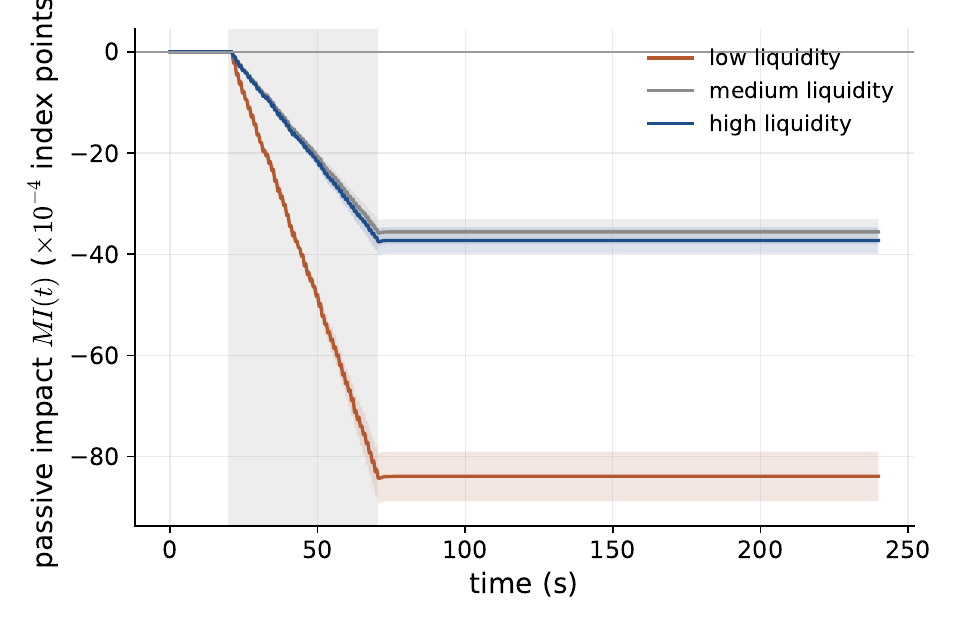}
  \qquad
  \includegraphics[width=0.42\textwidth]{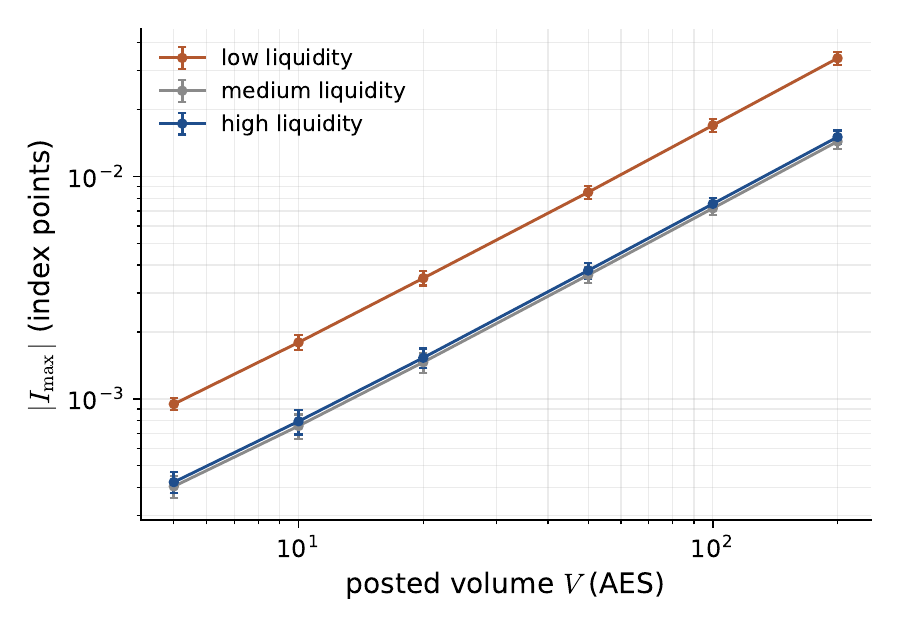}
  \caption{Passive impact by liquidity regime. Left: mean impact curve for $V=50$ AES over $50$ s. Right: peak impact against posted volume.}
  \label{fig:cf-regimes}
\end{figure}

\begin{figure}[htb!]

  \centering
  \includegraphics[width=0.8\textwidth]{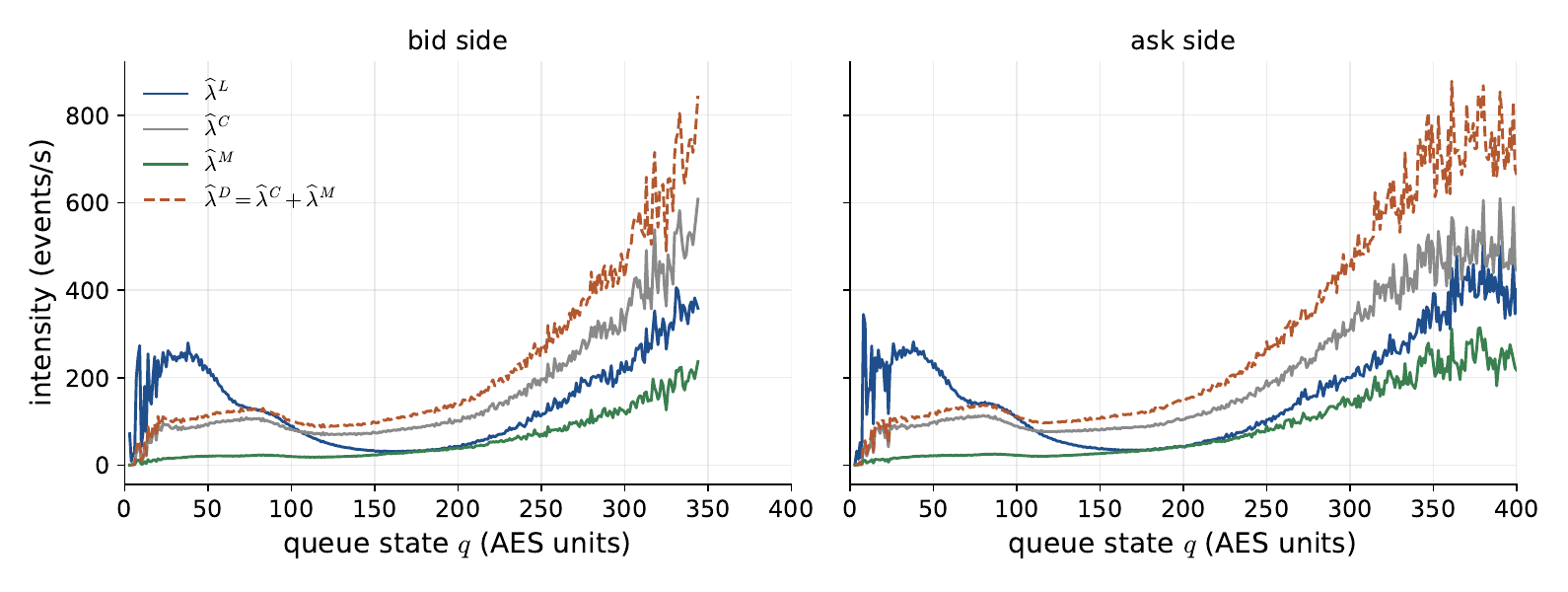}
  \caption{Queue-reactive intensities, ES 2025, aggregated three-level queue in AES units, one panel per side: insertions $\widehat\lambda^L$, cancellations $\widehat\lambda^C$, market depletions $\widehat\lambda^M$ and total depletions $\widehat\lambda^D$ (events per second, $q\le400$).}
  \label{fig:qrm-intensities-clean}
\end{figure}

\begin{figure}[htb!]
  \centering
  \includegraphics[width=0.8\textwidth]{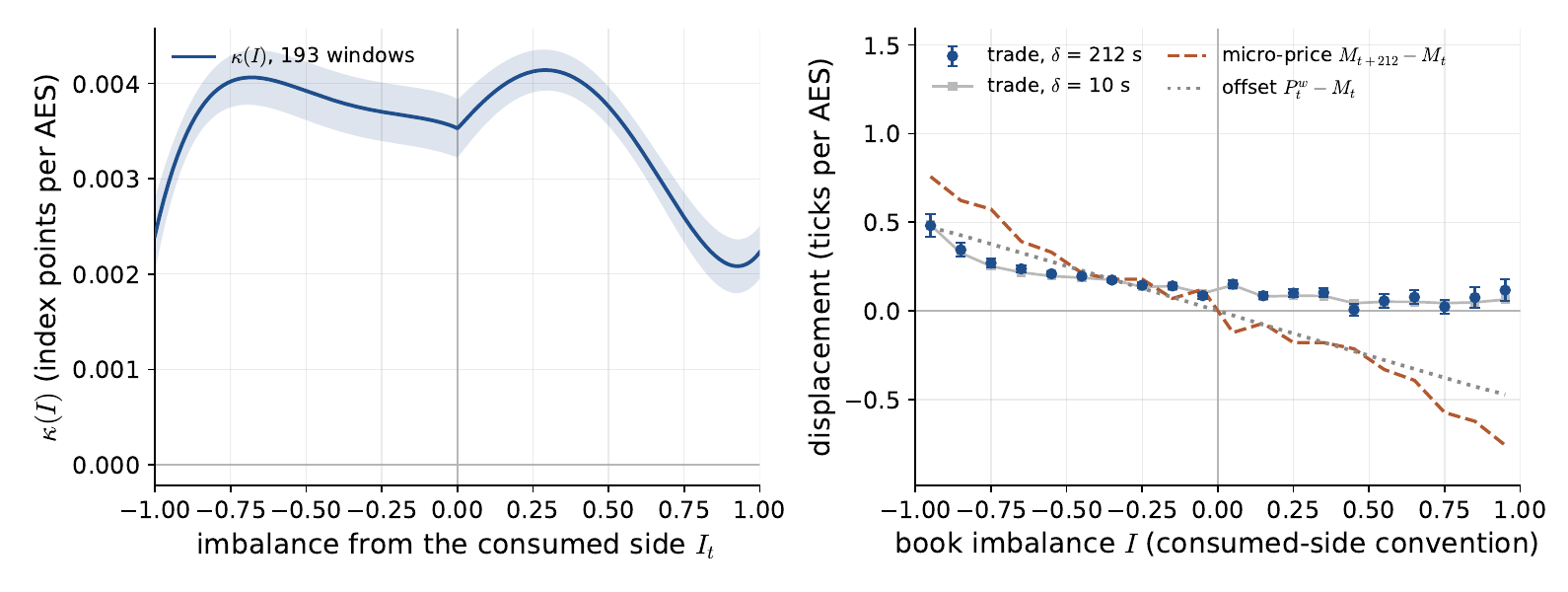}
  \caption{Left: fitted $\kappa(\iota)$ against the level-one imbalance seen from the consumed side, with a two-standard-error band. Right: displacement of the mid per AES unit of traded volume $212$ s and $10$ s after a trade, micro-price adjustment $\mathbb E[P^{\mathrm{mid}}_{t+212}-P^{\mathrm{mid}}_t\,|\,\iota]$ (dashed) and micro-price offset $P^{\mathrm{wmid}}_t-P^{\mathrm{mid}}_t$ (dotted).}
  \label{fig:kappa-imbalance}
\end{figure}

\end{document}